\documentclass[11pt,superscriptaddress,twocolumn,aps,nofootinbib,tightenlines,showpacs,floats]{revtex4-1}
\usepackage[usenames,dvipsnames]{color} \usepackage[margin=0.7 in]{geometry}
\usepackage{graphicx,amsmath,multirow,amssymb,mathtools,subfigure,slashed,float,color,booktabs,tikz,afterpage}

   \DeclareMathOperator{\gev}{GeV}          
     \newcommand{\cL}{{\cal L}} \newcommand{\cM}{{\cal M}}  \newcommand{\cO}{{\cal O}}   
\newcommand{\ep}{\epsilon} 
\newcommand{\half}{\frac12}

\newcommand{\pL}{\left(} \newcommand{\pR}{\right)} \newcommand{\bL}{\left[} \newcommand{\bR}{\right]}   \newcommand{\mL}{\left|} \newcommand{\mR}{\right|}
\newcommand{\beq}{\begin{equation}} \newcommand{\eeq}{\end{equation}}
\newcommand{\bea}{\begin{eqnarray}} \newcommand{\eea}{\end{eqnarray}}
\newcommand{\alg}[1]{\begin{align} \begin{split} #1 \end{split}  \end{align}}

\newcommand{\Eq}[1]{Eq.~\ref{#1}} \newcommand{\Eqs}[2]{Eqs.~\ref{#1} and~\ref{#2}} \newcommand{\Eqm}[2]{Eqs.~\ref{#1} through \ref{#2}}
  
\newcommand{\Fig}[1]{Fig.~\ref{#1}} 

   \newcommand{\Appm}[2]{Appendices~\ref{#1} through \ref{#2}} 
\def\lsim{\mathrel{\raise.3ex\hbox{$<$\kern-.75em\lower1ex\hbox{$\sim$}}}}
\def\gsim{\mathrel{\raise.3ex\hbox{$>$\kern-.75em\lower1ex\hbox{$\sim$}}}}

\begin{document}
\title{Simplified Dark Matter Models for the Galactic Center Gamma-Ray Excess}
\author{Asher Berlin}
\affiliation{Department of Physics, University of Chicago, Chicago, IL 60637}
\author{Dan Hooper}
\affiliation{Department of Astronomy and Astrophysics, University of Chicago, Chicago, IL 60637}
\affiliation{Center for Particle Astrophysics, Fermi National Accelerator Laboratory, Batavia, IL 60510}
\author{Samuel D. McDermott}
\affiliation{Center for Particle Astrophysics, Fermi National Accelerator Laboratory, Batavia, IL 60510}
\affiliation{Michigan Center for Theoretical Physics, Ann Arbor, MI 48109}
\date{\today}

\begin{abstract} 
\noindent 

Motivated by the gamma-ray excess observed from the region surrounding the Galactic Center, we explore 
particle dark matter models that could potentially account for the spectrum and normalization of this signal. Taking a model-independent approach, we consider an exhaustive list of tree-level diagrams for dark matter annihilation, and determine which could account for the observed gamma-ray emission while simultaneously predicting a thermal relic abundance equal to the measured cosmological dark matter density. We identify a wide variety of models that can meet these criteria without conflicting with existing constraints from direct detection experiments or the Large Hadron Collider (LHC).  The prospects for detection in near future dark matter experiments and/or the upcoming 14 TeV LHC appear quite promising. 

\end{abstract}

\pacs{95.35.+d, 95.85.Pw; FERMILAB-PUB-14-060-A, MCTP-14-07}

\maketitle

\section{Introduction}

Over the past several years, a gamma-ray excess from the region surrounding the Galactic Center has been identified in the data of the Fermi Gamma-Ray Space Telescope, with features similar to that expected from annihilating dark matter (DM) particles~\cite{Goodenough:2009gk,Hooper:2010mq,Hooper:2011ti,Abazajian:2012pn,Hooper:2013rwa,Gordon:2013vta,Huang:2013pda,Abazajian:2014fta,Daylan:2014rsa}. Unlike many of the other potential DM signals that have been reported~\cite{Finkbeiner:2004us,Hooper:2007kb,Adriani:2008zr,Weniger:2012tx,Su:2012ft,Bernabei:2008yi,Bernabei:2010mq,Aalseth:2010vx,Aalseth:2011wp,Angloher:2011uu,Agnese:2013rvf,Chang:2008aa}, however, DM interpretations of this gamma-ray excess have become increasingly compelling as the signal has become better measured and characterized. Recent analysis has shown this excess to be robust and highly statistically significant, exhibiting a spectrum and angular distribution that is in good agreement with that expected from the annihilations of $\sim$30 GeV DM particles~\cite{Daylan:2014rsa}. Assuming a DM profile with a local density of 0.3 GeV/cm$^3$, the overall normalization of the signal requires that the DM annihilates with a cross section of $\sigma v \simeq (1.7-2.3)\times 10^{-26}$ cm$^3$/s~\cite{Daylan:2014rsa}, remarkably similar to the value anticipated for a thermal relic~\cite{Steigman:2012nb}. And unlike other astrophysical observations which have received attention as possible detections of DM (the cosmic-ray positron excess, for example~\cite{Hooper:2008kg}), no plausible astrophysical interpretation for the gamma-ray excess has been proposed.\footnote{Although a population of several thousand millisecond pulsars has been discussed as a possible origin of the observed gamma-ray excess~\cite{Hooper:2010mq,Abazajian:2010zy,Hooper:2011ti,Abazajian:2012pn,Hooper:2013rwa,Gordon:2013vta}, the more recent determination that this signal extends to beyond at least $10^{\circ}$ from the Galactic Center~\cite{Daylan:2014rsa,Hooper:2013rwa} strongly disfavors this interpretation~\cite{Hooper:2013nhl}.}

In this paper, we attempt to identify the varieties of DM models that could be responsible for the observed gamma-ray excess. Taking a model-independent and bottom-up approach, we construct an exhaustive list of tree-level diagrams for DM annihilation into Standard Model (SM) fermions (see also Ref.~\cite{DiFranzo:2013vra}). By considering tree-level diagrams, instead of effective operators~\cite{Kumar:2013iva,Goodman:2010yf,Bai:2010hh,Goodman:2010ku,Fox:2011fx,Beltran:2010ww,Beltran:2008xg}, we avoid a number of potentially important pitfalls~\cite{Busoni:2013lha,Goodman:2011jq,Buchmueller:2013dya,Shoemaker:2011vi}. For instance, while resonances can be important in determining the annihilation cross section and relic density of the DM, these effects are ``integrated out" in the effective operator approach. By studying the set of tree-level diagrams with all possible combinations of charge- and flavor-conserving renormalizable dimension-four, and super-renormalizable dimension-three, operators compatible with Lorentz invariance, we are able to take a holistic and general view of the types of DM models that could potentially produce the gamma-ray excess observed from the region of the Galactic Center.\footnote{For previous studies which have considered DM models for the Galactic Center excess from an effective field theory perspective, see Refs.~\cite{hooperon,Huang:2013apa}.}

For any given model, we impose the following requirements:
\begin{enumerate}
\item In order to generate the observed spectral shape of the gamma-ray excess, we require that the DM consists of either a $\sim$35 GeV particle that annihilates mostly to $b\bar{b}$ or a $\sim$25 GeV particle that annihilates approximately democratically to SM fermions~\cite{Daylan:2014rsa}. \label{enum:0}
\item To accommodate the observed intensity of the gamma-ray excess, we require that the DM annihilates in the low-velocity limit with a cross section of $\langle \sigma v \rangle = (0.77-3.23) \times 10^{-26}$ cm$^3$/s or $\langle \sigma v \rangle = (0.63-2.40) \times 10^{-26}$ cm$^3$/s for the two cases described in Criterion~\ref{enum:0}, respectively~\cite{Daylan:2014rsa}. These ranges take into account the uncertainty in the local DM density~\cite{Iocco:2011jz}. The necessary cross sections are doubled in the case that the DM is not self-conjugate. \label{enum:1}
\item We require that the thermal relic density of the DM satisfies $\Omega_{\rm DM} = 0.268^{+0.013}_{-0.010}$, in accordance with measurements from WMAP and Planck~\cite{Ade:2013zuv}. \label{enum:2}
\item We require that the elastic scattering cross sections of the DM with nuclei are consistent with the constraints from LUX~\cite{Akerib:2013tjd} and other direct detection experiments.\label{enum:3}
\item We require that no constraints from the LHC or other accelerator experiments are violated.~\label{enum:4}
\end{enumerate}
Criteria~\ref{enum:1} and~\ref{enum:2} roughly correspond to the requirement that the DM is a thermal relic whose annihilations proceed largely through $s$-wave processes. Criterion~\ref{enum:3} roughly requires that any coherent (spin-independent) DM scattering with nuclei must be suppressed, such as by powers of momentum or relative velocity. In evaluating Criterion~\ref{enum:4}, we consider mono-jet and mono-$W/Z$ constraints from the LHC, mono-$b$ projections, as well as accelerator constraints on various classes of particles that might mediate the interactions of the DM.

For the DM and its mediator, we consider any combination of spin-0, -1/2, and -1 particles with interactions of the following general forms:
\alg{ \label{schematic}
\cL_s \supset& \pL {\rm \overline{DM}~DM~mediator} \pR + \pL {\rm \overline{SM}~SM~mediator} \pR,
\\ \cL_t \supset& \pL {\rm \overline{DM}~SM~mediator} \pR + \pL {\rm \overline{SM}~DM~mediator} \pR.
}
These refer to $s$-channel or $t$-~and $u$-channel annihilation diagrams, respectively. We will continue to use this terminology even when talking about elastic scattering processes for which the Feynman diagrams are oriented differently. For the purpose of avoiding ambiguities regarding the labels for the DM and mediating particles, we adopt the conventions shown in Table~\ref{EFT}.

We constrain the interactions of the mediator with DM and SM fermions only by the requirement that Lorentz invariance is respected at every vertex. We then consider all allowed combinations of scalar ($1$), pseudoscalar ($\gamma^5$), vector ($\gamma^{\mu}$), and axial ($\gamma^{\mu} \gamma^5$) interactions. We do not attempt to construct an ultraviolet completion for any model, leaving such exploration for future work.

\begin{table}[t]
\centering
   \begin{tabular}{| c || c | c | c |}
   \hline
    & Scalar & Fermion & Vector \\ \hline \hline
    DM & $\phi$ & $\chi$ & $X^\mu$ \\ \hline 
    Mediator & $A$ & $\psi$ & $V^\mu$ \\ \hline
    SM (fermions) & $-$ & $f$ & $-$ \\ \hline
        \end{tabular}
     \caption{The particle notation used throughout this study.}
\label{EFT}
\end{table}

The remainder of this article is structured as follows. In Secs.~\ref{fermionsection} and~\ref{scalarsection}, we consider fermionic (spin-$1/2$) and bosonic (spin-0 or spin-1) DM, respectively, annihilating through $s$-channel Feynman diagrams. In each case, we determine which combination of spins and interaction types can satisfy the five criteria described in this section. In Sec.~\ref{tchannel}, we consider cases in which the DM annihilates through the $t$-channel exchange of a colored and charged mediator. In Sec.~\ref{collider}, we discuss constraints from collider experiments on the mass and couplings of the particles that mediate the DM's interactions. In Sec.~\ref{directD}, we discuss the prospects for operating and upcoming direct detection experiments. In Sec.~\ref{conclusions} we summarize our results and conclusions. This paper contains an extensive set of appendices which include, among other information, the full expressions for the DM annihilation and elastic scattering cross sections used in this study. 

%%%%%

\section{Fermionic Dark Matter}
\label{fermionsection}
\begin{table*}[t]
\centering
   \begin{tabular}{| c || c | c | c | c |}
   \hline
   \bf{ \emph{DM bilinear}} &\multicolumn{4}{|c|}{\bf{\emph{SM fermion bilinear}}} \\ \hline \hline
  \bf{ \emph{ fermion DM}}   & $\bar f f$ & $\bar f \gamma^5 f$ & $\bar f \gamma^\mu f$ & $\bar f \gamma^\mu\gamma^5 f$ \\ \hline
     $\bar \chi  \chi$ & \textcolor{black}{$\sigma v \sim v^2$, $\sigma_{\rm SI}$ $\sim 1$} & \textcolor{black}{$\sigma v \sim v^2$, $\sigma_{\rm SD}\sim q^2$} & $-$ & $-$ \\ \hline
     $\bar \chi \gamma^5 \chi$ & \textcolor{Green}{$\boldsymbol{\sigma v \sim 1}$, $\boldsymbol{\sigma_{\rm SI}}$  $\boldsymbol{\sim q^2}$}& \textcolor{Green}{$\boldsymbol{\sigma v \sim 1}$, $\boldsymbol{\sigma_{\rm SD}\sim q^4}$} & $-$ & $-$ \\ \hline
     $\bar \chi \gamma^\mu \chi$ (Dirac only) & $-$ & $-$ & \textcolor{blue}{$\sigma v \sim 1$, $\sigma_{\rm SI}$ $\sim 1$} & \textcolor{Green}{$\boldsymbol{\sigma v \sim 1}$, $\boldsymbol{\sigma_{\rm SD}\sim v_\perp^2}$} \\ \hline
     $\bar \chi \gamma^\mu\gamma^5 \chi$ & $-$ & $-$ & \textcolor{black}{$\sigma v \sim v^2$, $\sigma_{\rm SI} \sim v_\perp^2$} & \textcolor{Green}{$\boldsymbol{\sigma v \sim 1}$, $\boldsymbol{\sigma_{\rm SD}\sim 1}$} \\ \hline \hline 
     \end{tabular}
     \caption{A summary of the annihilation and elastic scattering behavior for all tree-level, $s$-channel annihilation diagrams, for cases in which the DM is a fermion (see Eqs.~\ref{lag1} and~\ref{lag2}). Because Majorana DM cannot couple to a vector current, this table encodes 14 (rather than 16) possible simplified models. Only those scenarios in which the low-velocity annihilation cross section is not suppressed ($\sigma v \sim 1$) can the DM potentially account for the observed gamma-ray excess.  For elastic scattering, we indicate whether the constraint on the spin-independent (SI) or spin-dependent (SD) cross section is currently more restrictive, and whether that cross section is unsuppressed ($\sim$1), or is suppressed by powers of momentum or velocity.  Any entry with a ``$-$'' symbol indicates that there is no particle representation that at tree-level can mediate the interaction indicated. We use \textcolor{Green}{\bf{green}} to indicate a model that satisfies all of our criteria, and \textcolor{blue}{blue} to indicate a model that allows for unsuppressed annihilation, but is ruled out by direct detection constraints. Models presented in \textcolor{black}{black} cannot account for the observed gamma-ray excess.}
\label{tab:cross1}
\end{table*}

In this section, we consider DM in the form of a (Dirac or Majorana) fermion, $\chi$, annihilating through the $s$-channel exchange of a spin-$0$ mediator, $A$:
\begin{equation}
\mathcal{L} \supset  \left[ a \, \bar{\chi}( \lambda_{\chi s} + \lambda_{\chi p} i \gamma^5 ) \chi +  \bar{f}( \lambda_{f s} + \lambda_{f p} i \gamma^5 ) f \right] A,  \\ 
\label{lag1}
\end{equation}
or through the $s$-channel exchange of a spin-$1$ mediator, $V_{\mu}$:
\begin{equation}
\mathcal{L} \supset \left[ a\,\bar{\chi} \gamma^\mu ( g_{\chi  v } + g_{\chi a} \gamma^5 ) \chi + \bar{f} \gamma^\mu ( g_{f  v } + g_{f a} \gamma^5 ) f \right] V_\mu. \\
\label{lag2}
\end{equation}
In each case, the couplings are defined such that $a=1$ $(1/2)$ for DM in the form of a Dirac (Majorana) fermion. For Majorana fermions, $g_{\chi v}$ is required to be zero. We will return to the case of $t$-channel annihilations in Sec.~\ref{tchannel}. 

The basic results of this section are summarized in Table~\ref{tab:cross1}. Of the fourteen linearly independent combinations that link the DM with SM fermions (counting Dirac and Majorana DM separately), there are eight in which the low-velocity annihilation cross section is not suppressed. We denote these models in the table with the shorthand $\sigma v \sim 1$. These models are capable of accounting for the observed gamma-ray excess.

\begin{figure*}[!t]
\includegraphics[width=3.41in]{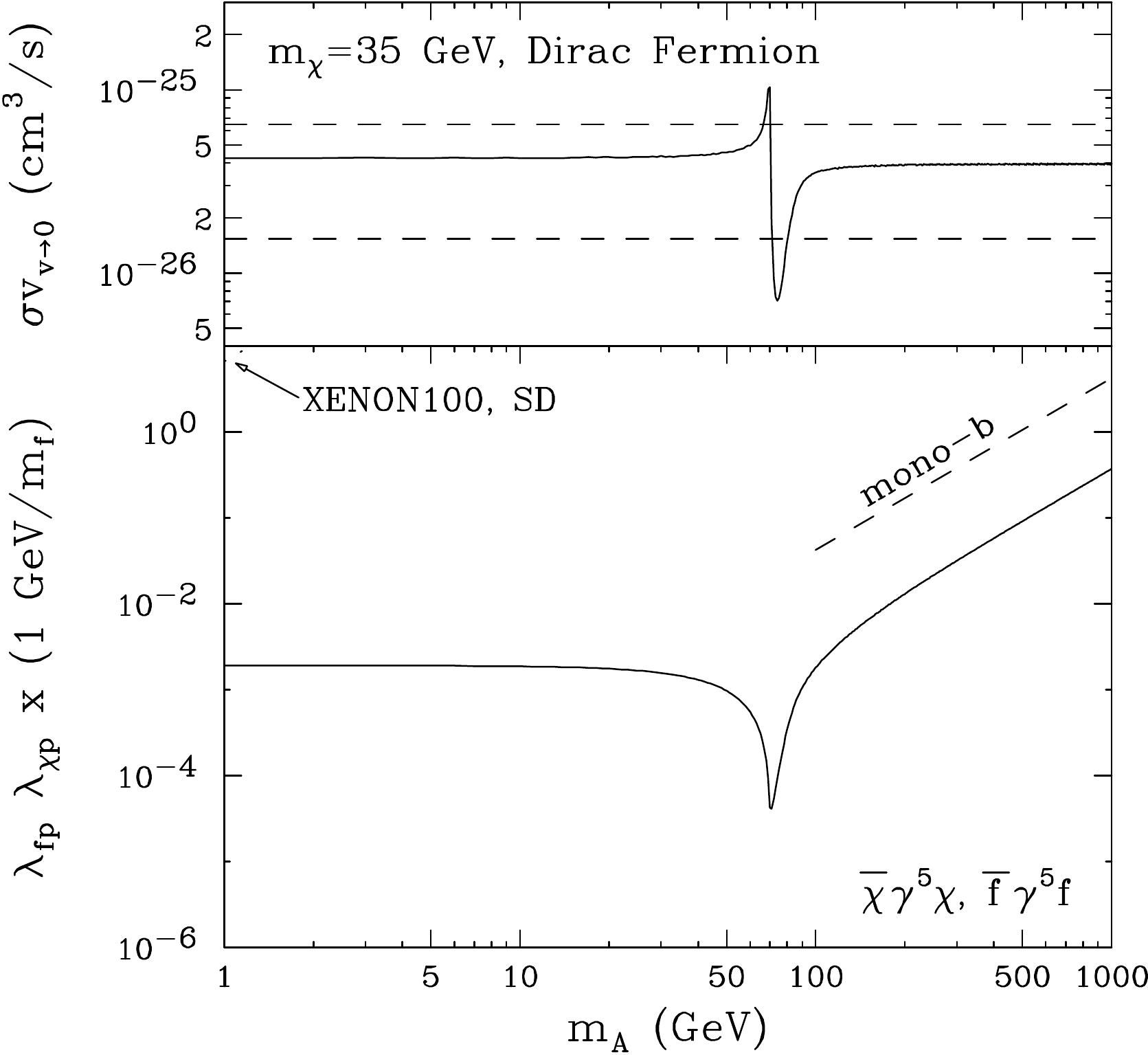}
\hspace{0.5cm}
\includegraphics[width=3.41in]{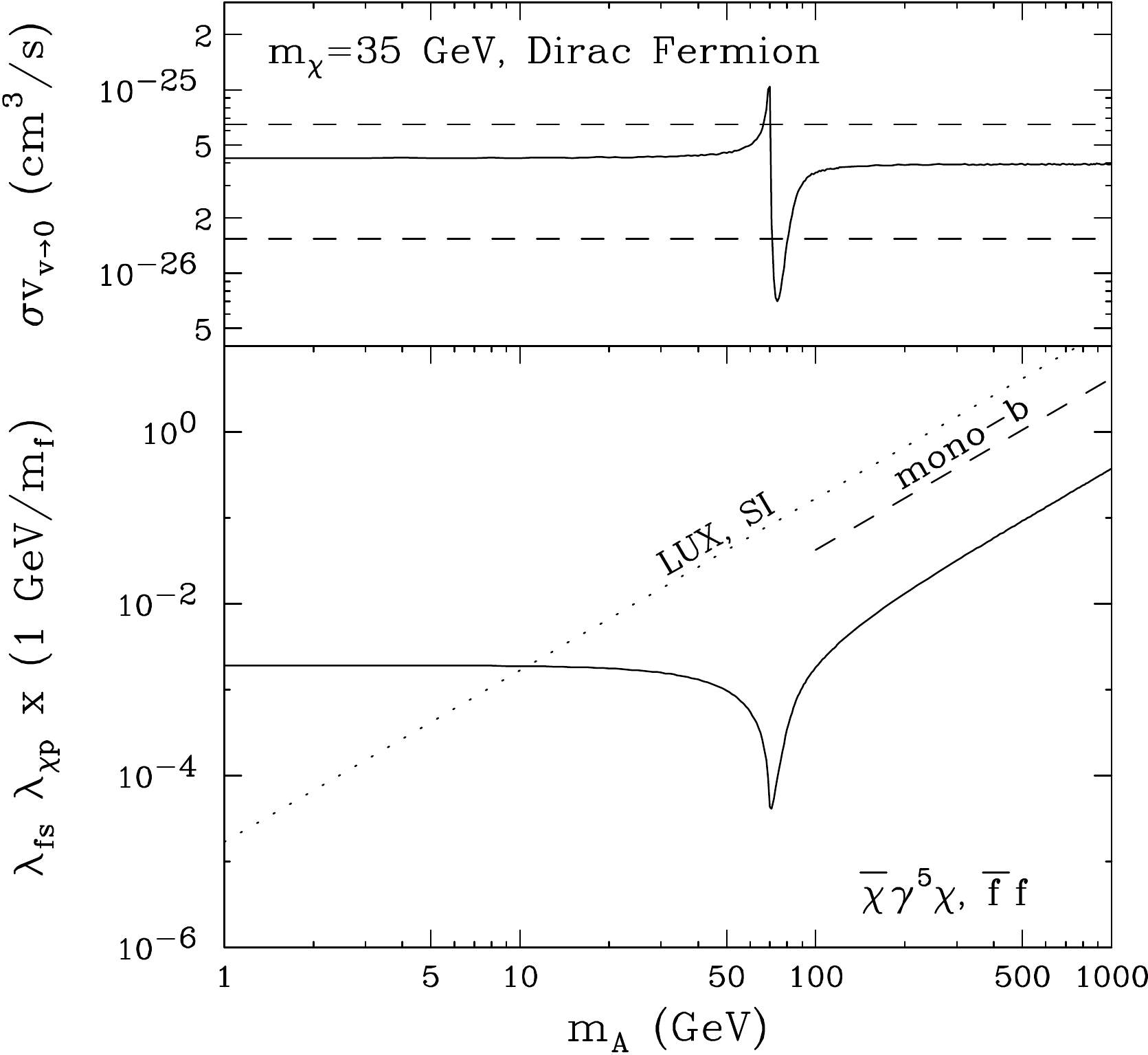} \\
\vspace{0.5cm}
\includegraphics[width=3.41in]{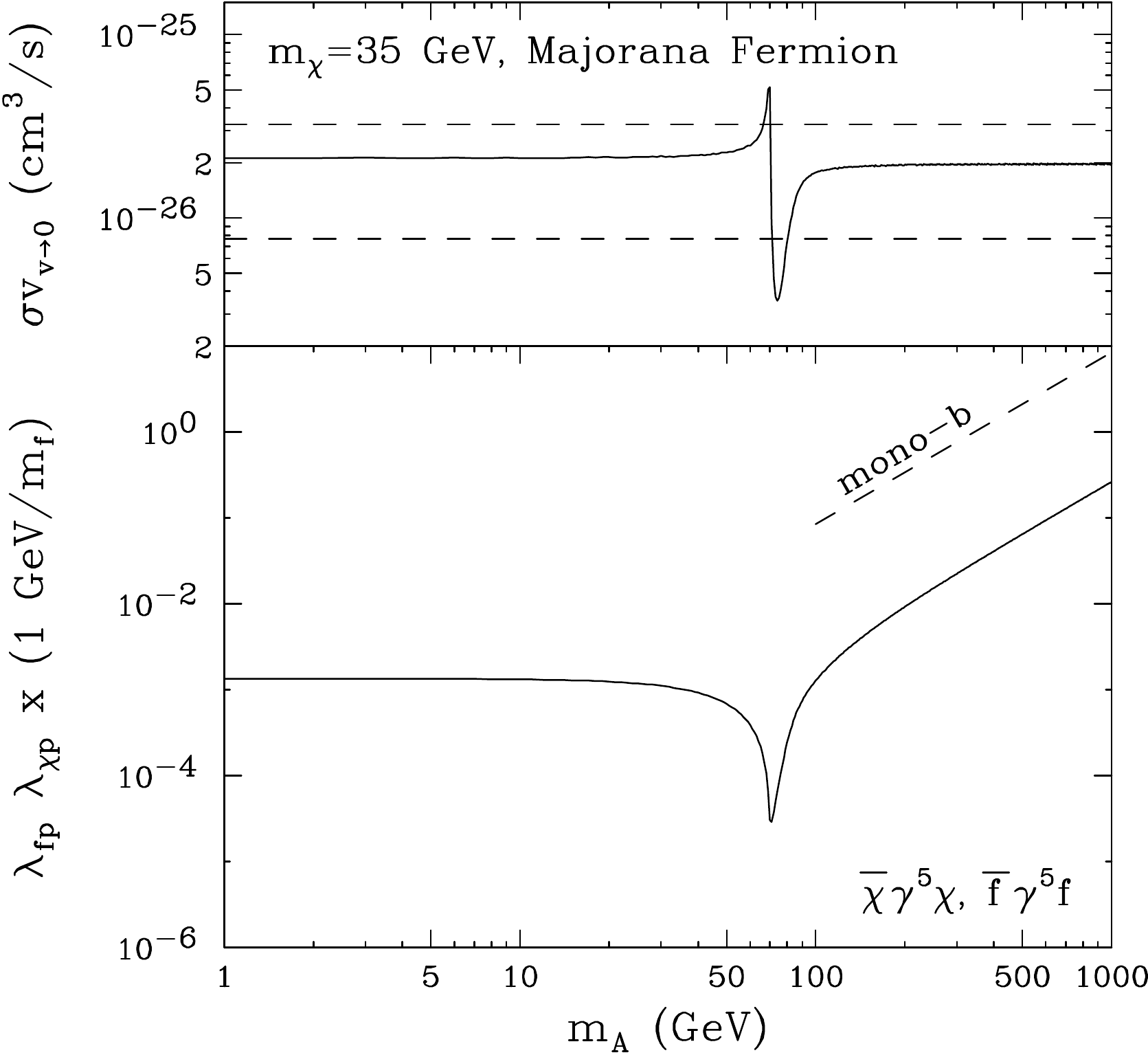}
\hspace{0.5cm}
\includegraphics[width=3.41in]{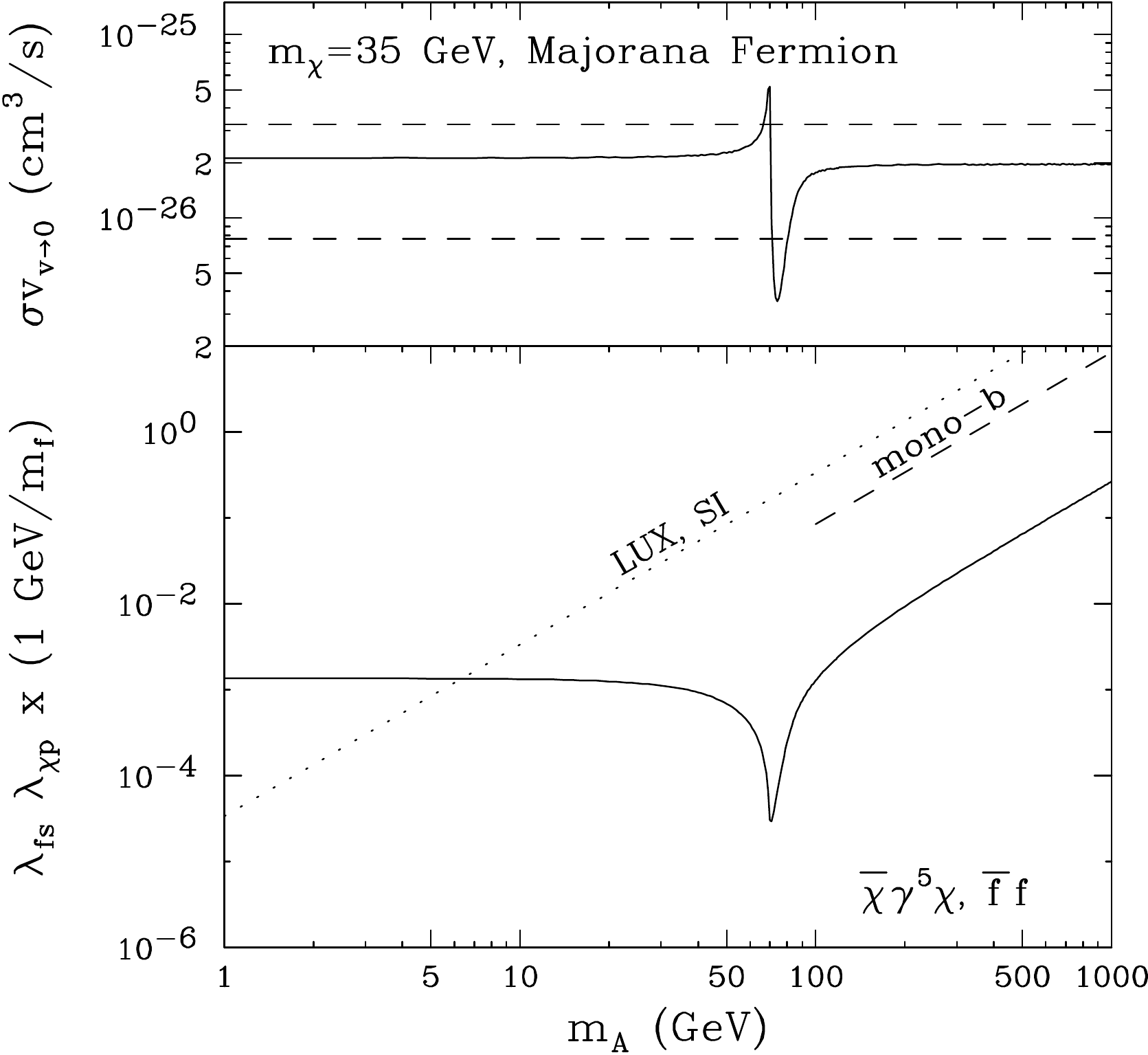}
\caption{The results for fermionic DM annihilating through the $s$-channel exchange of a spin-$0$ mediator. The upper frames correspond to the case of a Dirac fermion with either pseudoscalar-pseudoscalar (left) or pseudoscalar-scalar interactions (right). The lower frames denote the same interactions, but for the DM as a Majorana fermion. In the lower portion of each frame, the solid line represents the coupling strength required (as a function of the mediator mass, $m_A$) to produce a thermal relic abundance in agreement with the measured cosmological DM density (see Appendix~\ref{relic}). In the upper portion of each frame, we show the low-velocity annihilation cross section predicted, which must fall between the two horizontal dashed lines if the normalization of the gamma-ray excess is to be accommodated. Throughout this study, we have taken a value of 1 GeV for the width of the mediator, although the precise value of this quantity has little impact on our conclusions. In the lower portion of each frame, the dotted line denotes the current constraint from direct detection experiments (if not shown, the direct detection constraint is too weak to appear within the boundaries of the plot). The dashed line represents the projected constraint from LHC mono-$b$ searches~\cite{Lin:2013sca}, under the (possibly tenuous) assumption that effective field theory is valid in this application. For mediating particles heavier than $\sim$10 GeV, neither direct detection experiments nor the LHC constrains any of the models shown.}
\label{fermionscalar}
\end{figure*}

\begin{figure*}[!t]
\includegraphics[width=3.4in]{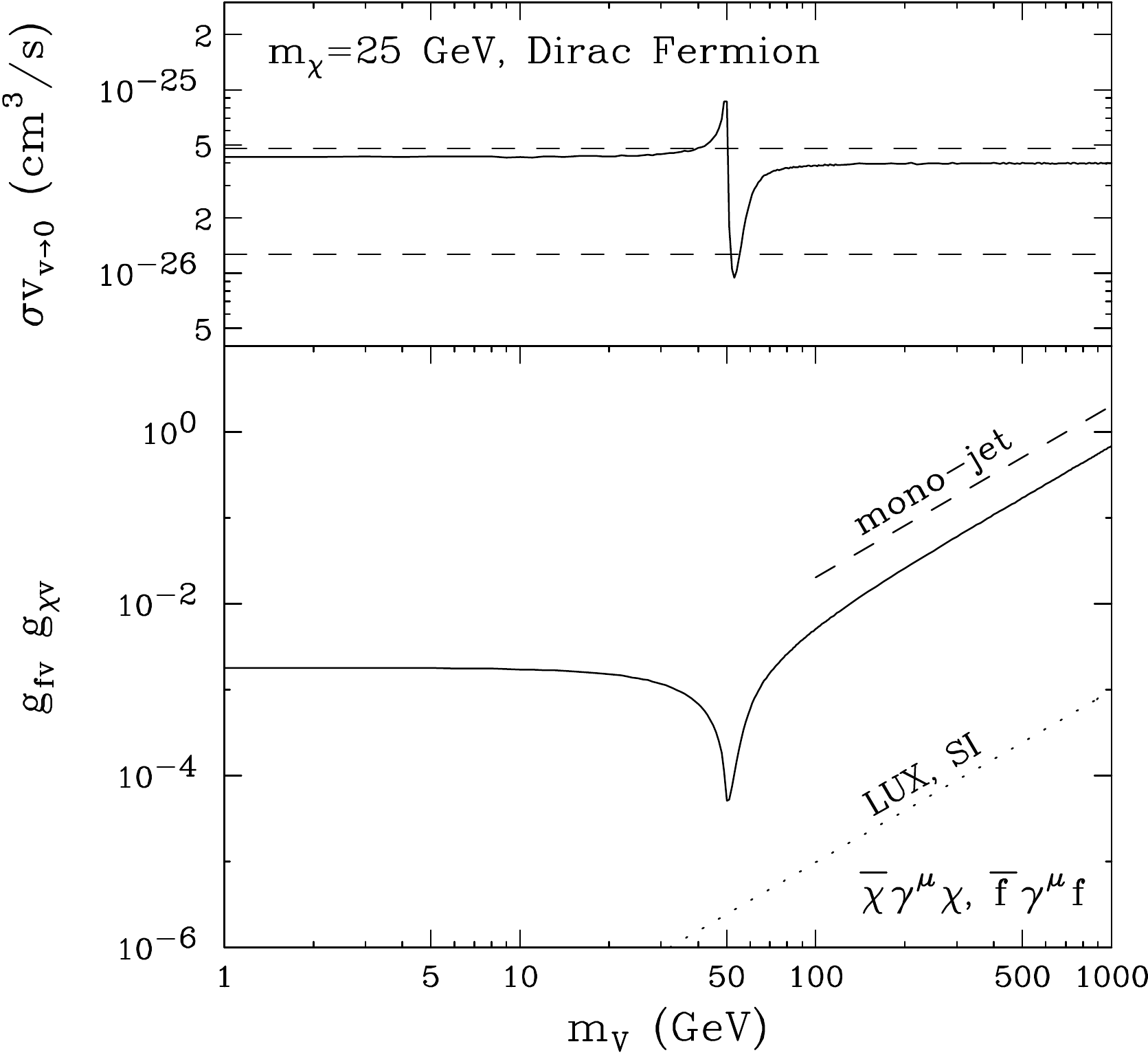}
\hspace{0.5cm}
\includegraphics[width=3.4in]{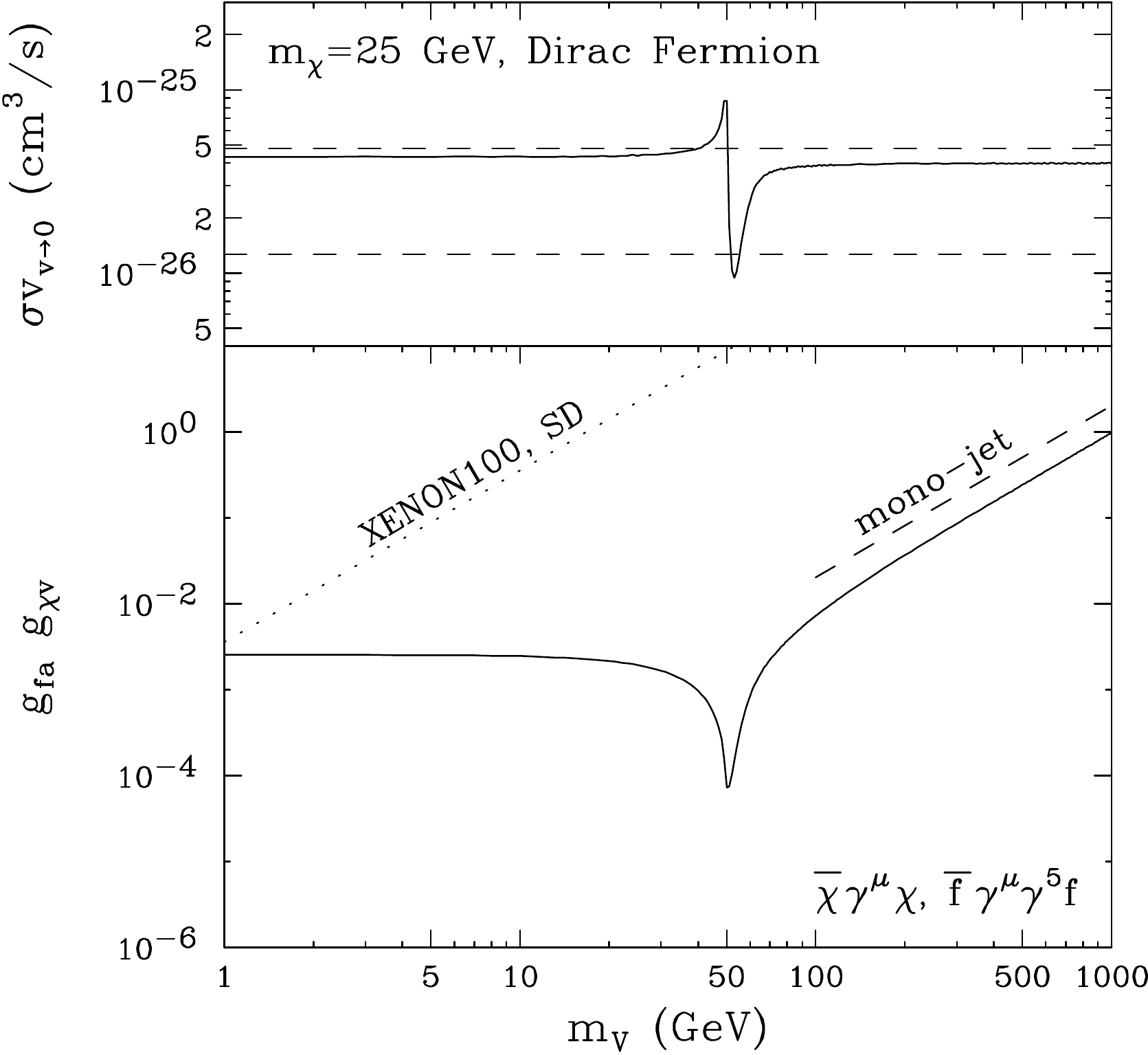} \\
\vspace{0.5cm}
\includegraphics[width=3.4in]{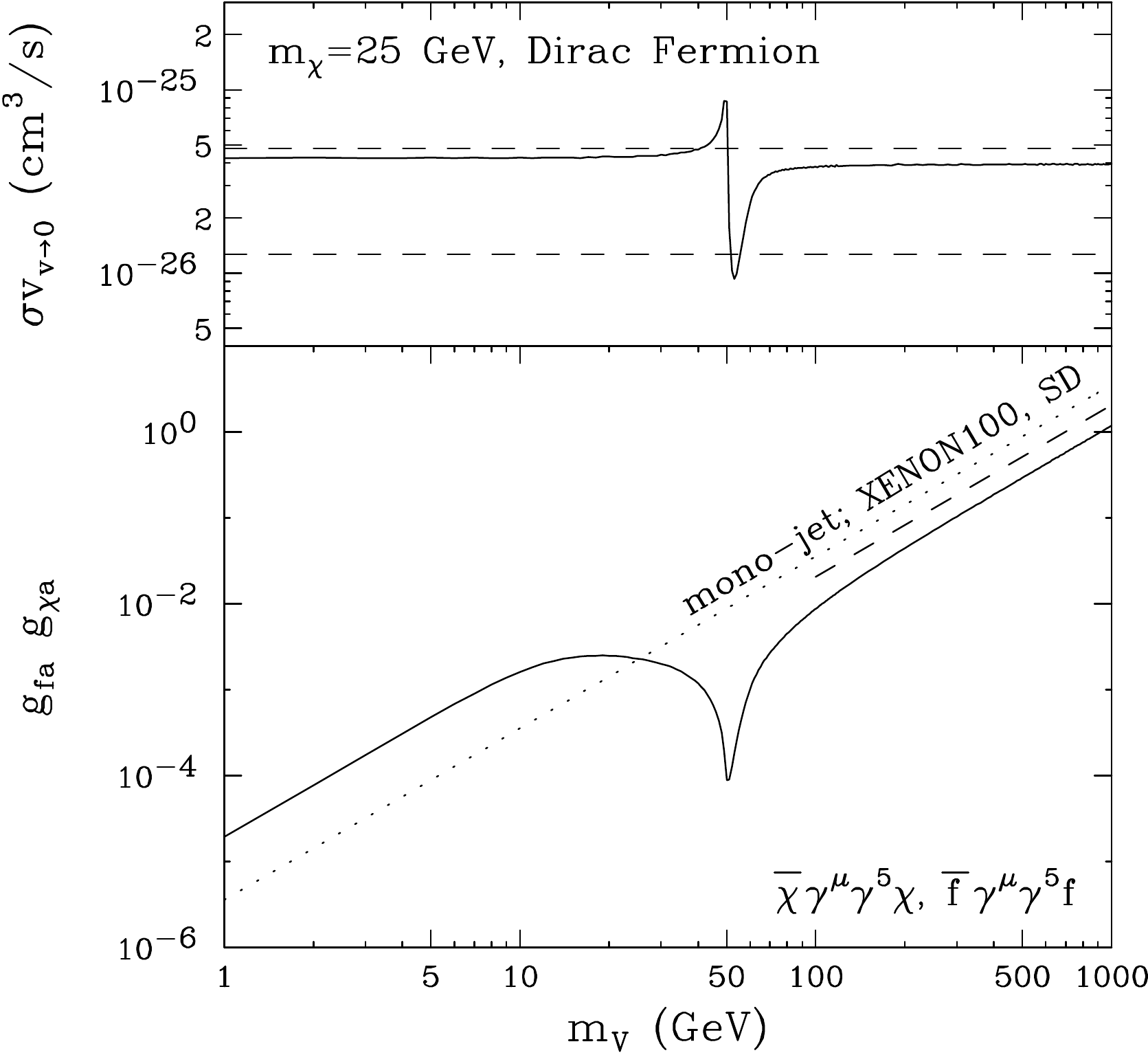}
\hspace{0.5cm}
\includegraphics[width=3.4in]{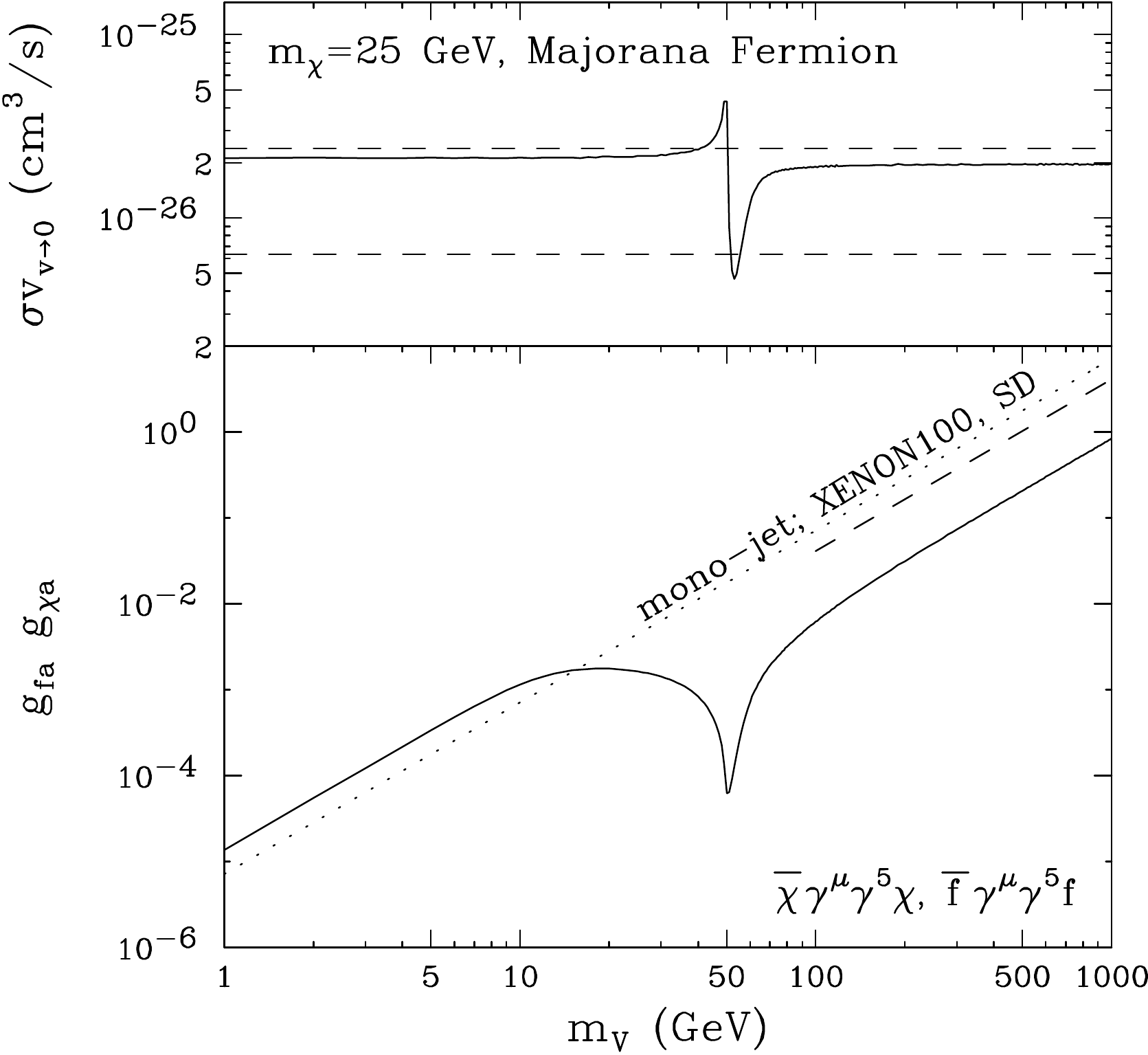}
\caption{Similar to as shown in Fig.~\ref{fermionscalar}, but for fermionic DM annihilating through the $s$-channel exchange of a spin-$1$ mediator. The upper frames correspond to the case of a Dirac fermion with either vector-vector (left) or vector-axial interactions (right). The lower frames denote the cases of a Dirac (left) or Majorana (right) fermion interacting through axial-axial interactions. In the lower portion of each frame, the dashed lines denote the constraint from LHC mono-jet searches~\cite{monojet}, under the (possibly tenuous) assumption that effective field theory is valid in this application. Only in the case of a Dirac fermion with vector-vector interactions (upper left) do direct detection constraints rule out any of the models shown (although XENON100 does restrict $m_V \gsim 20$ GeV in the case of axial-axial interactions).}
\label{fermionvector}
\end{figure*}

In Figs.~\ref{fermionscalar} and~\ref{fermionvector}, we show additional information for each of these eight interaction combinations.  In the lower portion of each frame, we show as a function of the mediator mass the product of the couplings that is required in order to produce a thermal relic density in agreement with the measured cosmological DM abundance (for the relevant cross sections, see~\Appm{diracscalar}{majoranavector}). In the upper portion of each frame, we show the low-velocity annihilation cross section that is predicted for that choice of couplings. If the solid curve falls between the two horizontal dashed lines, the model in question can account for the overall normalization of the Galactic Center's gamma-ray excess. In generating these plots we have assumed that spin-$1$ mediators couple equally to all SM fermions, and that spin-$0$ mediators couple to SM fermions proportionally to their mass (as motivated by minimal flavor violation~\cite{D'Ambrosio:2002ex}). Unless otherwise stated, we will maintain these assumptions throughout this paper.

We also assume that all DM annihilations proceed to pairs of SM fermions. If the mass of the mediator is less than that of the DM particles, however, annihilations could potentially be dominated instead by the production of mediator pairs. The fraction of DM annihilations that yield non-SM particles depends on the ratio of the mediator's couplings to the DM and to SM fermions. While we consider the exploration of such scenarios to be beyond the scope of the present study, we acknowledge that such models provide an additional degree of freedom that could allow them to account for the Galactic Center's gamma-ray excess.

Also shown in Figs.~\ref{fermionscalar} and~\ref{fermionvector} are the current constraints from direct detection experiments (shown as dotted lines). For the details of the calculation of the DM's elastic scattering cross section with nuclei, see Appendix~\ref{directappendix}. For the instances in which the spin-independent cross section provides the dominant constraint, we apply the results of the LUX Colloraboration~\cite{Akerib:2013tjd}. For those in which spin-dependent scattering with neutrons is more restrictive, we compare our results to the constraints of XENON100~\cite{Aprile:2013doa}.\footnote{In cases in which the DM's spin-dependent cross section with protons is much greater than that with neutrons, COUPP could potentially provide the most stringent limit~\cite{Behnke:2012ys}.}  At present, the most stringent of these constraints only rules out one of the eight scenarios shown, in which the DM is a Dirac fermion annihilating through a spin-$1$ mediator with vector interactions with both the DM and with SM fermions.

We also show in these figures the projected constraints (95\% CL) from mono-$b$~\cite{Lin:2013sca} and actual constraints from mono-jet~\cite{monojet} plus missing energy searches at the LHC (dashed lines).\footnote{The ATLAS Collaboration's search for hadronically decaying mono-W and mono-Z plus missing energy events has a sensitivity that is comparable to that of their monojet search~\cite{Aad:2013oja}. We do not additionally plot these limits here.} Although these constraints do not rule out any of the models under consideration, it is possible that data taken after the upcoming energy upgrade could be sensitive to such scenarios. We caution, however, that these constraints are derived under the assumptions of effective field theory, whose applicability to the problem at hand is far from clear~\cite{Busoni:2013lha,Goodman:2011jq,Buchmueller:2013dya,Shoemaker:2011vi}. In particular, these constraints are calculated under the assumption that the mass of the mediator is well above that of the parton-level center-of-mass energy of collisions at the LHC.  We expect such constraints to be conservative for mediator masses in the range of roughly 350 GeV to 2 TeV, in some cases underestimating their sensitivity by an order one factor. For lighter mediator masses, in contrast, the effective field theory approach can very significantly overestimate the sensitivity of such searches~\cite{Buchmueller:2013dya}. With this in mind, we plot these constraints down to mediator masses of 100 GeV, and ask the reader to remain aware of their limitations. In Sec.~\ref{collider}, we will discuss other collider constraints, such as those resulting from dijet and heavy Higgs searches.

\begin{table*}[!]
\centering
   \begin{tabular}{| c || c | c | c | c |}
   \hline
   \bf{ \emph{DM bilinear}} &\multicolumn{4}{|c|}{\bf{\emph{SM fermion bilinear}}} \\ \hline \hline
  \bf{ \emph{ scalar DM}}    & $\bar f f$ & $\bar f \gamma^5 f$ & $\bar f \gamma^\mu f$ & $\bar f \gamma^\mu\gamma^5 f$ \\ \hline
     $\phi^{\dagger}  \phi$ & \textcolor{blue}{$\sigma v \sim 1$, $\sigma_{\rm SI} \sim 1$} & \textcolor{Green}{$\boldsymbol{\sigma v \sim 1}$, $\boldsymbol{\sigma_{\rm SD}\sim q^2}$} & $-$ & $-$  \\ \hline
     $\phi^{\dagger}  \overset{\leftrightarrow}{\partial_{\mu}}  \phi$ (complex only) & $-$ & $-$ & \textcolor{black}{ $\sigma v \sim v^2$, $\sigma_{\rm SI} \sim 1$} & \textcolor{black}{$\sigma v \sim v^2$, $\sigma_{\rm SD}\sim v_\perp^2$} \\ \hline \hline 
  \bf{ \emph{ vector DM}}     & $\bar f f$ & $\bar f \gamma^5 f$ & $\bar f \gamma^\mu f$ & $\bar f \gamma^\mu\gamma^5 f$ \\ \hline
     $X^\mu X_\mu^{\dagger}$ & \textcolor{blue}{$\sigma v \sim 1$, $\sigma_{\rm SI}  \sim 1$} & \textcolor{Green}{$\boldsymbol{\sigma v \sim 1}$, $\boldsymbol{\sigma_{\rm SD}\sim q^2}$} & $-$ & $-$ \\ \hline
 $X^\nu \partial_\nu X_\mu^{\dagger}$ & $-$ & $-$ & \textcolor{black}{$\sigma v \sim v^2$, $\sigma_{\rm SI} \sim q^2 \cdot v_\perp^2$} &  \textcolor{black}{$\sigma v \sim v^2$, $\sigma_{\rm SD}\sim q^2$} \\  \hline
          \end{tabular}
     \caption{A summary of the annihilation and elastic scattering behavior for all tree-level, $s$-channel annihilation diagrams, for cases in which the DM is a real or complex scalar or a real or complex vector (see Eqs.~\ref{lag3}-\ref{lag6}). Only in those scenarios in which the low-velocity annihilation cross section is not suppressed ($\sigma v \sim 1$) can the DM potentially account for the observed gamma-ray excess.  For elastic scattering, we indicate whether constraints on the spin-independent (SI) or spin-dependent (SD) cross section is currently more restrictive, and whether that cross section is unsuppressed ($\sim 1$), or is suppressed by powers of momentum or velocity.  Any entry with a ``$-$'' symbol indicates that there is no particle representation that at tree-level can mediate the interaction indicated. We use \textcolor{Green}{\bf{green}} to indicate a model that satisfies all of our criteria, \textcolor{blue}{blue} to indicate a model that allows for unsuppressed annihilation, but is ruled out by direct detection constraints. Models presented in \textcolor{black}{black} cannot provide the observed gamma-ray excess.}
\label{tab2}
\end{table*}

\begin{figure}[!t]
\includegraphics[width=3.4in]{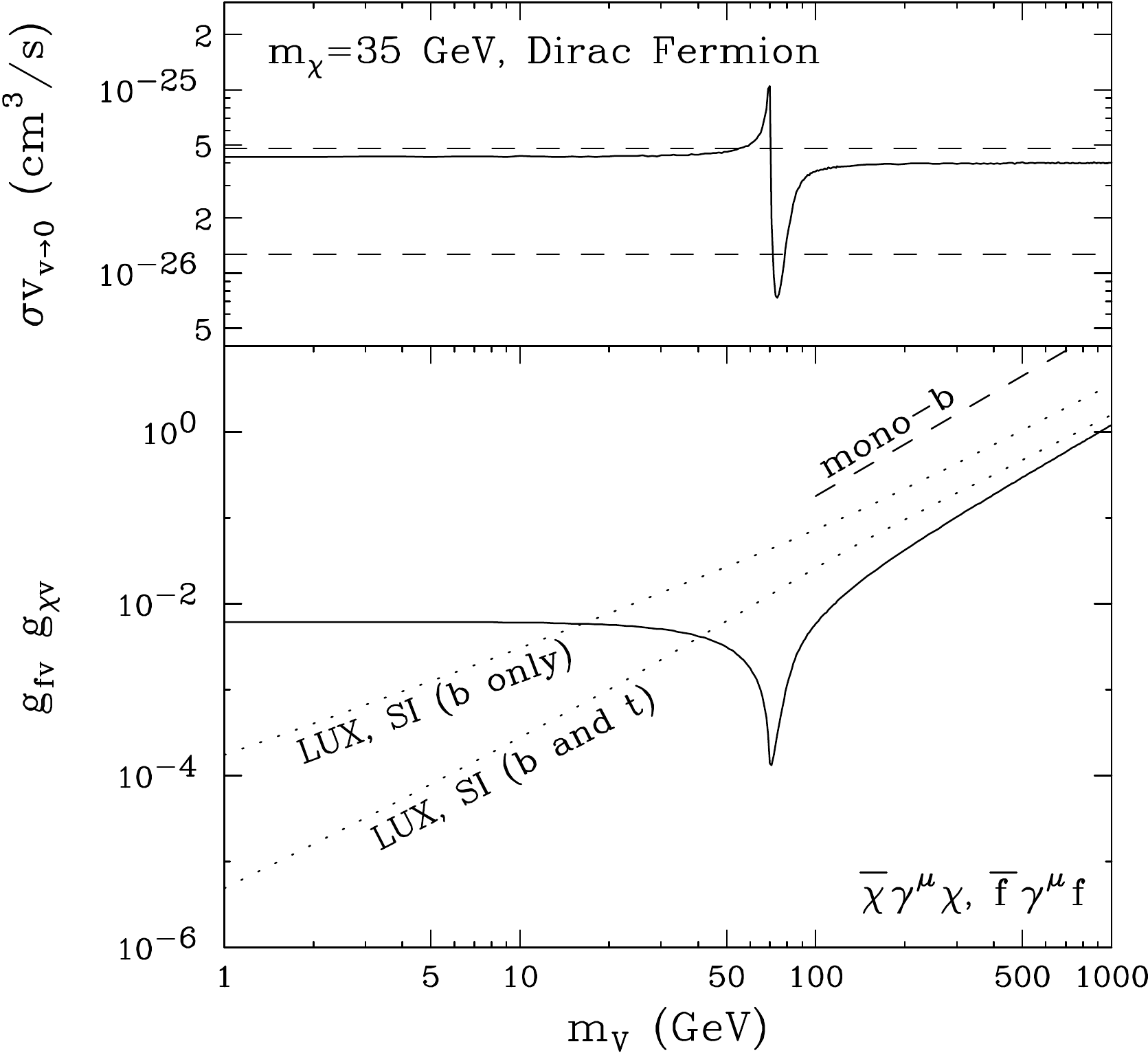}
\caption{As in the upper left frame of Fig.~\ref{fermionvector}, but for a spin-$1$ mediator that only couples to third generation SM fermions. In contrast to the case of democratic couplings, this choice makes it possible to evade current constraints from LUX and other direct detection experiments.}
\label{3gen}
\end{figure}

In generating Fig.~\ref{fermionvector}, we assumed that the mediator couples democratically to all SM fermions. If we instead consider DM annihilations that are mediated by a spin-$1$ particle with vector couplings to only third generation fermions, then the elastic scattering cross section will be additionally loop-suppressed. As loops with two gluons do not contribute in the case of vector interactions~\cite{Kaplan:1988ku,Ji:2006vx}, the dominant contribution comes from diagrams in which the bottom loop is coupled to the nucleus through a photon \cite{Kopp:2009et}. The suppression associated with this diagram allows this variation of the vector-mediated case to evade current constraints from direct detection (see Fig.~\ref{3gen}).  Even with this suppression, however, this elastic scattering cross section is still fairly large and will likely fall within the reach of future observations by LUX~\cite{Akerib:2012ys} and XENON1T~\cite{Aprile:2012zx}.

Summarizing the results of this section, of the fourteen interaction combinations shown in Table~\ref{tab:cross1} (counting Majorana and Dirac DM separately), we found eight to be capable of producing the Galactic Center's gamma-ray excess. Of these eight cases, only one (Dirac fermions interacting through a spin-$1$ mediator with a pair of vector interactions) is currently ruled out by direct detection constraints. Even this case, however, is consistent with the results of such experiments if the mediator only couples to third generation fermions (in addition to the DM). The other viable scenarios each include mediators with either pseudoscalar\footnote{The pseudoscalar case was recently considered in Ref.~\cite{Boehm:2014hva}. } or axial couplings. In these models, the DM is able to efficiently annihilate at low-velocities while also evading otherwise stringent constraints from direct detection experiments. The LHC's (actual) mono-jet and (projected) mono-$b$ constraints do not yet conflict with any of the models considered in this section.

\section{Bosonic Dark Matter}
\label{scalarsection}

We begin this section by considering DM in the form of a real or complex scalar, $\phi$, annihilating through the $s$-channel exchange of a spin-$0$ mediator, $A$:
\beq
\mathcal{L} \supset \left[a\, \mu_\phi | \phi |^2 + \bar{f} ( \lambda_{fs} + \lambda_{fp} i \gamma^5 ) f \right] A .
\label{lag3}
\eeq
or through the $s$-channel exchange of a spin-$1$ mediator, $V_\mu$:
\beq
\mathcal{L} \supset \left[i \, g_\phi  \phi^\dagger \overset{\leftrightarrow}{\partial_{\mu}}  \phi  + \bar{f} \gamma_{\mu}( g_{f v} + g_{f a}  \gamma^5 ) f \right]  V^{\mu}. \label{LscmxV}
\eeq
Here, $a=1$ $(1/2)$ for DM in the form of a complex (real) scalar. 

\begin{figure*}[t!]
\includegraphics[width=3.4in]{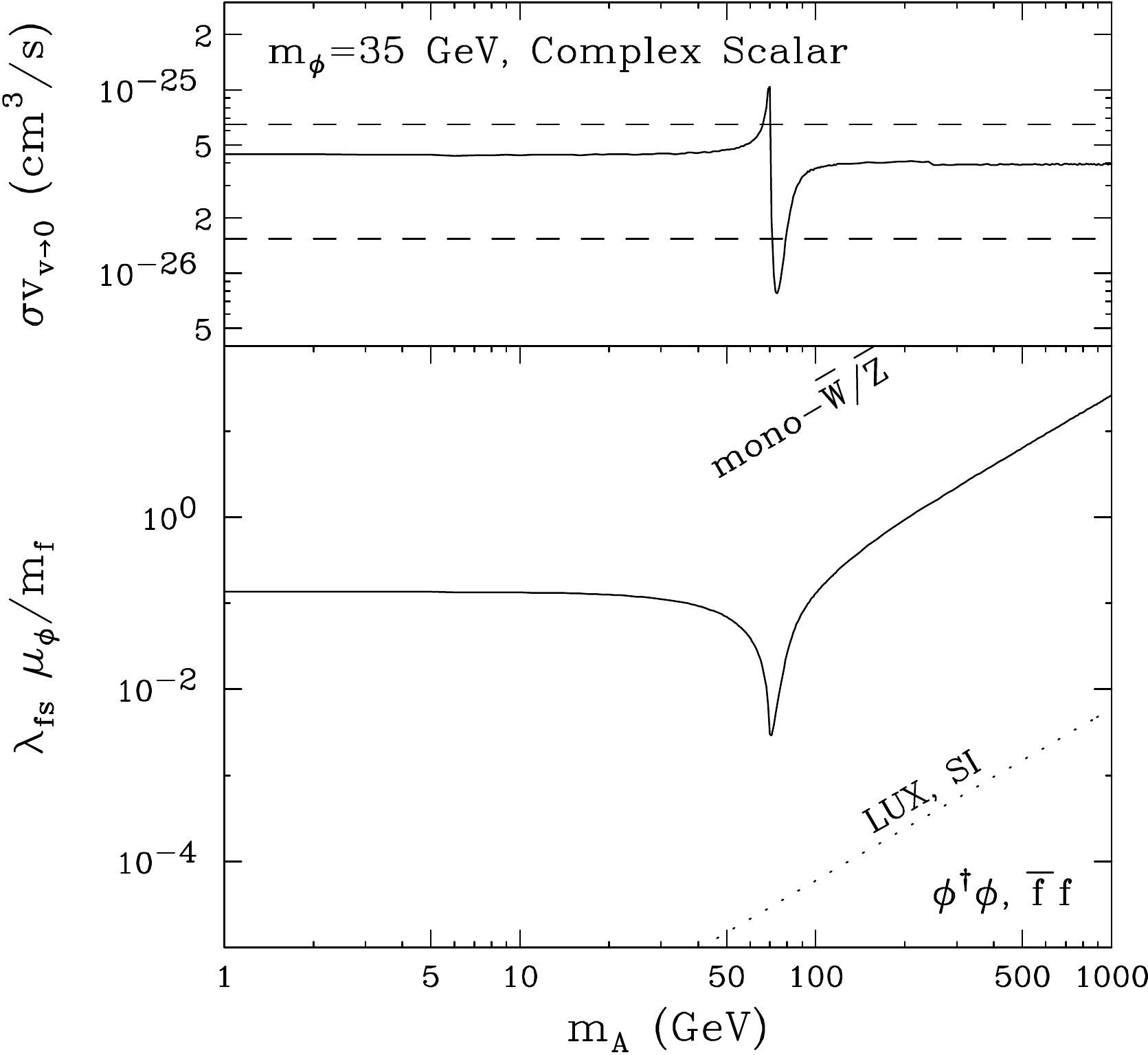}
\hspace{0.5cm}
\includegraphics[width=3.4in]{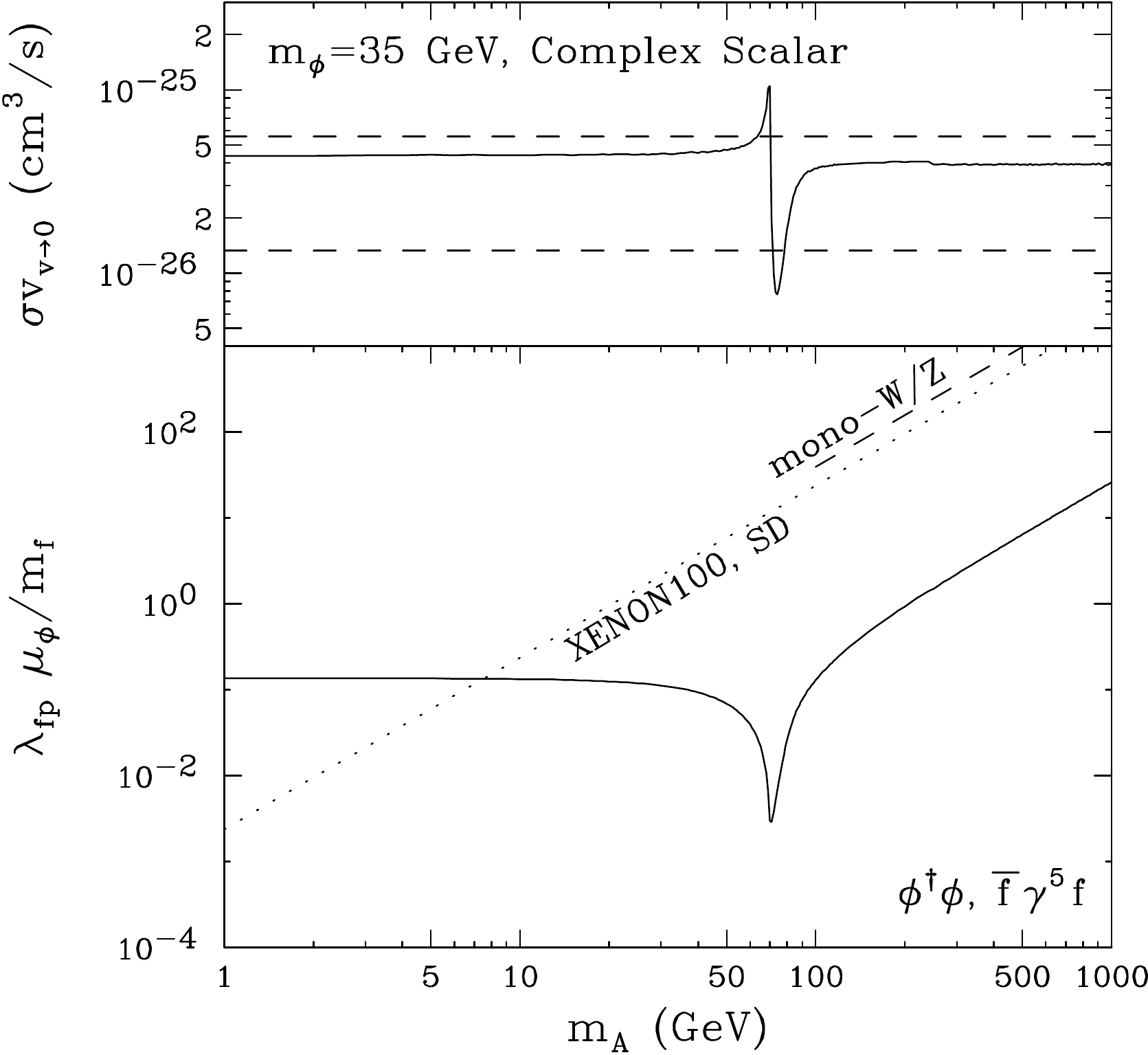} \\
\vspace{0.5cm}
\includegraphics[width=3.4in]{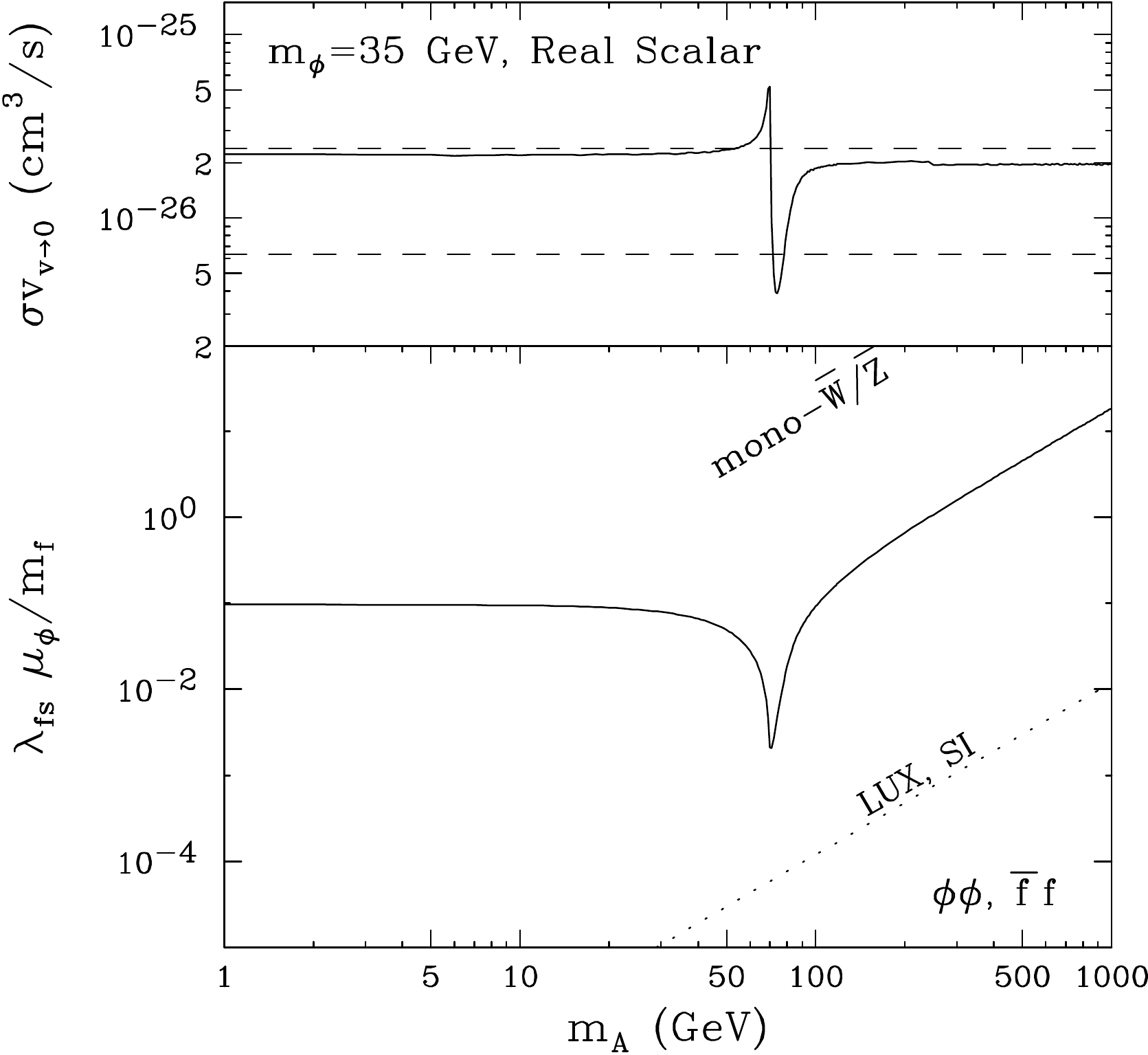}
\hspace{0.5cm}
\includegraphics[width=3.4in]{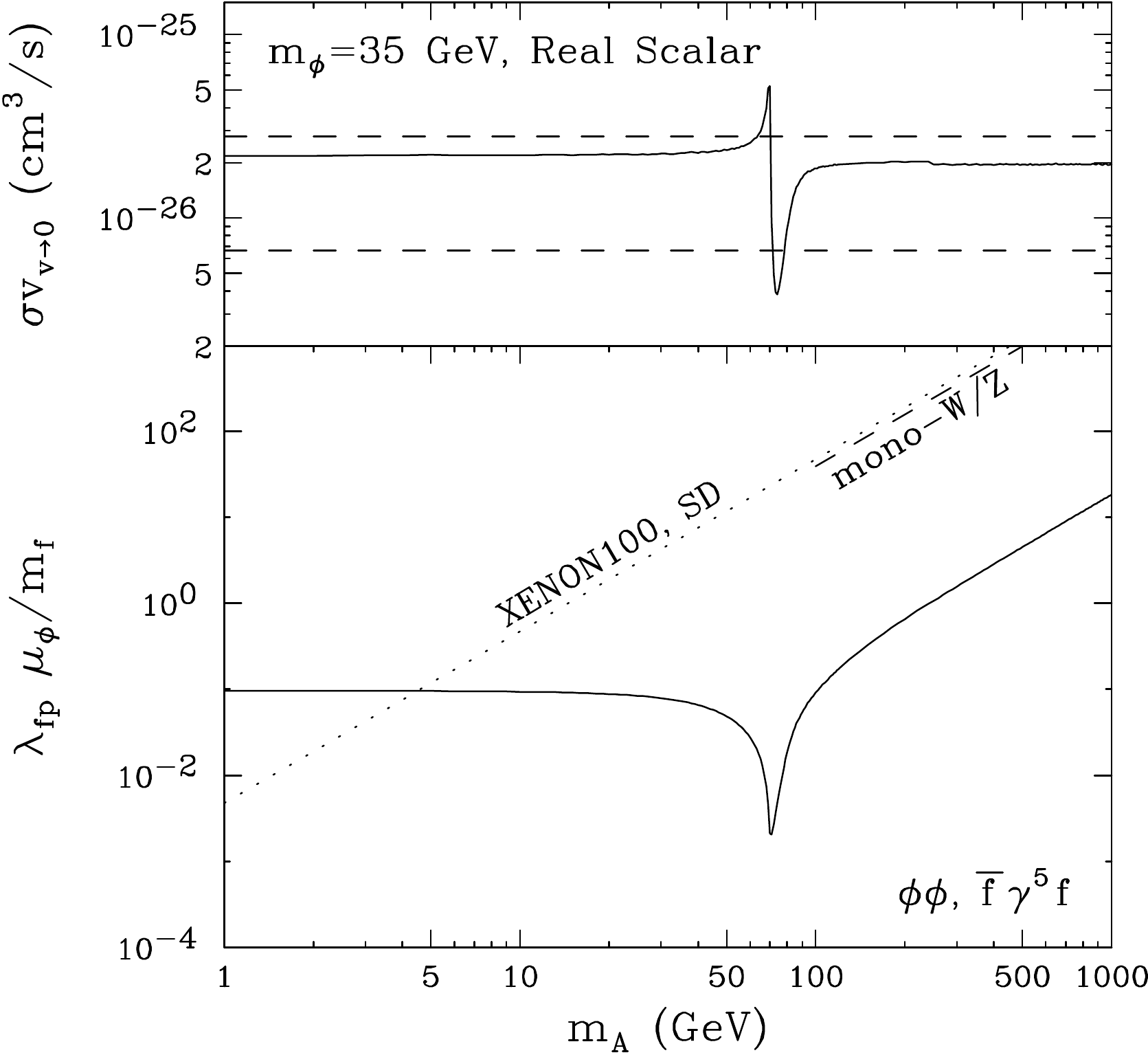}
\caption{Similar to as shown in the previous figures, but for scalar DM annihilating through the $s$-channel exchange of a spin-$0$ mediator. The upper frames correspond to the case of a complex scalar with either scalar (left) or pseudoscalar interactions (right). The lower frames denote the cases of a real scalar interacting through scalar (left) or pseudoscalar (right) interactions. In the lower portion of each frame, the dashed lines denote the constraint from LHC mono-$W/Z$ searches, under the (possibly tenuous) assumption that effective field theory is valid in this application. Direct detection constraints exclude the case of a either a complex or real scalar with scalar interactions (left frames).}
\label{scalarscalar}
\end{figure*}

We see from Table~\ref{tab2} and Fig.~\ref{scalarscalar} that in the case of scalar DM, there are only four $s$-channel models that are capable of generating the gamma-ray excess: a complex or real scalar, annihilating through a spin-$0$ mediator with either scalar or pseudoscalar couplings to SM fermions. Models which are mediated by spin-$1$ particles, in contrast, predict velocity suppressed annihilation cross sections (see \Appm{cszero}{rsone}). Furthermore, constraints from direct detection experiments rule out the scenarios in which the DM annihilates through a spin-$0$ mediator with scalar interactions with SM fermions. Again, the plots shown in this section assume that spin-$1$ mediators couple equally to all SM fermions, and that spin-$0$ mediators couple to SM fermions proportionally to their mass.

\begin{figure*}[t!]
\includegraphics[width=3.4in]{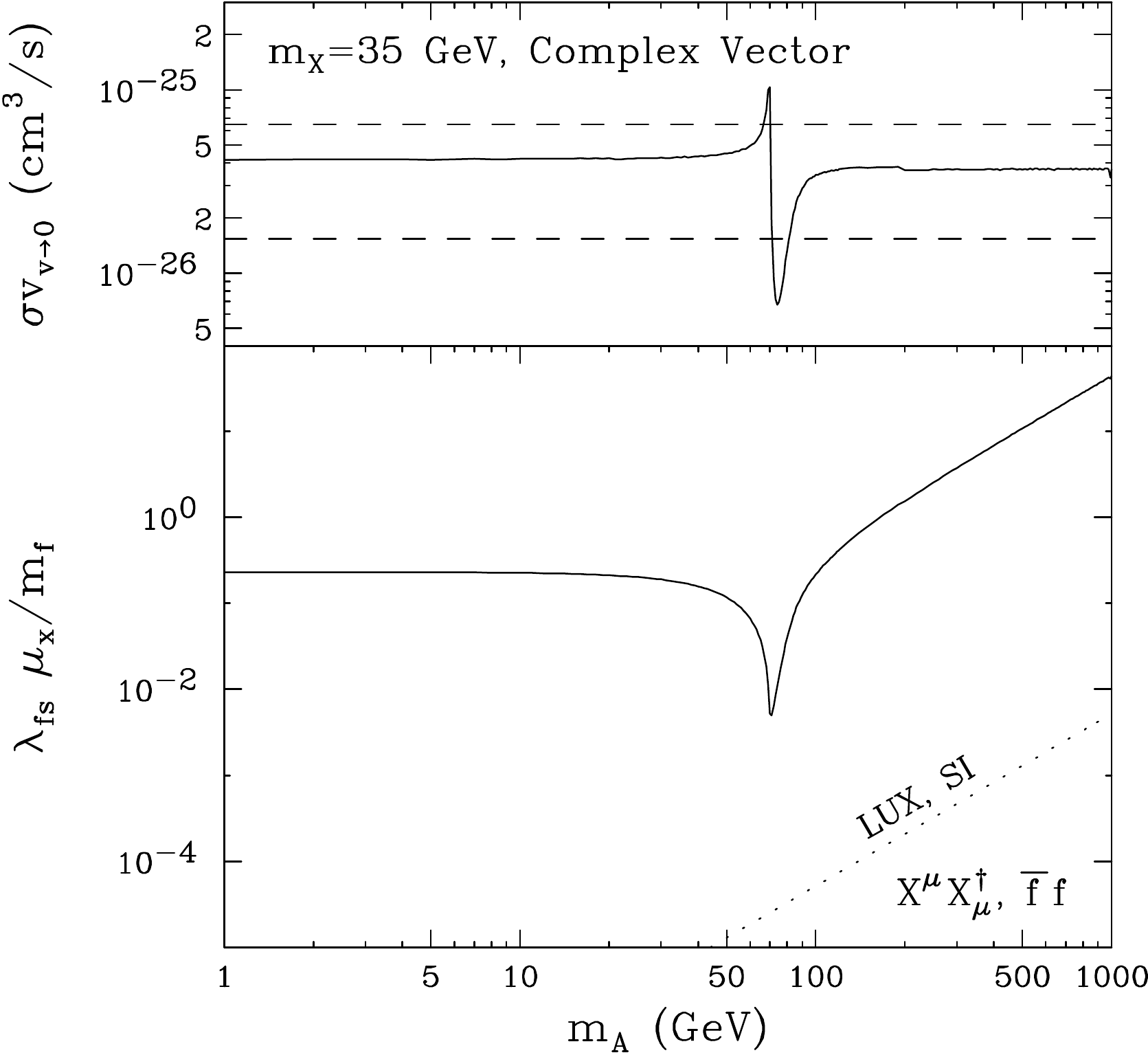}
\hspace{0.5cm}
\includegraphics[width=3.4in]{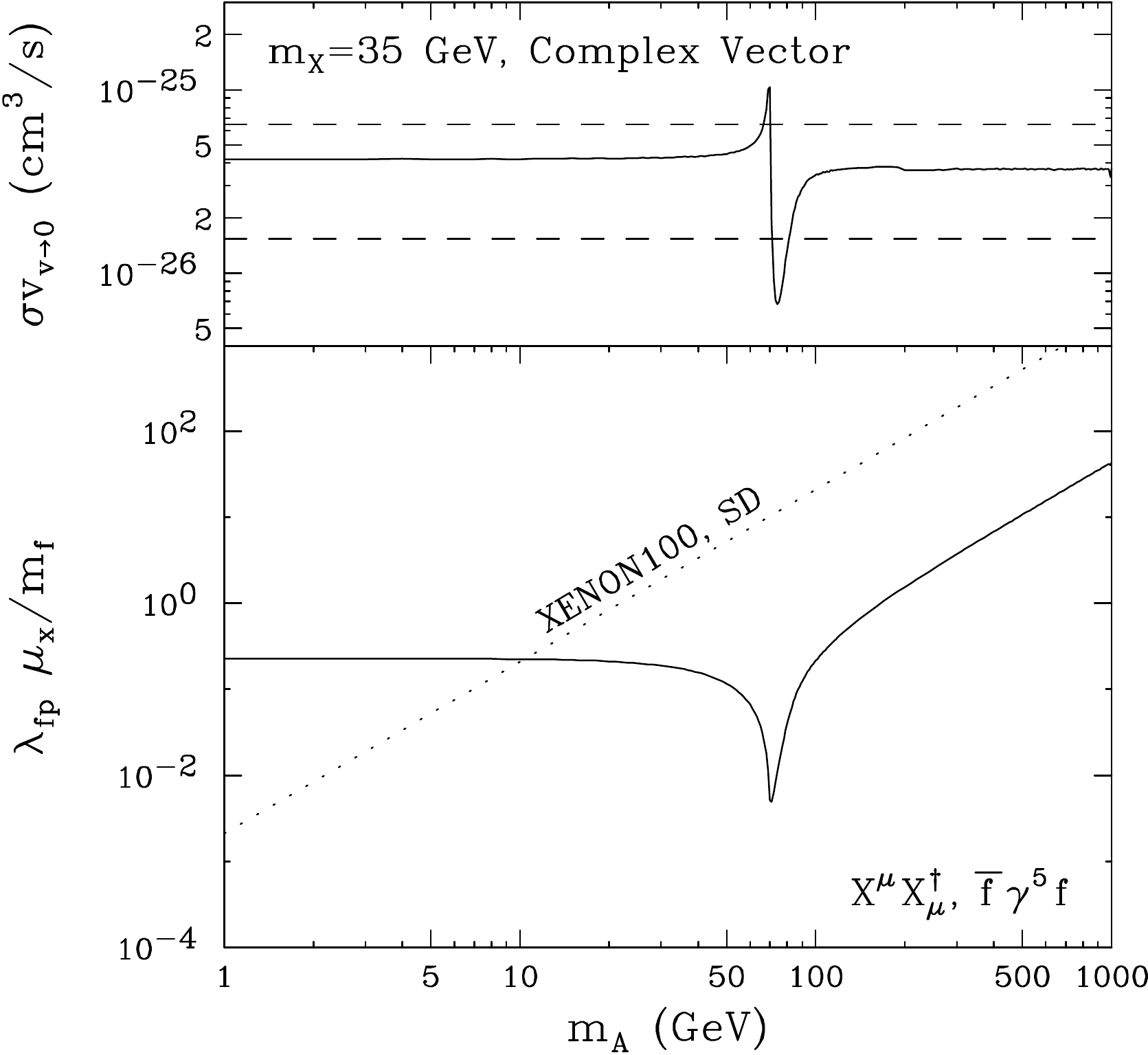} \\
\vspace{0.5cm}
\includegraphics[width=3.4in]{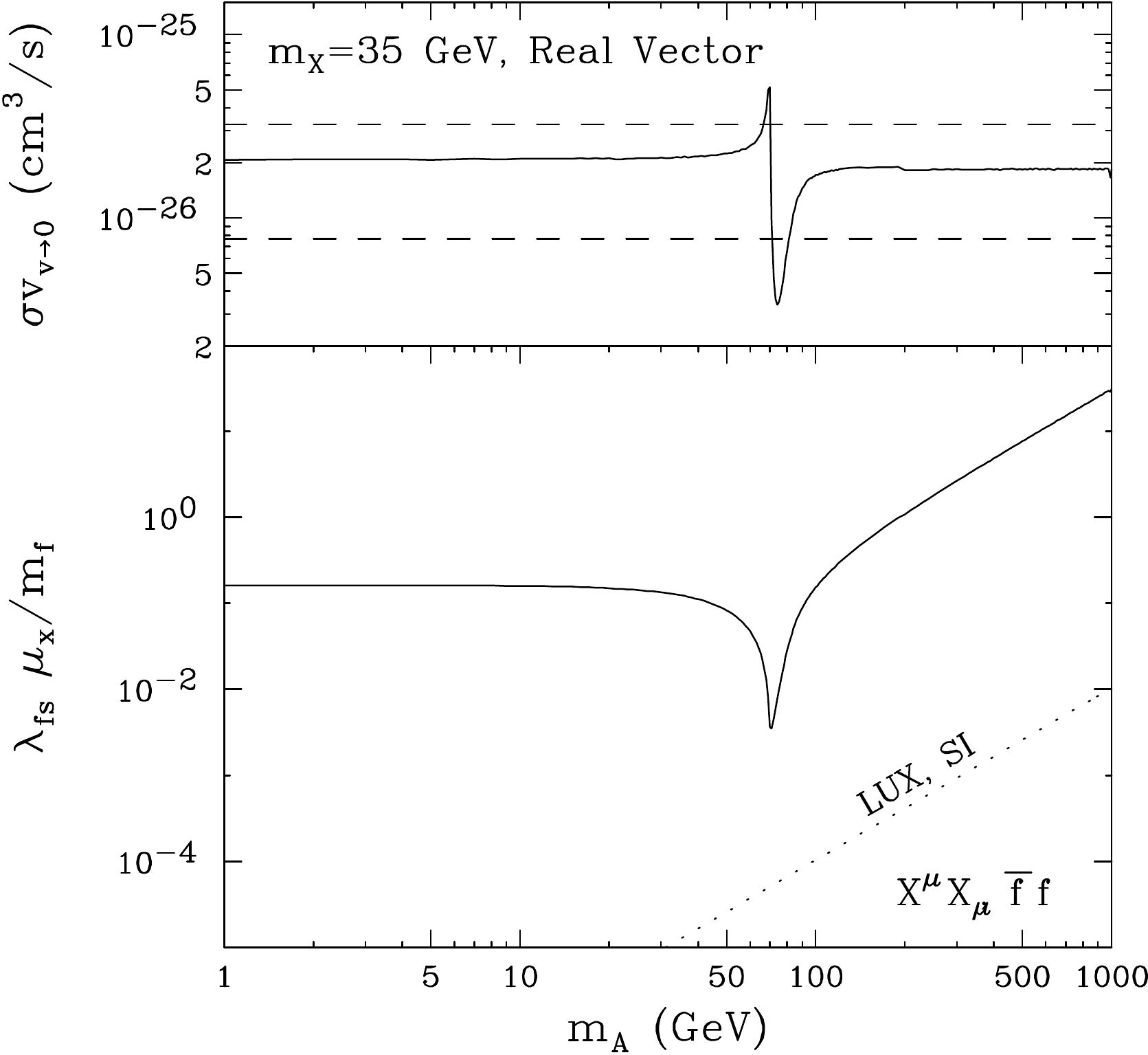}
\hspace{0.5cm}
\includegraphics[width=3.4in]{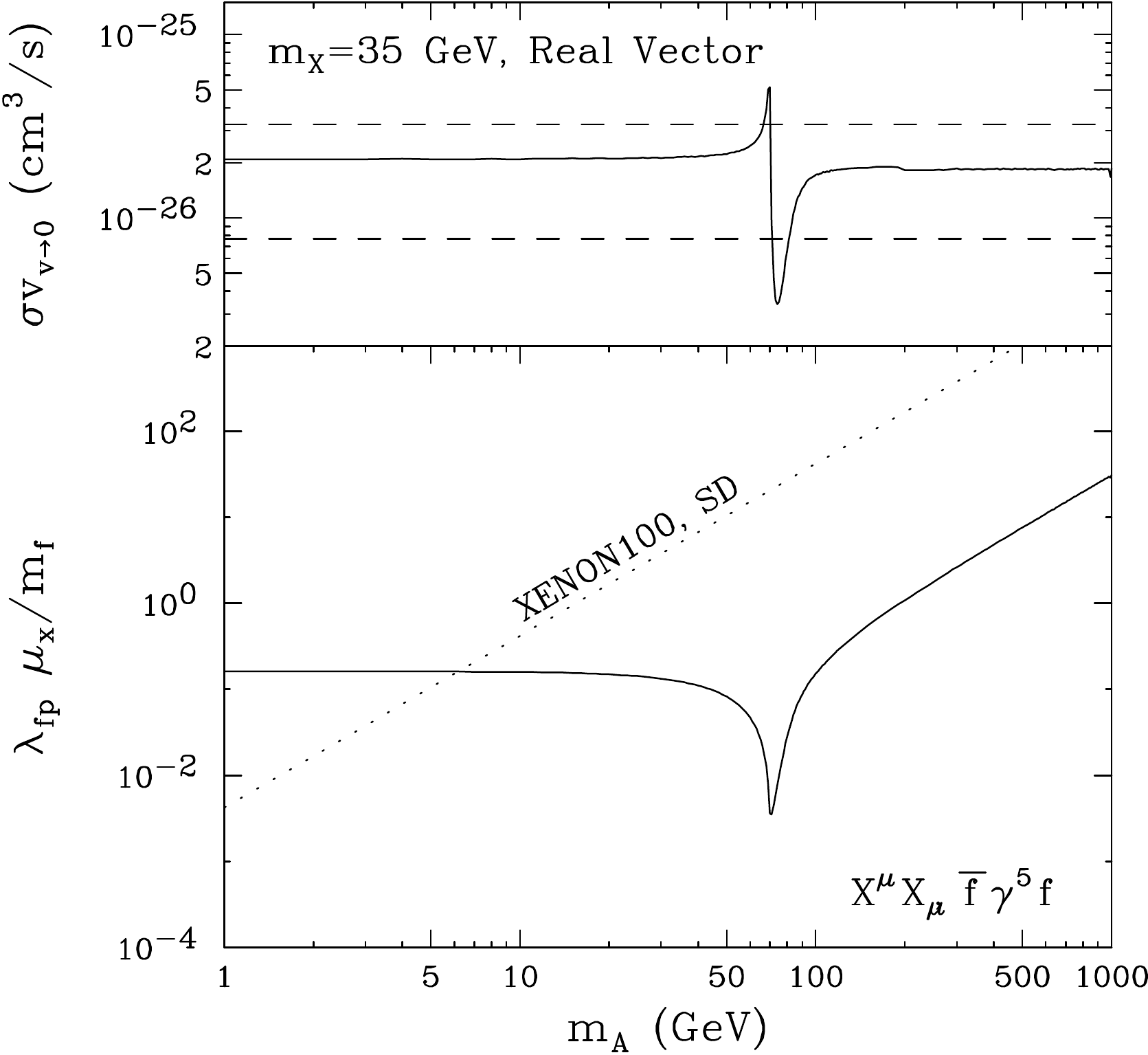}
\caption{Similar to as shown in the previous figures, but for vector DM annihilating through the $s$-channel exchange of a spin-$0$ mediator. The upper frames correspond to the case of a complex vector with either scalar (left) or pseudoscalar interactions (right). The lower frames denote the cases of a real vector interacting through scalar (left) or pseudoscalar (right) interactions. Direct detection constraints exclude the case of a either a complex or real vector with scalar interactions (left frames).}
\label{vectorscalar}
\end{figure*}

Next, we consider DM in the form of a complex or real vector, $X_\mu$, interacting either through the exchange of a spin-0 mediator:
\beq
\mathcal{L} \supset \left[a\, \mu_X X^\mu X_\mu^\dagger  + \bar{f}( \lambda_{f s} + \lambda_{f p} i \gamma^5 ) f \right] A, 
\label{lag5}
\eeq
or a spin-1 mediator:
\begin{eqnarray}
\label{lag6}
\mathcal{L} \supset \Big[a\, g_X\big(X^{\dagger \nu} \partial_\nu X^\mu &+& \text{h.c.} \big)  \\
&+& \bar{f} \gamma^\mu \left( g_{f v} + g_{f a} \gamma^5 \right) f  \Big] V_\mu, \nonumber
\end{eqnarray}
where $a=1$ $(1/2)$ for DM as a complex (real) vector. 

The conclusions regarding vector DM are very similar to those found for scalar DM. This can be seen by comparing the upper and lower portions of  Table~\ref{tab2}. Again, we find four cases with an annihilation cross section that is not velocity-suppressed: those in which the DM annihilates through the $s$-channel exchange of a spin-$0$ mediator (see \Appm{cvzero}{rvone}). Again, two of these four cases are compatible with direct detection constraints: those with pseudoscalar, rather than scalar, interactions. We show the results for these models in Fig.~\ref{vectorscalar}.

%%%%

To date, mono-$b$ projected constraints have only been presented for the case of fermionic DM. For this reason, the figures in this and the following section do not include such constraints. Such constraints should be qualitatively similar for the cases of scalar or vector DM as they are for the fermion case. In particular, we do not expect current mono-$b$ projections to restrict any of the models under consideration. For scalar DM, however, we do plot the constraints (90\% CL) from hadronic mono-W/Z plus missing energy searches by the ATLAS Collaboration~\cite{Aad:2013oja}. We remind the reader that the same caveats associated with the validity of effective field theory hold for this channel as in the cases of mono-jet and mono-$b$ searches.

It is possible that some of these statements could be modified somewhat in a case in which DM annihilations proceed through a finely-tuned resonance. For instance,  if there existed a scalar with a mass of $\sim$70 GeV and a narrow width ($\Gamma \ll 1$ GeV), it might also be possible for scalar DM to efficiently annihilate through that mediator while also evading direct detection constraints~\cite{Okada:2013bna}. From the top portions of the upper left frames of Figs.~\ref{scalarscalar} and \ref{vectorscalar}, however, we see that in this case the low-velocity annihilation cross section is pulled away from the required range of values, making it unlikely that resonance annihilation is responsible for the observed gamma-ray excess.

To summarize this section, we find that DM in the form of a scalar or a vector could account for the gamma-ray excess only if it annihilates through a spin-$0$ mediator with pseudoscalar interactions. All other $s$-channel annihilation diagrams lead to either a velocity-suppressed annihilation cross section, or predict an elastic scattering cross section with nuclei that is in conflict with direct detection constraints.

\begin{table*}[t]
\centering
   \begin{tabular}{| c | c | c | c |}
   \hline
%   \bf{ \emph{DM bilinear}} &\multicolumn{4}{|c|}{\bf{\emph{SM fermion bilinear}}} \\ \hline \hline
  \bf{ \emph{DM}}     & \bf{\emph{Mediator}} & \bf{\emph{Interaction}}  & \bf{\emph{Assessment}}  \\ \hline \hline
     \rm{Dirac Fermion} & Spin-$0$ & $1\pm \gamma^5$ & \textcolor{Green}{$\boldsymbol{\sigma v \sim 1}$, LHC OK} \\ \hline
     \rm{Dirac Fermion} & Spin-$1$ & $\gamma^{\mu}(1\pm \gamma^5)$ & \textcolor{Green}{$\boldsymbol{\sigma v \sim 1}$, LHC OK} \\ \hline
          \rm{Majorana Fermion} & Spin-$0$ & $1\pm \gamma^5$ & $\sigma v \sim v^2$ \\ \hline
     \rm{Majorana Fermion} & Spin-$1$ & $\gamma^{\mu}(1\pm \gamma^5)$ & $\sigma v \sim v^2$ \\ \hline     
     \rm{Real Scalar} & Spin-$1/2$ & $1\pm \gamma^5$ & \textcolor{blue} {$\sigma v \sim 1$, LHC Excluded} \\ \hline
     \rm{Complex Scalar} & Spin-$1/2$ & $1\pm \gamma^5$ & $\sigma v \sim v^2$\\ \hline
          \rm{Real Vector} & Spin-$1/2$ & $\gamma^{\mu}(1\pm \gamma^5)$ & \textcolor{Green}{$\boldsymbol{\sigma v \sim 1}$, LHC OK} \\ \hline
               \rm{Complex Vector} & Spin-$1/2$ & $\gamma^{\mu}(1\pm \gamma^5)$ & \textcolor{Green}{$\boldsymbol{\sigma v \sim 1}$, LHC OK} \\ \hline \hline   
     \end{tabular}
     \caption{A summary of the annihilation and elastic scattering behavior for all tree-level, $t$-channel annihilation diagrams which do not lead to a scalar elastic scattering cross section with nuclei. Only those scenarios in which the low-velocity annihilation cross section is not suppressed ($\sigma v \sim 1$) can the DM account for the observed gamma-ray excess.  We use \textcolor{Green}{\bf{green}} to indicate a model that satisfies all of our criteria, \textcolor{blue}{blue} to indicate a model that allows for unsuppressed annihilation, but that is ruled out by LHC constraints. All of the models shown evade current constraints from direct detection. Models presented in \textcolor{black}{black} are not capable of generating the observed gamma-ray excess.}
\label{tab:tchannel}
\end{table*}

\begin{figure*}[t!]
\includegraphics[width=3.4in]{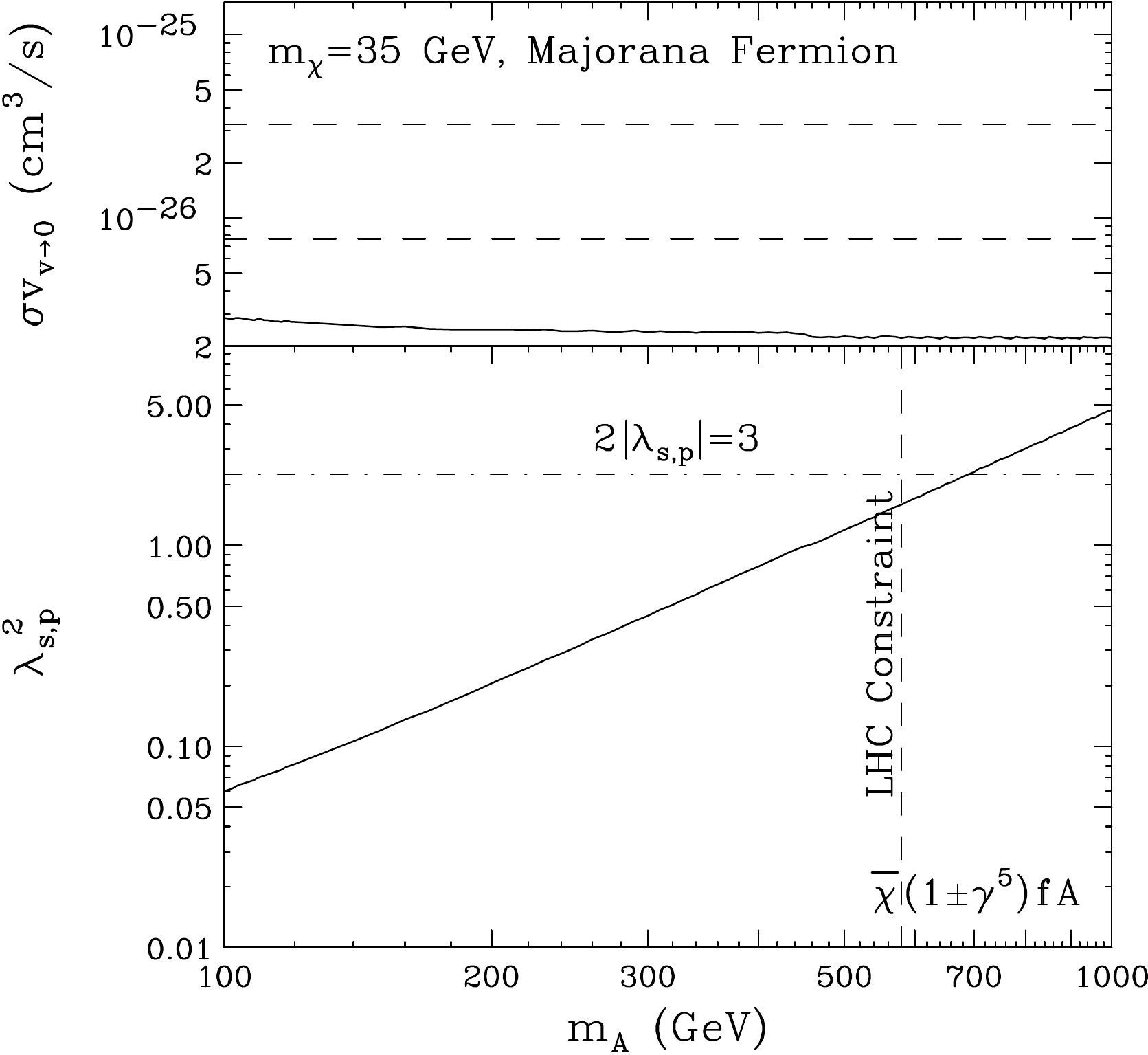}
\hspace{0.5cm}
\includegraphics[width=3.4in]{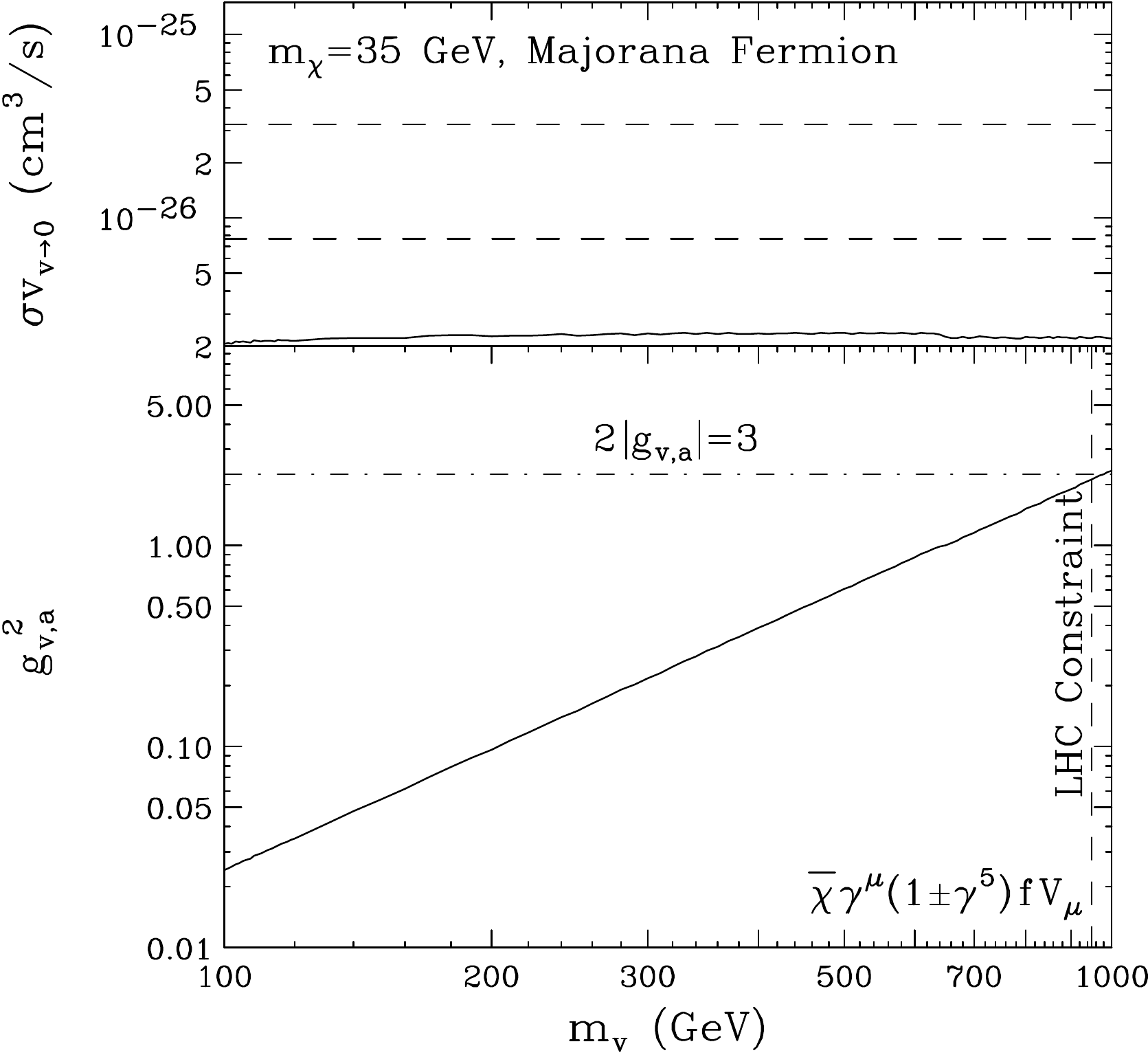} \\
\vspace{0.5cm}
\includegraphics[width=3.4in]{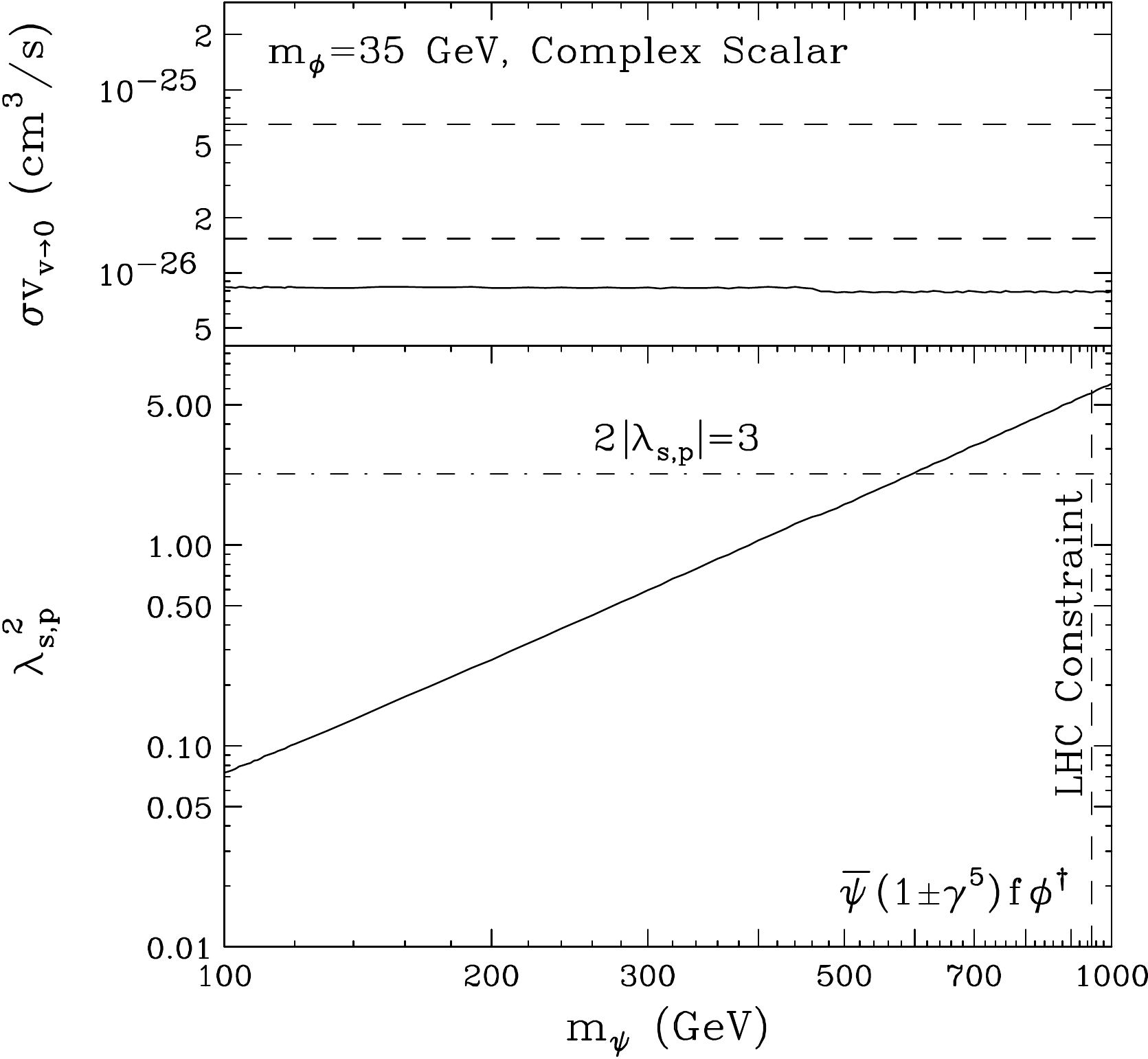}
\hspace{0.5cm}
\includegraphics[width=3.4in]{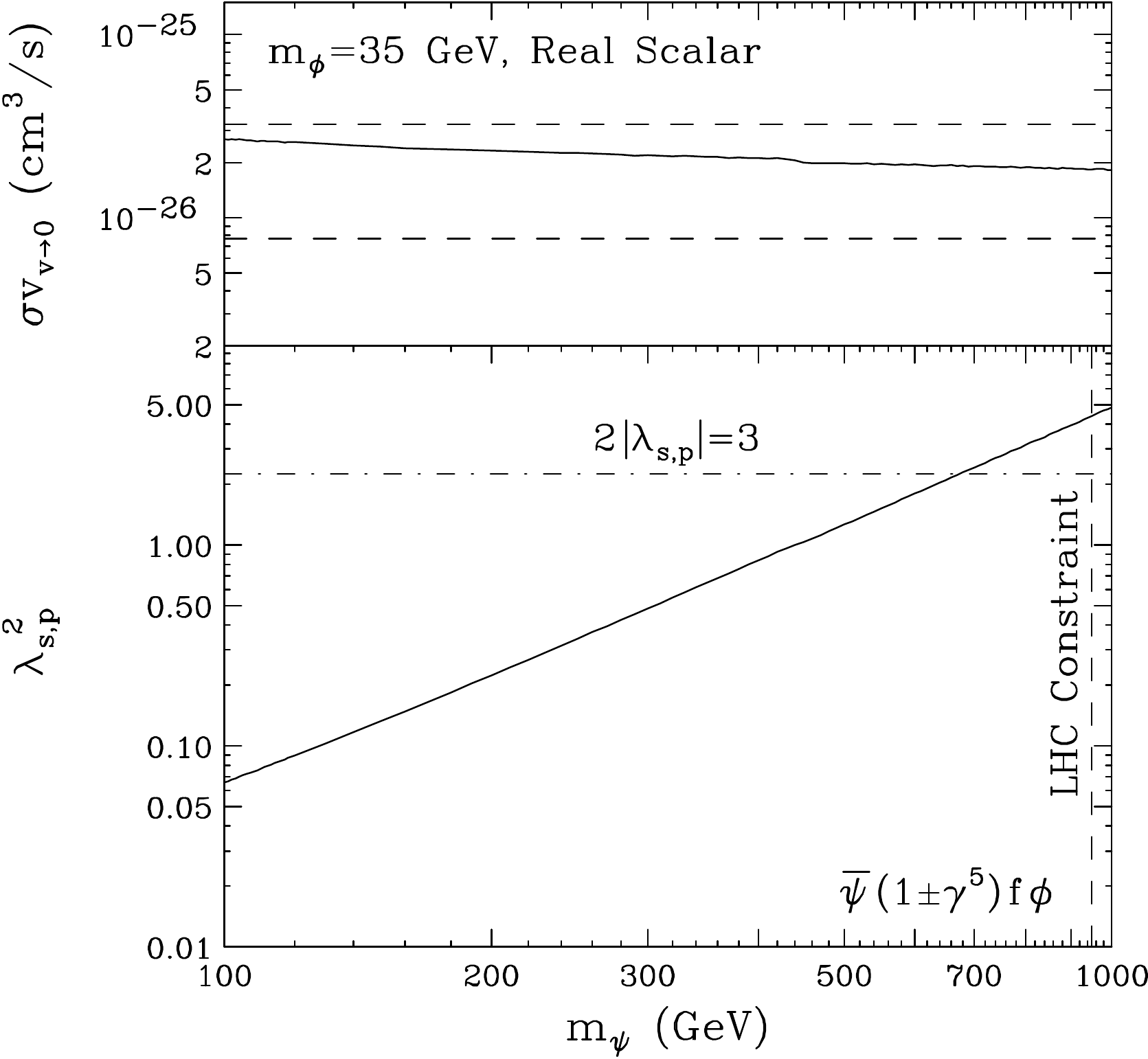}
\caption{Similar to previous figures, but for DM annihilating through $t$-channel Feynman diagrams. The four cases shown in this figure {\it cannot} account for the observed gamma-ray signal. The upper frames show the results for Majorana DM interacting through combinations of scalar and pseudoscalar (left) or vector and axial (right) interactions, while in the lower frames, we show the results for complex (left) or real (right) scalar DM interacting through a combination of scalar and pseudoscalar couplings. In the first three of these cases, the low-velocity annihilation cross section is too low to produce the gamma-ray excess.  In the real scalar case, the constraint from sbottom searches at the LHC (dashed) cannot be evaded without non-perturbative couplings (dot-dashed).}
\label{tchannelplotsbad} 
\end{figure*}

\begin{figure*}[t!]
\includegraphics[width=3.4in]{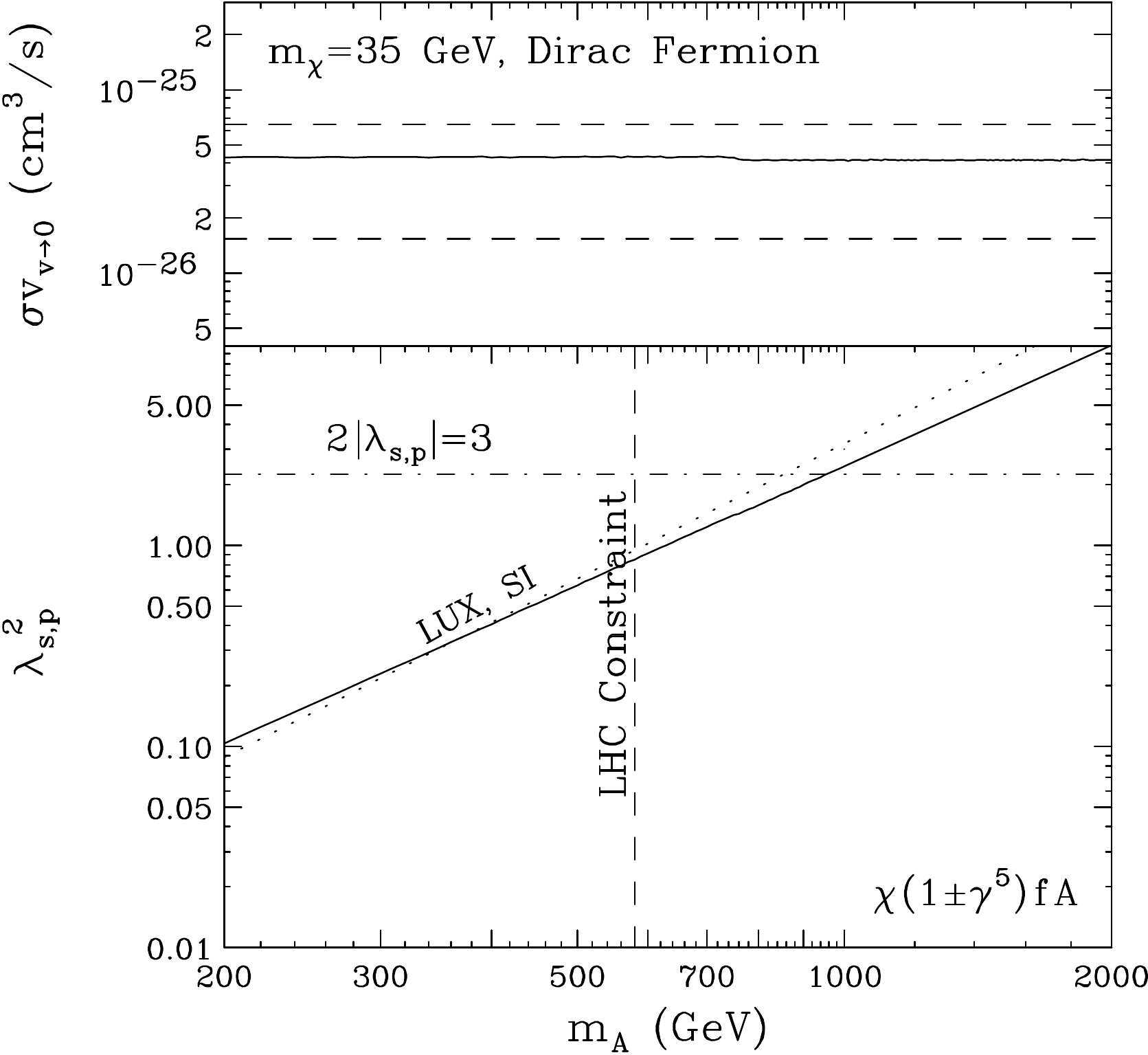}
\hspace{0.5cm}
\includegraphics[width=3.4in]{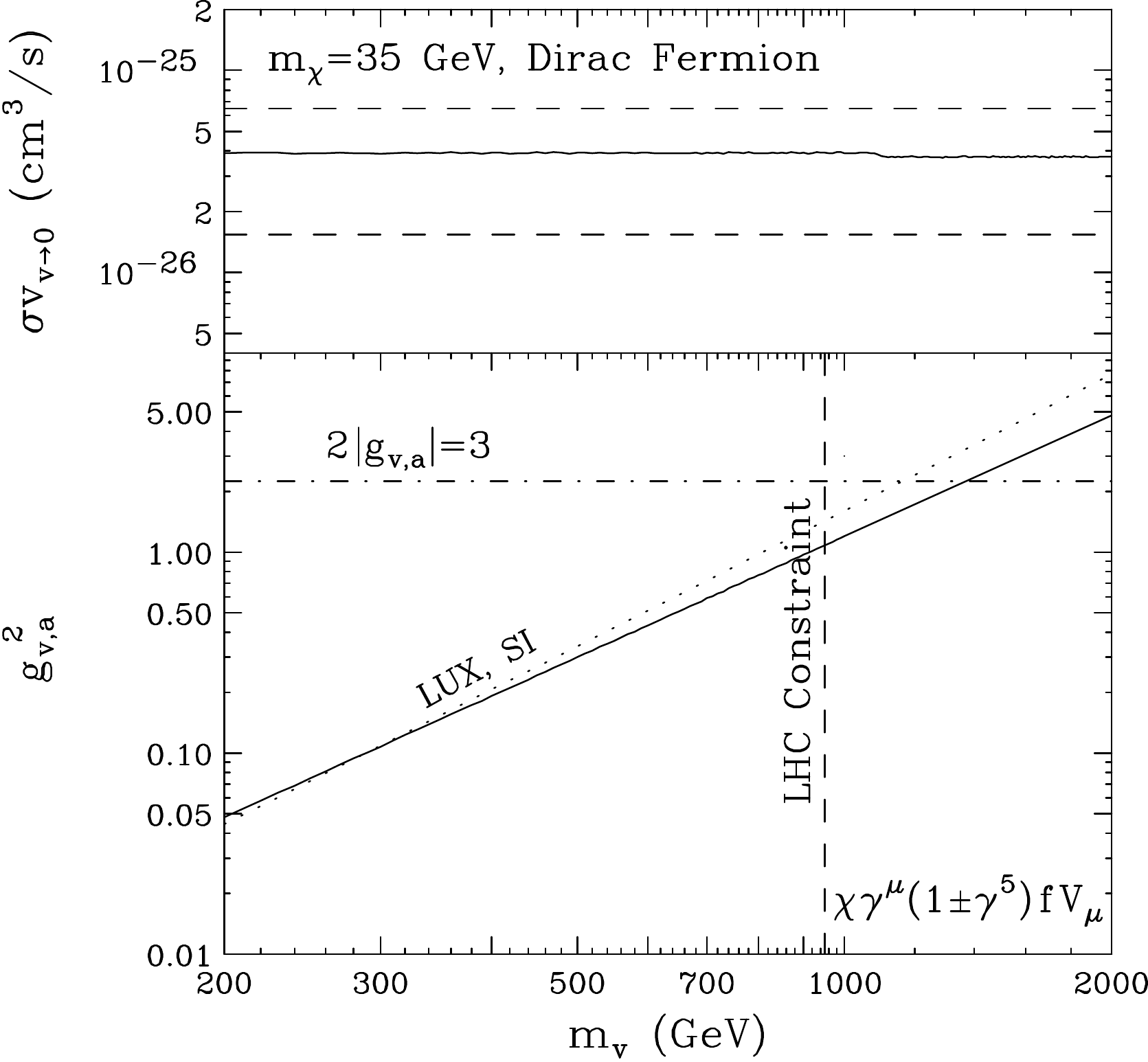} \\
\vspace{0.5cm}
\includegraphics[width=3.4in]{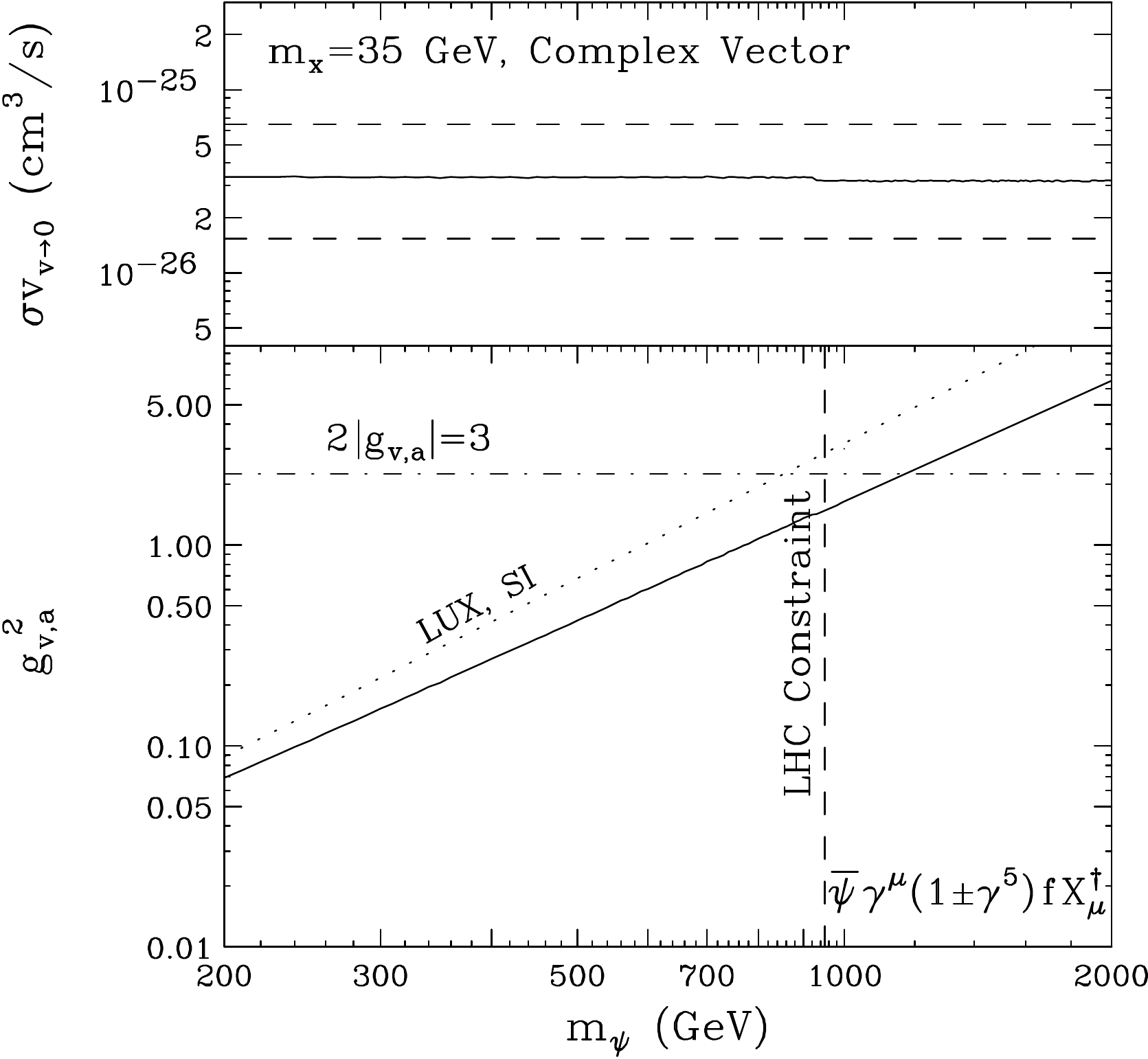}
\hspace{0.5cm}
\includegraphics[width=3.4in]{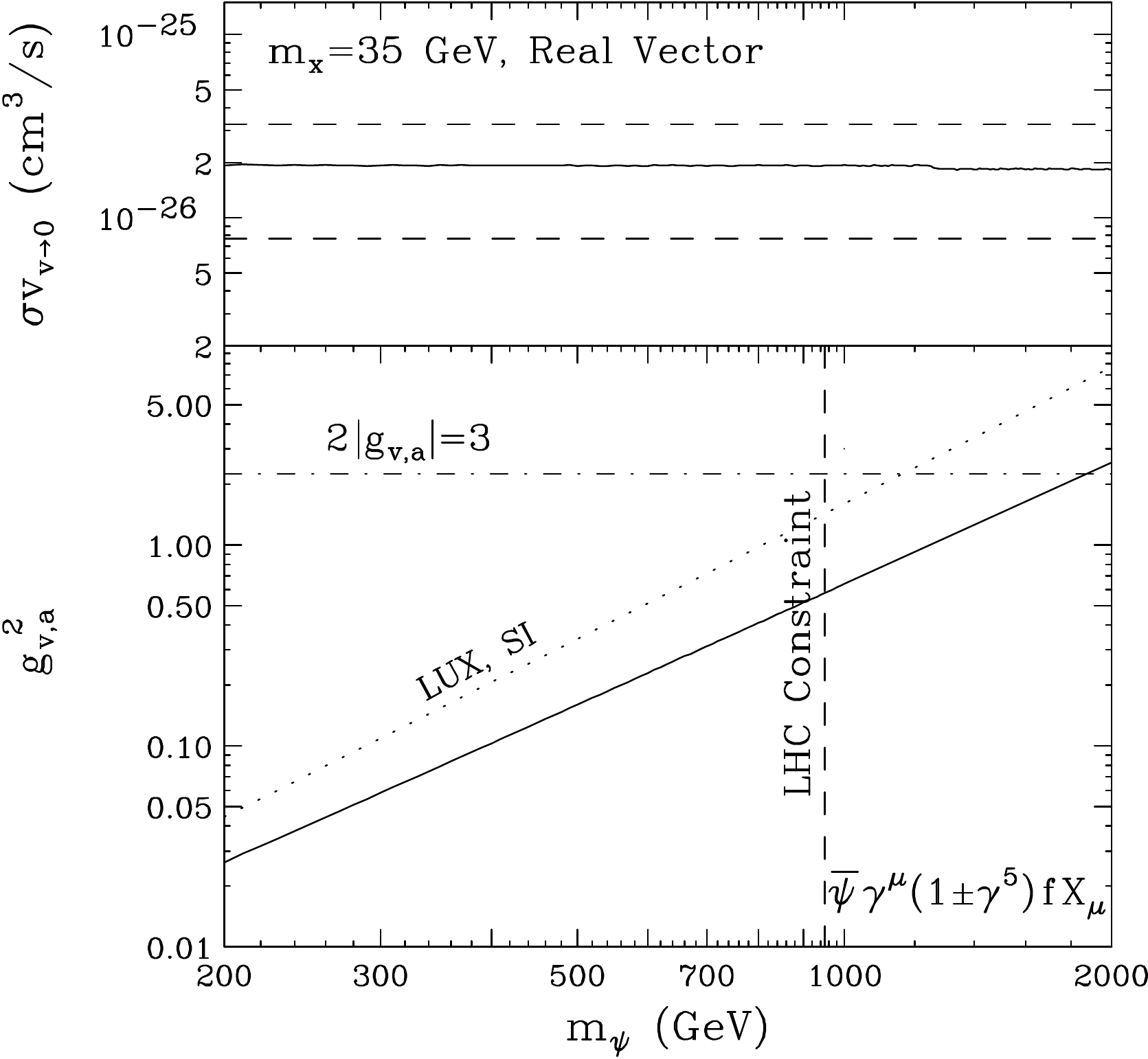}
\caption{Similar to previous figures, but for DM annihilating through $t$-channel Feynman diagrams. The four cases shown in this figure {\it can} account for the observed gamma-ray signal, without violating any constraints from the LHC or direct detection experiments. The upper frames show the results for Dirac DM interacting through combinations of scalar and pseudoscalar (left) or vector and axial (right) interactions, while in the lower frames, we show the results for complex (left) or real (right) vector DM interacting through a combination of vector and axial couplings. In the lower portion of each frame, the vertical line denotes the limit on the mass of the colored mediator, based on the results of sbottom searches at the LHC. The dot-dashed lines denote the approximate point at which the couplings become non-perturbative. Although direct detection and LHC constraints are each near the sensitivity required to test these scenarios, they do not rule out any of the models shown.}
\label{tchannelplots}
\end{figure*}

\section{Dark Matter Annihilating Through $t$-Channel Diagrams}
\label{tchannel}

In this section, we shift our focus to DM that is described by the $t$-channel Lagrangian of Eq.~\ref{schematic}. More specifically, we consider Fermionic DM, $\chi$, that annihilates into SM fermions via the $t$-channel exchange of either a spin-$0$ mediator, $A$:
\beq \label{LtdirS}
\mathcal{L} \supset  \bar{\chi} ( \lambda_s + \lambda_p \gamma^5 ) f A +  \bar{f} ( \lambda_s - \lambda_p  \gamma^5 ) \chi A^\dagger, 
\eeq
or a spin-$1$ mediator, $V_\mu$:
\beq
\mathcal{L} \supset  \bar{\chi} \gamma^\mu ( g_{\chi v} + g_{\chi a}\gamma^5 ) f V_\mu + \bar{f} \gamma^\mu ( g_{\chi v} + g_{\chi a}\gamma^5 ) \chi V_\mu^\dagger.
\eeq

We also consider the case of a spin-$1/2$ mediator, $\psi$, with either a scalar DM particle, $\phi$:
\beq
\mathcal{L} \supset  \bar{\psi} \left( \lambda_s + \lambda_p \gamma^5 \right) f \phi^\dagger +  \bar{f} \left( \lambda_s - \lambda_p \gamma^5 \right) \psi \phi,
\eeq
or vector DM, $X_\mu$:
\beq
\mathcal{L} \supset  \bar{\psi} \gamma^\mu \left( g_v+g_a \gamma^5 \right) f X_\mu^\dagger + \bar{f}  \gamma^\mu \left( g_v+g_a \gamma^5 \right) \psi X_\mu.
\eeq

These models are different from $s$-channel scenarios in three important ways. First, the $t$-channel mediator is required to carry the same quantum numbers as the final state quarks, and thus is both colored and charged. As a result, the mediator can be pair produced via QCD, making constraints from the LHC significantly more restrictive~\cite{Agrawal:2011ze,Chang:2013oia,An:2013xka,Papucci:2014iwa}. Second, direct detection constraints are in most cases much more difficult to evade, particularly in the case in which the DM or the mediator couples to first generation quarks~\cite{Agrawal:2011ze,Chang:2013oia,An:2013xka}. Finally, there are two diagrams that contribute to the scattering \cite{Agrawal:2011ze}, as opposed to a single diagram in the $s$-channel case, although the magnitude of the scattering cross-section is roughly the same. There are important caveats that apply to the first two of these these statements.

Just as in the case of $s$-channel annihilation models, we evaluate the elastic scattering cross section of the DM with nuclei by integrating out the mediator. In the $t$-channel case,  we then perform a Fierz transformation to convert the resulting contact operator into a sum of the $s$-channel interactions described in the preceding sections. If we start with any single interaction form in isolation (scalar, pseudoscalar, axial, or vector), this procedure invariably generates a non-negligible amount of all possible interaction forms~\cite{Agrawal:2010fh}. In particular, in each of these cases, we find an unsuppressed scalar contact interaction. As demonstrated in Figs.~\ref{scalarscalar} and~\ref{vectorscalar}, scalar contact interactions with couplings proportional to quark mass already significantly exceed the constraints from LUX.  This excludes the majority of $t$-channel models that we can consider.

The exception to this conclusion arises when one considers interactions which include {\it both} scalar and pseudoscalar couplings, or vector {\it and} axial couplings. In particular, a $t$-channel annihilation diagram with interactions of the form $1+ \gamma^{5}$ (as obtained for $\lambda_s=\lambda_p$) leads to an effective operator that is a sum of vector and axial interactions, such as $(1/2) \bar{\chi} \gamma^{\mu}(1-\gamma^5) \chi \, \bar{f} \gamma^{\mu}(1+\gamma^5) f$. Similarly, a $t$-channel annihilation diagram with an interaction of the form $\gamma^{\mu}(1+\gamma^5)$ (corresponding to $g_v= g_a$) transforms to yield an effective operator of the form $\gamma^{\mu}(1-\gamma^5)$. In both of these cases, no scalar term appears, and the leading direct detection constraint comes from the vector interaction.

If this coupling applies to all quarks, the vector interaction would still induce a very large scattering cross section, incompatible with direct detection constraints. If the $t$-channel mediator couples only to $b$-quarks, however, the elastic scattering cross section will be loop-suppressed, allowing it to evade the current limits. In particular, the elastic scattering cross section in this case is dominated by the exchange of a photon between the bottom-mediator loop and the nucleus. This behavior is found, for example, in the flavored DM models of Refs.~\cite{Agrawal:2011ze,Batell:2013zwa}.

In Table~\ref{tab:tchannel}, we summarize the characteristics of the eight $t$-channel models with interaction forms capable of suppressing the DM's scalar elastic scattering cross section with nuclei. Of these eight models, four provide a viable explanation for the gamma-ray excess. In Fig.~\ref{tchannelplotsbad}, we show the results for the four of these models that {\it cannot} account for this signal. In three of these cases (Majorana fermion DM with interactions of the forms $(1+\gamma^5)$ or $\gamma^{\mu}(1+\gamma^5)$, and complex scalar DM with a $(1+\gamma^5)$ interaction), the low-velocity annihilation cross section is too small to provide the observed gamma-rays. In the fourth case (real scalar DM with a $(1+\gamma^5)$ interaction), the LHC constraint on heavy bottom partners (as derived in Ref.~\cite{Chang:2013oia} from the results of the CMS sbottom search~\cite{Chatrchyan:2013lya}) can only be satisfied if the couplings are very large and non-perturbative. In each frame, the horizontal dot-dashed line represents the approximate point at which the couplings become non-perturbative (where the coefficients of an operator of the form $(1+ \gamma^5)/2$ or $\gamma^{\mu}(1+ \gamma^5)/2$ exceeds three).

\begin{figure*}[t!]
\includegraphics[width=3.45in]{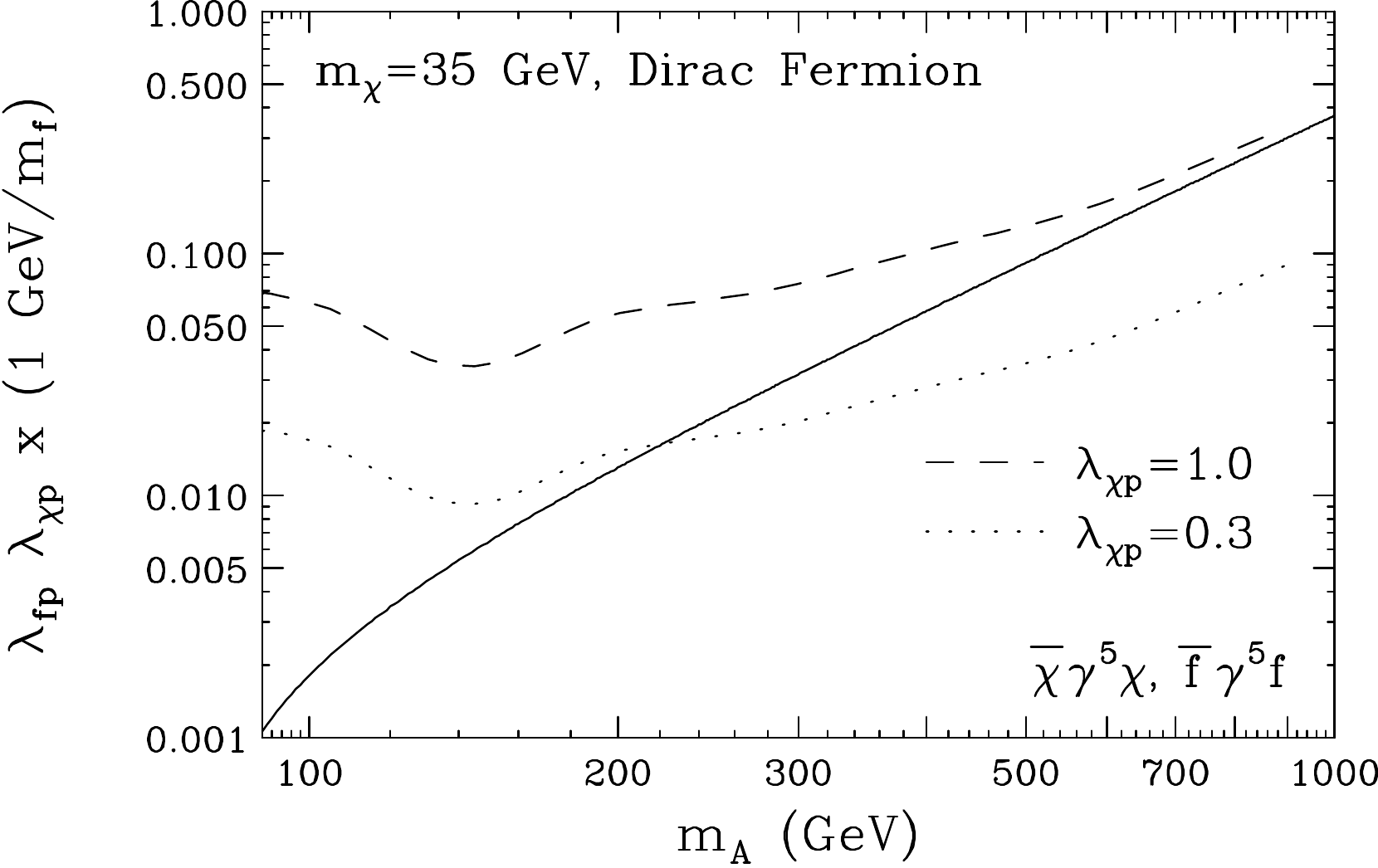}
\hspace{0.3cm}
\includegraphics[width=3.45in]{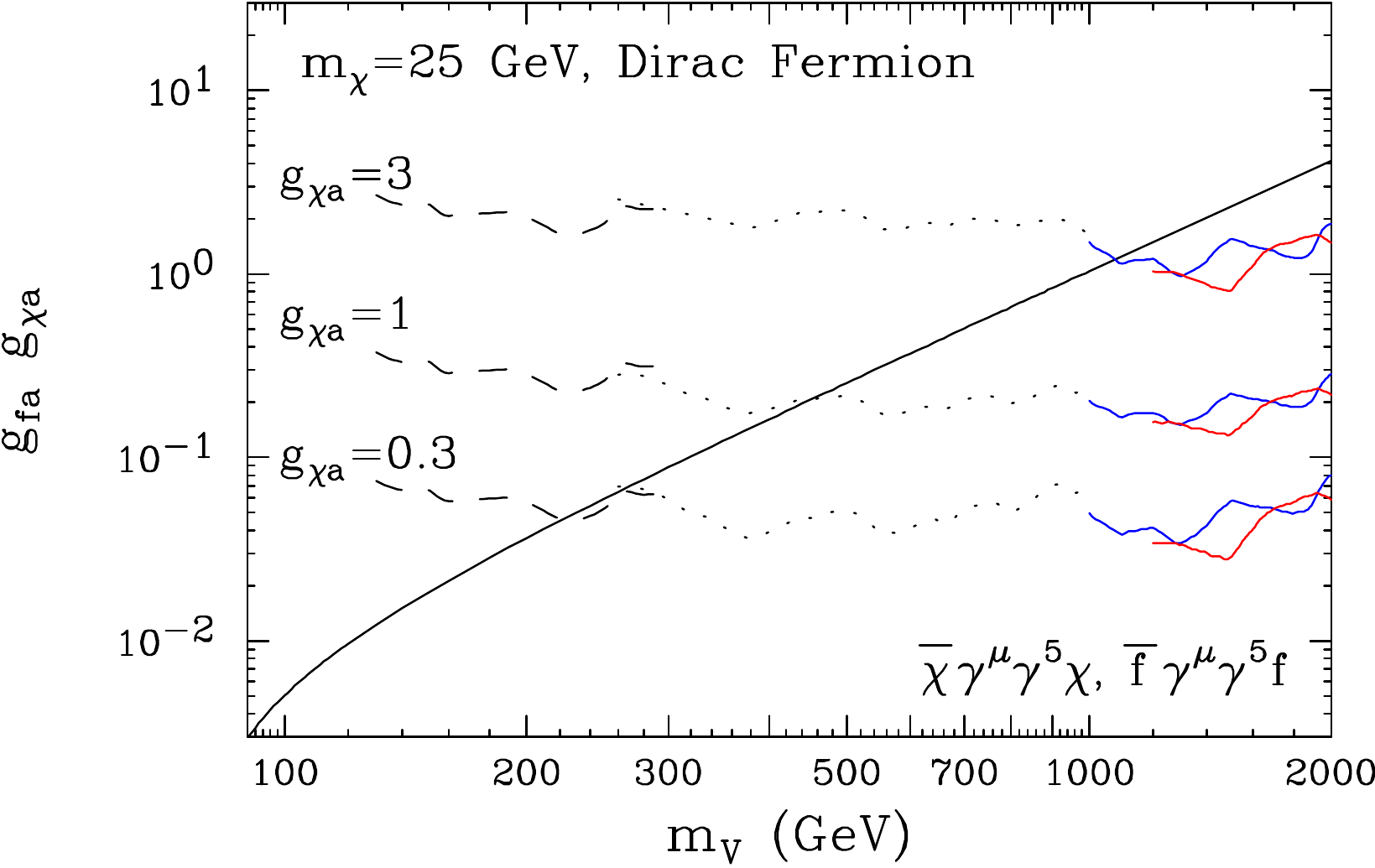} 
\caption{Constraints from the LHC and other colliders on the couplings of spin-$0$ or spin-$1$ particles that mediate the interactions of the DM. In the left frame, we plot the LHC's constraints on a spin-$0$ mediator~\cite{cmshiggs,Chatrchyan:2013qga,Aad:2011rv,Aad:2012cfr}, whereas in the right frame, we show the constraints on a spin-$1$ mediator from UA2~\cite{Alitti:1993pn} (dashes), CDF~\cite{Aaltonen:2008dn} (dotted), and CMS at 7 TeV~\cite{CMS:2012yf} (solid blue) and 8 TeV~\cite{CMSdijet} (solid red). In each frame, the black solid line represents the couplings required to generate a thermal relic abundance in agreement with the measured cosmological DM density.}
\label{colliderconstraints}
\end{figure*}

The four viable $t$-channel models are shown in Fig.~\ref{tchannelplots} (with cross sections given in Appendix~\ref{app:t}). For each of these scenarios, the predicted elastic scattering cross section is very close to the current constraints from LUX. These models should be definitively tested in the near future both by direct detection experiments and at the LHC.

\section{Additional Constraints from the LHC and other Colliders}
\label{collider}

Thus far in this paper, we have included discussion of the LHC constraints based on searches for mono-jet events (with and without $b$-tags), mono-$W/Z$ events, and sbottoms. In addition to such searches, collider experiments can directly search for and constrain the particles that mediate the DM's $s$-channel interactions.   In this section, we consider the impact of searches for vector and scalar particles at the LHC and at earlier collider experiments. 

The CMS~\cite{cmshiggs,Chatrchyan:2013qga} and ATLAS~\cite{Aad:2011rv,Aad:2012cfr} Collaborations have performed searches for the heavy, neutral, CP-even and CP-odd Higgs bosons predicted within the context of the minimal supersymmetric standard model (MSSM). The limits resulting from these searches can be directly translated into constraints on the mass and couplings of any spin-$0$ particle that might mediate the interactions of DM.  These constraints are shown in the left frame of Fig.~\ref{colliderconstraints}. For DM couplings of $\lambda_{\chi} \gsim 1$, these searches do not yet rule out any values of $m_A$. For smaller values of $\lambda_{\chi}$ (corresponding to larger values of $\lambda_{f}$), however, we can place an upper limit on $m_A$. For example, for couplings of $\lambda_{\chi}  \sim 0.3$, this bound constrains the mediator mass to be below $\sim$$\cO(250\gev)$.  We also point out that these constraints are dominated by the mediator's couplings to $\tau$ leptons. If we were to consider a model in which our spin-$0$ mediator coupled only to quarks, these constraints would be further weakened.

The LHC and other hadron colliders also provide constraints on spin-$1$ mediators through dijet searches. These constraints are summarized in the right frame of Fig.~\ref{colliderconstraints}, including limits from UA2~\cite{Alitti:1993pn}, CDF~\cite{Aaltonen:2008dn}, and CMS at both 7 TeV~\cite{CMS:2012yf} and 8 TeV~\cite{CMSdijet}. Again, these constraints do not rule out any of the scenarios considered in this paper. However, maintaining perturbativity in the dark sector does restrict the mass of any spin-$1$ mediator to be less than $\sim$1 TeV, with increasingly strong bounds for smaller DM couplings.  We note that if this spin-$1$ mediator also couples to electrons, then dilepton constraints could be somewhat more restrictive~\cite{Carena:2004xs}. As couplings to electrons do not play a significant role in the other aspects of this model, we do not directly consider these constraints here.

\begin{table*}[htdp]
\begin{center}
\begin{tabular}{|c||c|c||c|c|c|c|}
\hline {\bf \emph{Model}} &  \multirow{2}{*}{\bf{ \emph{DM}}} &  \multirow{2}{*}{ {\bf \emph{Mediator}}} &  \multirow{2}{*}{{\bf \emph{Interactions}}} & {\bf Elastic}  & \multicolumn{2}{|c|}{{\bf Near Future Reach?}}  \\ \cline{6-7} {\bf \emph{Number}} &&&& {\bf Scattering} & {\bf Direct} & {\bf LHC}  \\ 
\hline \hline 
1 & Dirac Fermion & Spin-$0$  &   $\bar{\chi}\gamma^5\chi$,  $\bar{f}f$&  $\sigma_{\rm SI} \sim (q/2m_{\chi})^2$  (scalar) & No & Maybe \\ \hline
1 & Majorana Fermion & Spin-$0$ &  $\bar{\chi}\gamma^5\chi$,  $\bar{f}f$&  $\sigma_{\rm SI} \sim (q/2m_{\chi})^2$  (scalar) & No & Maybe \\ \hline
2 & Dirac Fermion & Spin-$0$  &  $\bar{\chi}\gamma^5\chi$,  $\bar{f}\gamma^5f$&  $\sigma_{\rm SD} \sim (q^2/4 m_n m_{\chi})^2$  & Never & Maybe \\ \hline
2 & Majorana Fermion & Spin-$0$ &  $\bar{\chi}\gamma^5\chi$,  $\bar{f}\gamma^5f$ & $\sigma_{\rm SD} \sim (q^2/4 m_n m_{\chi})^2$ & Never & Maybe \\ \hline \hline
3 & Dirac Fermion & Spin-$1$ &    $\bar{\chi}\gamma^{\mu}\chi$, $\bar{b}\gamma_{\mu}b$ & $\sigma_{\rm SI} \sim$ loop (vector) & Yes & Maybe \\ \hline
\multirow{2}{*}{4} & \multirow{2}{*}{Dirac Fermion} & \multirow{2}{*}{Spin-$1$}    & \multirow{2}{*}{$\bar{\chi}\gamma^{\mu}\chi$,  $\bar{f}\gamma_{\mu}\gamma^5f$ }&$\sigma_{\rm SD} \sim (q/2 m_n)^2$ or & \multirow{2}{*}{Never} & \multirow{2}{*}{Maybe} \\ &&&&$\sigma_{\rm SD} \sim (q/2 m_{\chi})^2$& &\\ \hline
5 & Dirac Fermion & Spin-$1$ &    $\bar{\chi}\gamma^{\mu}\gamma^5\chi$,  $\bar{f}\gamma_{\mu}\gamma^5f$  & $\sigma_{\rm SD}\sim$ 1 & Yes & Maybe \\ \hline
5 & Majorana Fermion & Spin-$1$ &  $\bar{\chi}\gamma^{\mu}\gamma^5\chi$,  $\bar{f}\gamma_{\mu}\gamma^5f$ & $\sigma_{\rm SD}\sim$ 1 & Yes & Maybe \\
\hline \hline
6 & Complex Scalar & Spin-$0$  &    $\phi^{\dagger}\phi$,  $\bar{f}\gamma^5f$ &$\sigma_{\rm SD} \sim (q/2m_n)^2$   & No & Maybe \\ \hline
6 & Real Scalar        & Spin-$0$   &   $\phi^2 $,  $\bar{f}\gamma^5f$                     & $\sigma_{\rm SD} \sim (q/2m_n)^2$  &  No & Maybe \\ \hline
6 & Complex Vector & Spin-$0$  &  $B^{\dagger}_{\mu}B^{\mu}$,  $\bar{f}\gamma^5f$ & $\sigma_{\rm SD} \sim (q/2m_n)^2$  & No & Maybe\\ \hline
6 & Real Vector & Spin-$0$     & $B_{\mu}B^{\mu}$,  $\bar{f}\gamma^5f     $ &       $\sigma_{\rm SD} \sim (q/2m_n)^2$        &  No & Maybe\\ 
\hline \hline
7 & Dirac Fermion & Spin-$0$ ($t$-ch.)    & $ \bar \chi  ( 1 \pm \gamma^5) b$  &   $\sigma_{\rm SI} \sim$ loop (vector)  &    Yes & Yes\\ \hline
 7 & Dirac Fermion & Spin-$1$ ($t$-ch.)  & $ \bar \chi \gamma^\mu ( 1 \pm \gamma^5) b$ & $\sigma_{\rm SI} \sim$ loop (vector)  &  Yes & Yes\\ \hline
8 & Complex Vector & Spin-$1/2$ ($t$-ch.) & $X_{\mu}^{\dagger} \gamma^{\mu}(1\pm \gamma^5) b$ & $\sigma_{\rm SI} \sim$ loop (vector) &  Yes & Yes\\ \hline
8 & Real Vector & Spin-$1/2$ ($t$-ch.)& $X_{\mu} \gamma^{\mu}(1\pm \gamma^5) b$  & $\sigma_{\rm SI} \sim$ loop (vector) & Yes & Yes\\ 
\hline \hline
\end{tabular}
\end{center}
\caption{A summary of the simplified models identified in our study as capable of generating the observed gamma-ray excess without violating the constraints from colliders or direct detection experiments. In the last two columns, we indicate whether the model in question will be within the reach of near future direct detection experiments (LUX, XENON1T) or of the LHC. Models with an entry of ``Never'' predict an elastic scattering cross section with nuclei that is below the irreducible background known as the ``neutrino floor''. The ``Model Number" given in the first column provides the key for the model points shown in \Fig{direct}.}
\label{goodmodels}
\end{table*}%

\begin{figure*}[t!]
\includegraphics[width=4.5in]{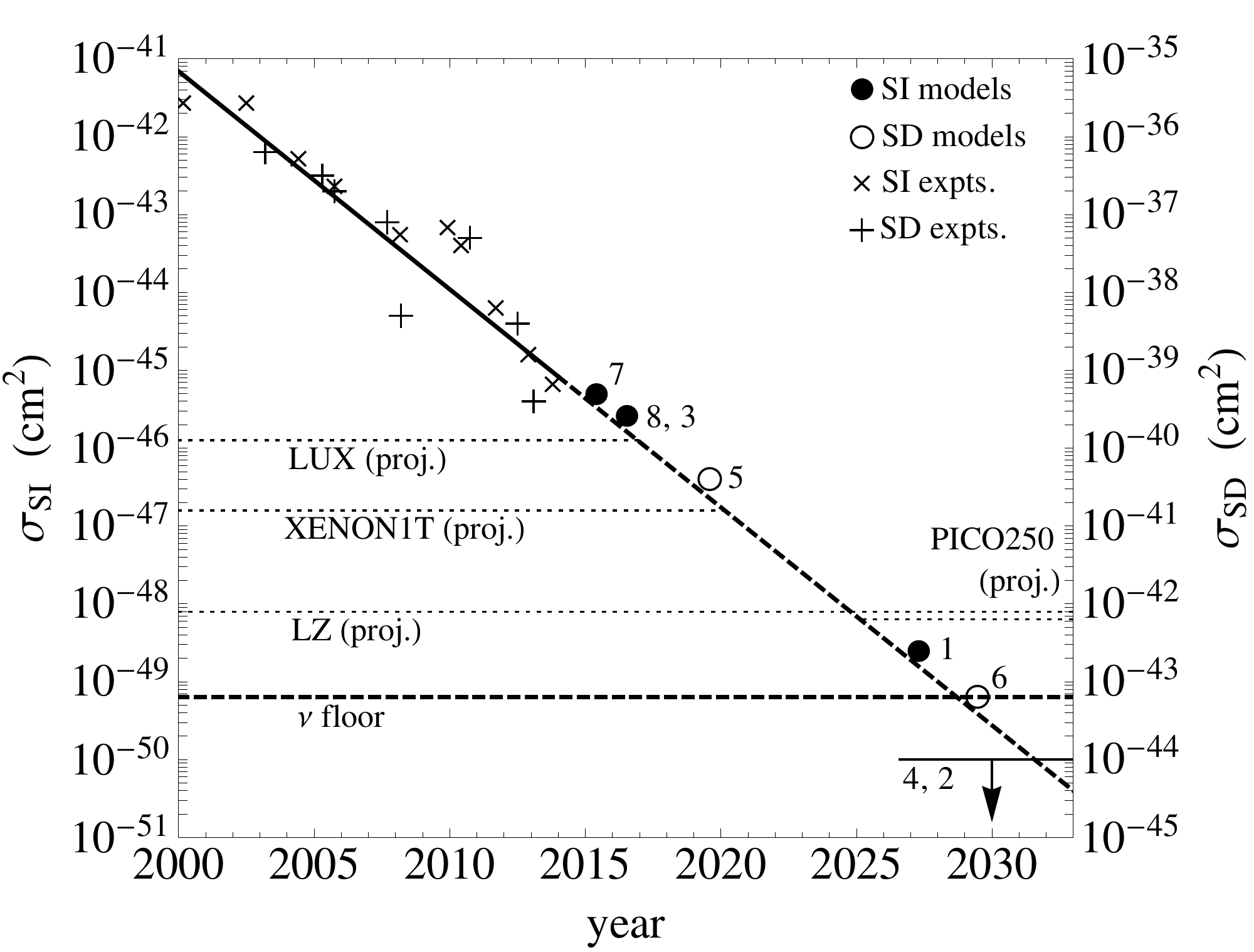}
\caption{The most stringent constraints on the DM elastic scattering cross-section \cite{Ahmed:2003su,Alner:2005kt,Akerib:2005za,Lee.:2007qn,Angle:2008we,Ahmed:2009zw,Ahmed:2008eu,Akerib:2005kh,Akerib:2004fq,Benoit:2002hf,Abusaidi:2000wg,Behnke:2010xt,Aprile:2012nq,Aprile:2011hi,Aprile:2010um,Behnke:2012ys,Angle:2007uj,Aprile:2013doa} from the past 14 years. We also shown an extrapolation of their future sensitivity. All of the models in which the DM annihilates through a $t$-channel Feynman diagram should be well within the reach of LUX~\cite{Akerib:2012ys} and XENON1T~\cite{Aprile:2012zx}. Fermionic DM that annihilates through a mediator with purely axial interactions is also expected to be within the reach of these experiments. In the more distant future, direct detection experiments also could become sensitive to several models in which the DM interacts via pseudoscalar couplings. See text for further details.}
\label{direct}
\end{figure*}

\section{Prospects for Direct Detection}
\label{directD}

In this paper, we have identified sixteen simplified models for DM annihilation that are capable of accounting for the observed gamma-ray excess without violating the constraints of colliders or direct detection experiments (these models are summarized in Table~\ref{goodmodels}). In this section, we discuss the prospects for future direct detection experiments to constrain or detect the DM particles associated with these models.

Roughly speaking, direct detection experiments are most sensitive to DM particles with a mass similar to that of the target nuclei. Such experiments are thus well suited to studying DM particles with masses in the range being considered here. In Fig.~\ref{direct}, we plot how the most stringent constraints on the DM elastic scattering cross section ($+$'s and $\times$'s) have evolved over the past 14 years, consistently improving at an exponential rate. Assuming that a similar rate of progress continues (as represented by the dashed line), we expect several of the models described in this study to be tested by direct detection experiments in the near future. In particular, all of the models in which the DM annihilates through a $t$-channel Feynman diagram should be well within the reach of LUX and XENON1T. Fermionic DM that annihilates through a mediator with purely axial interactions, or through a mediator with purely vector interactions with third generation quarks, is also expected to be probed by these ongoing and upcoming experiments.

In the more distant future, direct detection experiments could become sensitive to many more of the models listed in Table~\ref{goodmodels}. In particular, scalar or vector DM which annihilates through a spin-0 mediator with pseudoscalar couplings to SM fermions could eventually be detected, but would require extremely large detectors, beyond the next generation currently being planned (LZ, PICO250, etc.). Fermionic DM annihilating through a combination of pseudoscalar and scalar couplings could also be detected on this timescale. Extending direct detection sensitivity beyond that level, however, will be limited by the irreducible background induced by coherent neutrino scattering (known as the ``neutrino floor''). Due to this background, direct detection experiments would be unlikely to be able to detect fermionic DM annihilating through the exchange of a mediator with only pseudoscalar interactions, or through a spin-$1$ mediator with vector and axial couplings to the DM and SM fermions, respectively.

\section{Conclusions}
\label{conclusions}

In this study, we have taken a ``simplified model'' approach to determine which classes of dark matter models are capable of producing the gamma-ray excess observed from the region surrounding the Galactic Center. In doing so, we have identified 16 different models that can generate the observed excess without exceeding any of the constraints from direct detection experiments or from colliders (see Table~\ref{goodmodels}). These 16 models can be divided into the following three groups:
\begin{itemize}
\item{Models in which the dark matter (which could be spin-$0$, $1/2$, or $1$) annihilates through the exchange of a spin-$0$ particle with pseudoscalar interactions. Such a mediator could potentially be observed in future searches for heavy neutral Higgs bosons at the LHC.}
\item{Models in which the dark matter is a fermion that annihilates through the exchange of a spin-$1$ particle with axial couplings to standard model fermions, or with vector couplings to third generation standard model fermions. Assuming perturbative couplings, LHC constraints from dijet searches require that the mass of the mediator be {\it less} than $\sim$1 TeV.}
\item{Models in which the dark matter annihilates into b-quark pairs through the $t$-channel exchange of a colored and charged particle. Constraints from sbottom searches at the LHC restrict the mediator mass be greater than $\sim$600 GeV.  Both LUX and the LHC should be able to conclusively test this class of models in the near future.}
\end{itemize}

Upon reviewing this list of possibilities, it is clear that a wide range of simple dark matter models could be responsible for the Galactic Center's gamma-ray excess without running afoul of existing constraints. Moreover, the prospects for detecting the dark matter in these scenarios at either direct detection experiments or at the LHC appear to be quite promising. Of the 16 viable models identified in our study, LUX and XENON1T are expected to be sensitive to 7. Only 3 of these 16 models predict an elastic scattering cross section that will remain beyond the reach of future direct detection experiments due to the irreducible neutrino floor. Mono-jet searches, sbottom searches, and searches for heavy Higgs bosons at the LHC will further restrict the range of model parameters that remains viable. With 13-14 TeV data from the LHC, it will be possible to conclusively test several of the scenarios presented here. 

Many of the results presented in this study nicely illustrate the complementarity between indirect, direct, and collider searches for dark matter. Although future astrophysical observations (such as gamma-ray searches for dark matter annihilating in dwarf galaxies~\cite{Ackermann:2013yva} or future cosmic-ray anti-proton measurements~\cite{Cirelli:2013hv,Fornengo:2013xda}) may provide additional support for a dark matter interpretation of the Galactic Center gamma-ray excess, indirect detection signals alone are expected to determine little more than the mass and annihilation cross section of the particles that make up the dark matter, leaving many questions unanswered. Information from a combination of direct detection experiments and colliders will be needed if one is to identify the underlying interactions and particle content of the dark sector.

\bigskip

%

%%%%%%%%%%%%

%\noindent\makebox[\linewidth]{\rule{\columnwidth}{0.5pt}}
\section*{Acknowledgements}

We would like to thank Prateek Agrawal, Brian Batell, Jason Kumar, Tongyan Lin, and Felix Yu for helpful discussions. AB is supported by the Kavli Institute for Cosmological Physics at the University of Chicago through grant NSF PHY-1125897. SDM is supported by the Fermilab Fellowship in Theoretical Physics. Fermilab is operated by Fermi Research Alliance, LLC, under Contract No.~DE-AC02-07CH11359 with the US Department of Energy.

\section*{Appendix}

\appendix

\section{Cross Section Calculations}

\subsection{Annihilation}

The annihilation cross section for two DM particles resulting in final state fermions of identical mass, $m_f$, is given by:
\alg{
%\frac{d \sigma_{\rm ann}}{d \Omega} &= \frac{ \mL \overline \cM \mR^2}{64\pi^2 s}\frac{\sqrt{1-4m_f^2/s}}{\sqrt{1-4m_{\rm DM}^2/s}\\ 
\frac{d \pL \sigma_{\rm ann}  v \pR}{d \Omega} &= \frac{ \mL \overline \cM \mR^2}{32\pi^2 s}\sqrt{1-4m_f^2/s},
}
where $\Omega$ is the angle of the outgoing particles in the center-of-mass frame relative to the angle of the incoming particles, $v$ is understood to be the relative velocity between the dark matter particles in the center of mass frame, $\sqrt{s}$ is the center-of-mass energy, and $\mL \overline \cM \mR^2$ is the spin-averaged annihilation matrix element squared. 

The thermally averaged annihilation cross section, which is of interest for relic density and indirect detection expectations, is given by~\cite{Gondolo:1990dk}:
\beq \label{GGEqs}
\langle \sigma_{\rm ann}  v \rangle \simeq \frac{ 2 x^{3/2}}{\sqrt \pi} \int_0^\infty \pL \sigma v\pR_{\rm lab} \ep^{1/2} \exp\pL-x \ep \pR d\ep,
\eeq
where $x \equiv m/T$ and $\ep \equiv  p_{\rm rel}^2/{4m^2}=(v/2)^2 / [1-(v/2)^2]$. Far from resonances and particle production thresholds, and taking the limit $v\to0$, the annihilation cross section is well approximated by the first two terms of its Taylor series expansion. In this case, we write $\pL \sigma v\pR_{\rm lab} \simeq\pL \sigma v\pR_{\rm c.m.} \simeq a +bv^2 \simeq a+4 b \ep$, so the velocity-averaged cross section can be approximated by evaluating the integrals
\alg{
\frac{ 2 x^{3/2}}{\sqrt \pi} \int_0^\infty \ep^{1/2} \exp(- x \ep)d\ep&= 1 
\\ \frac{ 8 x^{3/2}}{\sqrt \pi} \int_0^\infty \ep^{3/2} \exp(- x \ep)d\ep &= \frac6x \simeq \frac94 v^2,
}
where the final equality assumes equipartition of energy in the non-relativistic limit. These assumptions give the standard relation
\beq \label{sigma v approx}
\langle \sigma_{\rm ann}  v \rangle \simeq  a + \frac94 bv^2,
\eeq
which is acceptable for understanding annihilations far from any resonances or thresholds. We emphasize that the physics of annihilation through an $s$-channel mediator that can be produced on shell is not captured by this expansion, and the full annihilation cross section may be needed to produce accurate results.

\subsection{Elastic Scattering and Direct Detection} 
\label{directappendix}

\begin{table*}[htdp]
\begin{center}
\begin{tabular}{|c|c||c|c|c|}
\hline \multirow{2}{*}{  $\langle S \rangle_{\rm DM}$} &  \multirow{2}{*}{ {\bf Type}} &  \multirow{2}{*}{{\bf Interaction}} & {\bf Elastic}  &  {\bf Kinematic} \\ &&& {\bf Scattering} & {\bf Suppression} \\ \hline
1/2 & Dirac  &   $\bar{\chi}\gamma^5\chi\bar{q}q$&  SI (scalar) &$(q/2m_{\chi})^2$ \\ \hline
1/2 & Majorana &  $\bar{\chi}\gamma^5\chi\bar{q}q$&  SI (scalar)& $(q/2m_{\chi})^2$ \\ \hline
1/2 & Dirac  &  $\bar{\chi}\gamma^5\chi\bar{q}\gamma^5q$&  SD& $(q^2/4m_n m_{\chi})^2$ \\ \hline
1/2 & Majorana &  $\bar{\chi}\gamma^5\chi\bar{q}\gamma^5q$ & SD &$(q^2/4m_n m_{\chi})^2$ \\ \hline
1/2 & Dirac &    $\bar{\chi}\gamma^{\mu}\chi\bar{q}\gamma_{\mu}q$ & SI (vector) & 1 \\ \hline
1/2 & Dirac    & $\bar{\chi}\gamma^{\mu}\chi\bar{q}\gamma_{\mu}\gamma^5q$ & SD & $(q/2m_n)^2$ or $(q/2 m_{\chi})^2$\\ \hline
1/2 & Dirac &    $\bar{\chi}\gamma^{\mu}\gamma^5\chi\bar{q}\gamma_{\mu}\gamma^5q$ & SD & 1 \\ \hline
1/2 & Majorana &  $\bar{\chi}\gamma^{\mu}\gamma^5\chi\bar{q}\gamma_{\mu}\gamma^5q$ & SD& 1 \\
\hline \hline
0 & Complex  &  $\phi^{\dagger}\phi\bar{q}q$  &       SI (scalar)& 1 \\ \hline
0 & Real    &   $\phi^2 \bar{q}q$          &          SI (scalar)& 1 \\ \hline
0 & Complex  &    $\phi^{\dagger}\phi\bar{q}\gamma^5q$ &SD (scalar)   &$(q/2m_n)^2$ \\ \hline
0 & Real       &$\phi^2 \bar{q}\gamma^5q$      &      SD (scalar)  & $(q/2m_n)^2$ \\
\hline \hline
1 & Complex   &  $B^{\dagger}_{\mu}B^{\mu}\bar{q}q$   &       SI (scalar) & 1\\ \hline
1 & Real    &   $B_{\mu}B^{\mu}\bar{q}q$         &           SI (scalar) & 1 \\ \hline
1 & Complex  &  $B^{\dagger}_{\mu}B^{\mu}\bar{q}\gamma^5q$ & SD      &       $(q/2m_n)^2$ \\ \hline
1 & Real     & $B_{\mu}B^{\mu}\bar{q}\gamma^5q     $ &       SD      &       $(q/2m_n)^2$\\ \hline
\end{tabular}
\end{center}
\caption{Direct detection suppression of various operators that can produce $s$-wave DM annihilation.}
\label{default}
\end{table*}%

In this appendix, we describe our calculations of the dark matter's elastic scattering cross section with nuclei, and its implications for direct detection. We focus on those scenarios with unsupressed low-velocity annihilation cross sections (those potentially able to account for the observed gamma-ray excess).  We will provide the analytic expression for each scattering cross section in the following appendix; here, we simply describe our method of calculation.

To begin, we make the simplifying assumption of low momentum transfer. As long as the mass of the mediating particle is well above the momentum transfer of a typical elastic scattering event ($m_{\rm med} \gsim 100$ MeV), we can safely approach direct detection within the context of effective field theory and integrate out this mediator.

Given this assumption, the direct detection interaction for dark matter from an $s$-channel Lagrangian will be simply described by a single dimension-five or -six operator,
\alg{
\cL_{\rm DD} &\supset \cO_{\rm DM} \cdot {\rm Prop}_{\rm med} \cdot \cO_q
\\ &~~\to \pL \frac1{m_{\rm med}^2} \cO_{\rm DM} \pR \cO_n
\\&~~~~~ \to \pL \frac1{m_{\rm med}^2} \cO_{\rm DM} \pR \cO_N,
}
where $\cO_q$ is the quark-level SM operator and $\cO_n$ is the nucleon-level operator. These nucleon-level operators are summed over to get nucleus-level operators, $\cO_N$. Spin-independent interactions sum coherently over the target nucleus, $N$, such that
\beq
\cO_N^{\rm SI} = Z \cO_p + (A-Z) \cO_n.
\eeq
Spin-dependent interactions, on the other hand, go like
\beq
\cO_N^{\rm SD} = \pL \frac{\langle S_p \rangle}{J_N} \cO_p+\frac{\langle S_n \rangle}{J_N} \cO_n \pR \langle J_N \rangle,
\eeq
where $J$ is the spin of the nucleus. After summing over the quarks and nucleons, we square the matrix element, average over incoming spins, and sum over outgoing spins. Ultimately, to derive cross sections, we must take the nonrelativistic limits for spinors. For spin-independent interactions, we have both $ \overline \xi_{\rm out}  \xi_{\rm in}  \to 2m_\xi$ and $ \overline \xi_{\rm out} \gamma^\mu \xi_{\rm in}  \to 2m_\xi \delta_0^\mu$, while for spin-dependent  interactions, we have $ \overline \xi_{\rm out}  \gamma^5 \xi_{\rm in}  \to 2 q_\xi^i \langle S^i_\xi \rangle$ and $ \overline \xi_{\rm out} \gamma^\mu \gamma^5 \xi_{\rm in}  \to 4 m_\xi \langle S^i_\xi \rangle \delta^\mu_i$. See Ref.~\cite{Kumar:2013iva} for further details.

The $t$-channel interactions can be constructed from the $s$-channel cases by Fierz transformations~\cite{Agrawal:2010fh}.\footnote{We remind the reader that by $t$- or $s$- channel, we refer to the diagram responsible for annihilation, as opposed to the diagram for elastic scattering.} 
Note that the DM and SM physics factorize into distinct pieces. Here we discuss the summation of the SM operator into pieces relevant for the nucleon, $n$. We will use tildes to denote that, in contrast with much of the literature, our couplings in the following discussion are all dimensionless.

Consider the interaction of the DM through a scalar-mediated force with quarks:
\alg{
\cO_{{\rm SI},s} =& \widetilde f_q  \bar f f
%\\&
\to 
\widetilde f_n \overline n n.
}
As scalar couplings generally scale with the mass of the interacting fermion, such interactions are typically dominated by the dark matter's couplings to the nucleon's strange quark content, or by the coupling to gluons through loops of heavy quarks ($c$, $b$, $t$). DM scattering off the nucleons is given by summing over their quark content, so we are interested instead in the coefficients $\widetilde f_n$:
\alg{ \label{scalar-couplings}
\frac{\widetilde f_n}{m_n} =& \sum_{q=u,d,s} f_{T_q}^n\,\frac{\widetilde f_q}{m_q} +\frac{2}{27} \, f_{TG} \sum_{q=c,b,t} \frac{\widetilde f_q}{m_q}
\\ & \to \frac{\lambda_s}\gev \pL  \frac79 \sum_{q=u,d,s} f_{T_q}^n +\frac29 \pR ,
}
where in the second line we take $\widetilde f_q = \lambda_s m_q/{\rm GeV}$ and $f_{TG}=1-f^n_{T_u}-f^n_{T_d}-f^n_{T_s}$. We adopt the standard values to describe the nuclear quark content: $f_{T_u}^p=0.020$ and $f_{T_d}^p=0.026$ (and the reverse for neutrons) and $f_{T_s}=0.043$, as favored by recent lattice QCD calculations~\cite{Junnarkar:2013ac}. This gives $f_{TG} \simeq 0.91$, which implies that heavy quark loops mediate much of this scattering. 

Consider now the interaction of the DM through a pseudoscalar-mediated force with quarks:
\alg{
\cO_{{\rm SD}, p} = & \widetilde t_q  \bar f \gamma^5 f
%\\& 
\to  \widetilde t_n  \overline n \gamma^5 n.
}
This interaction is spin-dependent and momentum suppressed \cite{Kumar:2013iva}. We will include the momentum suppression at the target level because the momentum transfer defined by $q^2=2\mu_{\chi N}^2 v^2 (1-\cos \theta)$ is set at the reduced DM-nucleus mass scale. The heavy quarks and gluons contribute negligibly to the spin content of the nucleons, so DM scattering off nucleons is given by summing over the light quarks. The coefficients $\widetilde t_n$ can be written as:
\alg{ \label{pseudoscalar-couplings}
\frac{\widetilde t_{n}}{m_{n}} =&  \sum_{q=u,d,s} \frac{\widetilde t_q}{m_q} f^{(5n)}_q  
\\& \to \frac{\lambda_{p} }{ \gev}  \sum_{q=u,d,s} f^{(5n)}_q ,
}
where in the second line we take $\widetilde t_q = \lambda_{p} m_q/{\rm GeV}$. For our exclusion curves, we take the standard values to describe the nucleon quark content, with \cite{Cheng:2012qr} $f_u^{(5p)}=0.43$, $f_d^{(5p)}=-0.84$, $f_u^{(5n)}=-0.42$, $f_d^{(5n)}=0.85$, $f_s^{(5p)}=-0.50$, and  $f_s^{(5n)}=-0.08$.

We are also interested in the interaction of the DM through a vector-mediated force:
\alg{
\cO_{{\rm SI},v} =& \widetilde b_q   \bar f \gamma_\mu f
%\\ &
\to \widetilde b_n   \overline n \gamma_\mu n.
}
This is spin-independent, and since the vector current is conserved only valence quarks contribute to the interaction. Thus, DM scattering off nucleons is trivial to write down. The coefficients $\widetilde b_n$ can be written as:
\alg{ \label{vector-couplings}
\widetilde b_p=2\widetilde b_u+\widetilde b_d, \qquad \widetilde b_n=\widetilde b_u+2\widetilde b_d.
}
We will take $\widetilde b_q = \lambda_{v}$ to be uniform for all quarks.

Finally, we discuss the interaction of the DM through an axial force with quarks:
\alg{
\cO_{{\rm SD},a} =& \widetilde d_q  \bar f  \gamma_\mu \gamma^5 f
%\\ &
 \to  \widetilde a_n  \overline n  \gamma_\mu \gamma^5 n.
}
This is spin-dependent but is not suppressed by any powers of momentum. The heavy quarks and gluons contribute negligibly to the spin content of the nucleon, so DM scattering off nucleons is given by summing over the light quarks. The coefficients that describe this scattering off nucleons are traditionally written as $a_n$:
\alg{ \label{axial-couplings}
\widetilde a_{n} =&   \sum_{q=u,d,s} \widetilde d_q \Delta^{(n)}_q  
\\& \to  \lambda_{a} \sum_{q=u,d,s} \Delta^{(n)}_q ,
}
where the $\Delta$'s are defined by $\Delta_ u^{(p)}=\Delta_ d^{(n)}=0.84$, $\Delta_ u^{(n)}=\Delta_ d^{(p)}=-0.43$, and $\Delta_ s^{(p)}=\Delta_ s^{(n)}=-0.09$~\cite{Cheng:2012qr}. We also have $\widetilde d_q = \lambda_{a}$, assumed to be uniform for all quarks.

\begin{widetext}
\section{$s$-Channel Interactions}\label{app:s}

We list all annihilation and direct detection cross sections for dark matter that interacts through $s$-channel Lagrangians. Simplified models with annihilation cross sections that are unsuppressed in the limit $v \to 0$ are presented in Secs.~\ref{diracscalar},~\ref{majS},~\ref{dirV},~\ref{majoranavector},~\ref{cszero},~\ref{rszero},~\ref{cvzero}, and \ref{rvzero}. For completeness, we also include simplified models that annihilate with $v^2$ suppression in Secs.~\ref{csone},~\ref{rsone},~\ref{cvone}, and \ref{rvone}.

\subsection{Dirac dark matter, spin zero mediator}
\label{diracscalar}

Consider the following interactions for a Dirac dark matter particle, $\chi$, and a spin-$0$ mediator, $A$:
\begin{equation}
\mathcal{L} \supset  \left[ \bar{\chi}( \lambda_{\chi s} + \lambda_{\chi p} i \gamma^5 ) \chi +  \bar{f}( \lambda_{f s} + \lambda_{f p} i \gamma^5 ) f \right] A.  \\ \nonumber
\end{equation}
The cross section for $\chi\bar{\chi}$ annihilation is given by:
\begin{equation}
\sigma = \frac{n_c}{16 \pi s \bL (s-m_A^2)^2 +m_A^2 \Gamma_A^2 \bR} \sqrt{\frac{1-4 m_f^2/s}{1-4 m_{\chi }^2/s}} \Big[\lambda_{f s}^2 (s-4 m_f^2)+ \lambda_{f p}^2 s\Big] \Big[\lambda_{\chi s}^2 (s-4 m_{\chi }^2)+ \lambda_{\chi p}^2 s\Big] \label{sigsdirS}. \\
\end{equation}
Where $n_c=$3 for quarks and 1 for leptons.  Expanding in powers of $v^2$ gives
\begin{eqnarray}
%\begin{align}
\sigma v  &\approx& \frac{n_c \lambda_{\chi p}^2 \sqrt{1-m_f^2/m_\chi^2} \Big[m_{\chi }^2 (\lambda_{f p}^2+\lambda_{f s}^2)-m_f^2 \lambda _{f s}^2\Big]}{2 \pi   \left(m_A^2-4 m_{\chi }^2\right)^2} \nonumber \\
&+&\frac{n_c v^2}{16 \pi  m_{\chi }^2 \left(4 m_{\chi }^2-m_A^2\right)^3 \sqrt{1-m_f^2/m_\chi^2}} \Bigg[ \lambda _{\chi p}^2 \bigg\{\lambda_{f p}^2 m_{\chi }^2 \Big(m_A^2 (m_f^2-2 m_{\chi }^2)+12 m_f^2 m_{\chi }^2-8 m_{\chi }^4\Big)  \nonumber \\
&+&\lambda_{f s}^2 (m_f^2-m_{\chi }^2) \Big(m_A^2 (m_f^2+2 m_{\chi }^2)-20 m_f^2 m_{\chi }^2+8 m_{\chi }^4\Big)\bigg\}  \label{sigVsdirS} \\
&-&2 \lambda _{\chi s}^2 (m_A^2-4 m_{\chi }^2) (m_{\chi }^2-m_f^2) \Big(m_{\chi }^2 (\lambda_{f p}^2+\lambda _{f s}^2)-m_f^2 \lambda _{f s}^2\Big)\Bigg]. \nonumber
%\end{align}
\end{eqnarray}
The mediator's width to SM fermions is given by:
\alg{
\Gamma_A \equiv  \sum_f \Gamma(A \to f\bar f) &= \sum_f  \frac{n_c m_A }{8 \pi S } \sqrt{1-\frac{4 m_f^2}{m_A^2} }\bL \lambda_{fs}^2 \left(1-\frac{4 m_f^2}{m_A^2}\right) +\lambda _{fp}^2\bR,
}
where $S \equiv$ 1 (2) for (in)distinguishable final state particles. Note that \Eq{sigVsdirS} is only $s$-wave if $\lambda_{\chi p} \neq 0$. In other words, the dark matter requires a pseudoscalar coupling to annihilate through the $s$-wave.

In the limit that $n_c \to3$, $m_f \to 0$, taking all couplings equal to $\lambda$ (so that, {\it e.g.}, we have interaction vertices proportional to projection operators), and carrying out the thermal average as in \Eq{sigma v approx}, we have
\beq
\langle \sigma v \rangle \simeq \frac{3\lambda^4 }{\pi} \frac{m_\chi^2}{ \pL m_A^2-4m_\chi^2 \pR^2} \bL1+ \frac98 \frac{v^2}{1-4m_\chi^2/m_A^2} \bR. 
\eeq
This differs by a factor of 16 from some standard references due to the factor of $1/2$ in the definition: $P_{L,R} = (1 \mp \gamma^5)/2$.

Considering the very low momentum exchange involved in DM scattering with nuclei, we can safely integrate out the mediator. If we take  $\lambda _{\chi p}=\lambda _{ f p}=0$, we reproduce the standard scalar-mediated spin-independent scattering cross section. Summing over the quark content of the nucleus as described above in \Eq{scalar-couplings} gives
\beq
\sigma^{\rm SI}_{(s,s)} \simeq \frac{\mu_{\chi N}^2 \lambda_{\chi ,s}^2}{\pi m_A^4} \bL  Z \widetilde f_p +  (A-Z) \widetilde  f_n \bR^2, \label{dirSIs}
\eeq
where $\widetilde f_{n}$ are dimensionless quantities defined  in \Eq{scalar-couplings}. They are related to the traditional dimensionful couplings by $f_{n}\equiv \lambda_{\chi ,s} \widetilde f_{n} /m_A^2 $. If instead we take $\lambda _{\chi s}=\lambda _{f s}=0$, we find a spin-dependent scattering cross section that is suppressed by four powers of the momentum:
\beq
\sigma^{\rm SI}_{(p,p)} \simeq \frac43 \pL \frac{2 \mu_{\chi N}^2 v^2}{4m_\chi m_N} \pR^2 \frac{ 4 \mu_{\chi N}^2 \lambda_{\chi ,p}^2}{\pi m_A^4} J_N(J_N+1) \pL \frac{\langle S_p \rangle}{J_N}  \widetilde t_p +  \frac{\langle S_n \rangle}{J_N}  \widetilde t_n \pR^2, \label{dirSIp}
\eeq
where $\widetilde t_{n}$ are dimensionless quantities defined  in \Eq{pseudoscalar-couplings}. They are related to the traditional dimensionful couplings by  $ t_{n} \equiv \lambda_{\chi p} \widetilde t_{n} /m_A^2 $.

We need not restrict ourselves to these interactions, of course. In addition to the scalar-scalar and pseudoscalar-pseudoscalar cross sections, we may write the mixed scalar-pseudoscalar and pseudoscalar-scalar cross sections. These are given by
\alg{
\sigma^{\rm SI}_{(p,s)} &\simeq \frac{\mu_{\chi N}^2 v^2}{2m_\chi^2} \frac{\mu_{\chi N}^2 \lambda_{\chi , p}^2}{\pi m_A^4} \bL  Z \widetilde f_p +  (A-Z) \widetilde  f_n \bR^2, \\
\sigma^{\rm SD}_{(s,p)} &\simeq  \frac{2 \mu_{\chi N}^2 v^2}{4 m_N^2}  \frac{ 4 \mu_{\chi N}^2 \lambda_{\chi , s}^2}{\pi m_A^4} J_N(J_N+1) \pL \frac{\langle S_p \rangle}{J_N}  \widetilde t_p +  \frac{\langle S_n \rangle}{J_N}  \widetilde t_n \pR^2, 
\label{dir0mixed}
}
with coefficients as defined above.

%\noindent\makebox[\linewidth]{\rule{\paperwidth}{0.5pt}} \textcolor{ForestGreen}{\checkmark}
\subsection{Majorana dark matter, spin zero mediator}
\label{majS}

For Majorana fermion DM and a spin-$0$ mediator with the following interactions:
\begin{align}
\mathcal{L} \supset  \left[ \frac{1}{2}\  \bar{\chi}( \lambda_{\chi s} + \lambda_{\chi p} i \gamma^5 ) \chi +  \bar{f}( \lambda_{f s} + \lambda_{f p} i \gamma^5 ) f \right] A,
\end{align}
the expressions for the annihilation and direct detection cross sections are identical as shown in \Eqm{sigsdirS}{dir0mixed}.

%\noindent\makebox[\linewidth]{\rule{\paperwidth}{0.5pt}} \textcolor{ForestGreen}{\checkmark}
\subsection{Dirac dark matter, spin one mediator} 
\label{dirV}

Consider the following interactions for a Dirac dark matter particle, $\chi$, and a spin-$1$ mediator, $V_{\mu}$:
\begin{equation}
\mathcal{L} \supset \left[ \bar{\chi} \gamma^\mu ( g_{\chi  v } + g_{\chi a} \gamma^5 ) \chi + \bar{f} \gamma^\mu ( g_{f  v } + g_{f a} \gamma^5 ) f \right] V_\mu. \\
\end{equation}
The annihilation cross section is given by:
\begin{eqnarray}
\sigma &=& \frac{n_c }{12 \pi  s \bL \left(s-m_v ^2\right)^2  + m_v^2 \Gamma_v^2 \bR } \sqrt{\frac{1-4 m_f^2/s}{1-4 m_{\chi }^2/s}} \Bigg[g_{f a}^2\bigg\{g_{\chi a}^2 \bigg(4 m_{\chi }^2 \bigg[m_f^2 \pL 7- \frac{6s}{m_v^2}+\frac{3s^2}{m_v^4} \pR-s \bigg]+s  \left(s-4 m_f^2\right)\bigg) \nonumber \\
&+&g_{\chi v}^2  (s-4 m_f^2) (2 m_{\chi}^2+s)\bigg\}+g_{fv}^2 (2 m_f^2+s) \Big(g_{\chi a}^2 (s-4 m_{\chi }^2)+g_{\chi v}^2 (2 m_{\chi }^2+s)\Big) \Bigg]. \label{sigsdirV}
\end{eqnarray}
 Expanding in powers of $v^2$ gives
\begin{eqnarray}
\sigma v &\approx& \frac{n_c \sqrt{1-m_f^2/m_\chi^2}}{2 \pi m_v^4  \left(m_v^2-4 m_{\chi }^2\right)^2} \Bigg[ g_{fa}^2 \left(m_f^2 g_{\chi a}^2 \left(m_v^2-4 m_{\chi }^2\right)^2+2 g_{\chi v}^2 m_v^4 \left(m_{\chi }^2-m_f^2\right)\right)+g_{fv}^2 g_{\chi v}^2 m_v^4 \left(m_f^2+2 m_{\chi }^2\right) \Bigg] \nonumber \\
&-& \frac{n_c v^2}{48 \pi  m_v^4 m_{\chi }^2 \sqrt{1-m_f^2/m_{\chi }^2} \left(4 m_{\chi }^2-m_v^2\right)^3} 
\Bigg[ g_{fa}^2 \bigg\{g_{\chi a}^2 \left(m_v^2-4 m_{\chi }^2\right) \Big(m_f^4 \left(-72 m_v^2 m_{\chi }^2+17 m_v^4+144 m_{\chi }^4\right) \nonumber \\
&+&m_f^2 \left(48 m_v^2 m_{\chi }^4-22 m_v^4 m_{\chi }^2-96 m_{\chi }^6\right)+8 m_v^4 m_{\chi }^4\Big)  \label{sigVsdirV}
   \\&-&2 g_{\chi v}^2 m_v^4 \left(m_f^2-m_{\chi }^2\right) \Big(4 m_{\chi }^2 \left(m_v^2-17
   m_f^2\right)+5 m_f^2 m_v^2+32 m_{\chi }^4\Big)\bigg\} \nonumber 
   +g_{fv}^2 m_v^4 \bigg\{ g_{\chi v}^2 \Big(8 m_{\chi }^4 \left(m_v^2-4
   m_f^2\right)
   \\&-&4 m_f^2 m_{\chi }^2 \left(17 m_f^2+m_v^2\right)+5 m_f^4 m_v^2+64 m_{\chi }^6\Big)-4 g_{\chi a}^2 \left(m_f^2 m_{\chi
   }^2+m_f^4-2 m_{\chi }^4\right) \left(m_v^2-4 m_{\chi }^2\right)\bigg\} \Bigg] \nonumber.
\end{eqnarray}
The mediator's width to SM fermions is
\beq
\Gamma_v \equiv \sum_f \Gamma(V \to f \bar f) =  \sum_f \frac{n_c m_v }{12 \pi   S}  \sqrt{1-\frac{4 m_f^2}{m_v^2}} \bL g_{fa}^2 \left(1-\frac{4 m_f^2}{m_v^2} \right)+g_{fv}^2 \left(1+ 2 \frac{m_f^2}{m_v^2} \pR \bR,
\eeq
where $S \equiv$ 1 (2) for (in)distinguishable final state particles.

In the limit that $n_c \to3$, $m_f \to 0$, and taking all couplings equal to $\lambda$, we have
\beq
\langle \sigma v \rangle \simeq \frac{6\lambda^4 }{\pi} \frac{m_\chi^2}{\pL m_v^2-4m_\chi^2 \pR^2} \bL1+ \frac{3v^2}4 \frac{1+2m_\chi^2/m_v^2}{1-4m_\chi^2/m_v^2} \bR. 
\eeq

Now consider very low momentum exchange for DM elastic scattering with nuclei, so that we we can integrate out the mediator. If we take $g _{\chi  a}=g_{f a}=0$, we can integrate out the mediator. Summing over the quark content of the nucleus as suitable for a scalar-mediated interaction, we can find the direct detection spin-independent cross section, as described above. This gives
\beq
\sigma^{\rm SI}_{(v,v)} \simeq \frac{\mu_{\chi N}^2 g_{\chi v}^2}{\pi  m_v^4} \bL Z \pL 2 \widetilde b_u+\widetilde b_d \pR + (A-Z) \pL \widetilde b_u+ 2 \widetilde b_d \pR \bR^2, \label{dirSIV}
\eeq
where $\widetilde b_q$ are dimensionless quantities related to the traditional dimensionful couplings by $b_q=g_{\chi v}\widetilde b_q/m_v^2$. If we instead take $g_{\chi v}=g _{f v}=0$, we can find the spin-dependent direct detection cross section, as described above. This gives
\beq
\sigma^{\rm SD}_{(a,a)} \simeq  \frac{ 4 \mu_{\chi N}^2 g_{\chi a}^2}{\pi m_v^4} J_N(J_N+1) \pL \frac{\langle S_p \rangle}{J_N}  \widetilde a_p +  \frac{\langle S_n \rangle}{J_N}  \widetilde a_n \pR^2, \label{dirSDa}
\eeq
where $\widetilde a_n$ are dimensionless quantities related to the traditional dimensionful couplings by $a_n= g_{\chi a} \widetilde a_n  /m_v^2$.

Once more, we may write the cross sections with mixed spin-1 vertices. We have
\alg{
\sigma^{\rm SI}_{(a,v)} & \simeq \frac{2 \mu^2_{\chi N} v^2}{m^2_{\chi}} \, \frac{\mu_{\chi N}^2}{\mu_{\chi n}^2} \, \frac{\mu_{\chi N}^2 g_{\chi a}^2}{\pi  m_v^4} \bL Z \pL 2 \widetilde b_u+\widetilde b_d \pR + (A-Z) \pL \widetilde b_u+ 2 \widetilde b_d \pR \bR^2
\\ 
\sigma^{\rm SD}_{(v,a)}  & \simeq  \frac{\mu_{\chi N}^4}{\mu_{\chi n}^2} \, \frac{ 2  g_{\chi v}^2 v^2}{\pi m_v^4} J_N(J_N+1) \pL \frac{\langle S_p \rangle}{J_N}  \widetilde a_p +  \frac{\langle S_n \rangle}{J_N}  \widetilde a_n \pR^2, \label{dir1mixed}
}
where the kinematic suppression carried by the product $g_{\chi v}^2 v^2$ is due to the fact that the momentum transfer does not sum coherently for this operator. This is often written as a dependence on $ v_\perp^2 \sim \mu_{\chi N}^2 v^2/ \mu_{\chi n}^2$.

%\noindent\makebox[\linewidth]{\rule{\paperwidth}{0.5pt}} \textcolor{ForestGreen}{\checkmark}
\subsection{Majorana dark matter, spin one mediator} 
\label{majoranavector}
Similar to the case above, we may write down the Lagrangian for a Majorana DM particle, $\chi$, that interacts with the SM via a spin-$1$ mediator, $V_{\mu}$:
\beq
\mathcal{L} \supset \left[ \frac{1}{2} g_{\chi a} \bar{\chi} \gamma^\mu \gamma^5 \chi + \bar{f} \gamma^\mu ( g_{f v} + g_{f a} \gamma^5 ) f \right] V_\mu.
\eeq
Note that $g_{\chi v} \to0$ since the vector coupling to a self-conjugate particle vanishes. With the factor of $1/2$ written here and the caveat noted, the Majorana case gives the identical annihilation and direct-detection cross sections as in the Dirac case. These are given in \Eqm{sigsdirV}{dir1mixed}.

%\noindent\makebox[\linewidth]{\rule{\paperwidth}{0.5pt}} \textcolor{ForestGreen}{\checkmark}
\subsection{Complex Scalar dark matter, spin zero mediator} 
\label{cszero}

Consider the following interactions for a complex scalar dark matter particle, $\phi$, and a spin-0 mediator, $A$:
\beq
\mathcal{L} \supset \left[ \mu_\phi | \phi |^2 + \bar{f} ( \lambda_{fs} + \lambda_{fp} i \gamma^5 ) f \right] A .
\eeq
The cross section is given by
\beq
\sigma = \frac{n_c \mu _{\phi }^2}{8 \pi  s \bL \left(s-m_A^2\right)^2 +m_A^2 \Gamma_A^2 \bR} \sqrt{\frac{1-4 m_f^2/s}{1-4 m_{\phi}^2/s}} \Bigg[\lambda _{f s}^2 (s-4 m_f^2)+s \lambda _{fp}^2\Bigg]  \label{sigscmxS}  ,\eeq
which can be expanded in powers of $v^2$ as
\alg{
\sigma v \approx& \frac{n_c \mu _{\phi }^2 \sqrt{1-m_f^2/m_\phi^2} \bL \lambda _{fp}^2+\lambda _{fs}^2 \pL 1 - m_f^2/m_\phi^2  \pR \bR }{4 \pi  \left(m_A^2-4 m_{\phi }^2\right)^2}  \\
&-\frac{n_c \mu _{\phi }^2 v^2 }{32 \pi m_\phi^4 \left(4 m_{\phi }^2-m_A^2\right)^3\sqrt{1-m_f^2/m_\phi^2}} \Bigg[\lambda_{fp}^2 m_{\phi }^2 \left(m_A^2 m_f^2-20 m_f^2 m_{\phi }^2+16 m_{\phi }^4\right) \label{sigVscmxS} \\
&+\lambda _{fs}^2 \left(m_{\phi }^2-m_f^2\right)\left(3 m_A^2 m_f^2-28 m_f^2 m_{\phi }^2+16 m_{\phi }^4\right)\Bigg] .
}

In the limit that $n_c \to3$, $m_f \to 0$, and taking all couplings equal to $\lambda$ (with $\mu_\phi \to m_\phi \lambda$), we have
\beq
\langle \sigma v \rangle \simeq \frac{3\lambda^4 }{2\pi} \frac{m_\phi^2}{\pL m_A^2 -4m_\phi^2 \pR^2} \bL1+  \frac92\frac{v^2}{m_A^2/m_\phi^2-4} \bR. 
\eeq

Now consider very low momentum exchange for DM elastic scattering with nuclei, so that we we can integrate out the mediator. If we take $\mu_\phi=\lambda_\phi m_\phi$, we can read off the scattering cross section by rescaling \Eqs{dirSIs}{dirSIp}. In the case with  $\lambda _{ fp}=0$, we have
\beq
\sigma^{\rm SI}_{(s)} \simeq \frac{\mu_{\phi N}^2 \lambda_{\phi }^2}{4\pi m_A^4} \bL  Z \widetilde f_p +  (A-Z) \widetilde  f_n \bR^2, \label{cmxSIs}
\eeq
whereas for $\lambda _{ fs}=0$, we find a $q^2$ momentum suppressed spin-dependent scattering cross section:
\beq
\sigma^{\rm SD}_{(p)} \simeq   \frac{2 \mu_{\phi N}^2 v^2}{4 m_N^2}  \frac{ \mu_{\phi N}^2 \lambda_{\phi}^2}{\pi m_A^4} J_N(J_N+1) \pL \frac{\langle S_p \rangle}{J_N}  \widetilde t_p +  \frac{\langle S_n \rangle}{J_N}  \widetilde t_n \pR^2, \label{cmxSIp}
\eeq
where $\widetilde t_{n}$ are dimensionless quantities defined in \Eq{pseudoscalar-couplings}. They are related to the traditional dimensionful couplings by  $ t_{n} \equiv \lambda_{\chi p} \widetilde t_{n} /m_A^2 $.

%\noindent\makebox[\linewidth]{\rule{\paperwidth}{0.5pt}} \textcolor{ForestGreen}{\checkmark}
\subsection{Real Scalar dark matter, spin zero mediator} 
\label{rszero}

Similar to the case above, we may write down the Lagrangian for a real scalar DM particle, $\phi$, that interacts with the SM via a spin-0 mediator, $A$:
\beq
\mathcal{L} \supset \left[ \frac{1}{2} \mu_\phi  \phi ^2 + \bar{f} ( \lambda_{fs} + \lambda_{fp} i \gamma^5 ) f \right] A
\eeq
With the factor of $1/2$ written here, this gives the identical annihilation and scattering cross sections as in the complex case, given in \Eqm{sigscmxS}{cmxSIp},

%\noindent\makebox[\linewidth]{\rule{\paperwidth}{0.5pt}} \textcolor{ForestGreen}{\checkmark}
\subsection{Complex Scalar dark matter, spin one mediator}
\label{csone}

Consider the following interactions for a complex scalar dark matter particle, $\phi$, and a spin-1 mediator, $V_\mu$:
\beq
\mathcal{L} \supset \left[i\, g_\phi  \phi^\dagger  \overset{\leftrightarrow}{\partial_{\mu}}  \phi  + \bar{f} \gamma_{\mu}( g_{f v} + g_{f a}  \gamma^5 ) f \right]  V^{\mu} \label{LscmxV}
\eeq
This has an annihilation cross section given by
\alg{
\sigma =&\frac{ n_c g_\phi^2 s }{12 \pi  \left[ \left( m_v^2-s\right)^2 + \Gamma_v^2 m_v^2\right]} \sqrt{1-\frac{4 m_\phi^2}s} \sqrt{1-\frac{4 m_f^2}s}\left[g_a^2 \left(1 - \frac{4 m_f^2}s \right)+g_v^2 \left(1+\frac{2 m_f^2}s\right)\right]
\label{sigscmxV},
}
which can be expanded in powers of $v^2$ as
\alg{
\sigma v \approx& \frac{n_c g_\phi^2 m_\phi^2 v^2}{6\pi}\frac{\sqrt{1-m_f^2/m_\phi^2}  }{ \left(m_v^2-4m_\phi^2\right)^2} \left[ g_a^2  \left( 1-\frac{m_f^2}{m_\phi^2}\right) + g_v^2  \left( 1+\frac{m_f^2}{2m_\phi^2}\right)\right]
\label{sigVscmxV}.
}

In the limit that $n_c \to3$, $m_f \to 0$, and taking all couplings equal to $\lambda$, we have
\beq
\langle \sigma v \rangle \simeq \frac{9\lambda^4 v^2}{4\pi} \frac{m_\phi^2}{ \pL m_v^2-4m_\phi^2 \pR^2}. 
\eeq

Now consider very low momentum exchange for DM elastic scattering with nuclei, so that we we can integrate out the mediator. Since $\phi^\dagger \overset{\leftrightarrow}{\partial_{\mu}} \phi$ is dominated by the timelike component, we have $g_\phi \phi^\dagger  \overset{\leftrightarrow}{\partial_{\mu}} \phi \to 2 g_\phi m_\phi \delta^0_\mu$. Thus, we can read off the scattering cross section by rescaling \Eqs{dirSIV}{dirSDa}. If we take $g _{ f a}=0$,  we have
\beq
\sigma^{\rm SI}_{(v)} \simeq \frac{\mu_{\phi N}^2 g_{\phi }^2}{4 \pi  m_v^4} \bL Z \pL 2 \widetilde b_u+\widetilde b_d \pR + (A-Z) \pL \widetilde b_u+ 2 \widetilde b_d \pR \bR^2. \label{cmxSIV}
\eeq
If we instead take $g _{f v}=0$, we can find the rescaled spin-dependent scattering cross section, in analogy with the above. This gives
\beq
\sigma^{\rm SD}_{(a)} \simeq   \frac{  \mu_{\phi N}^4 g_\phi^2 v^2}{2\pi \mu_{\phi n}^2  m_v^4} J_N(J_N+1) \pL \frac{\langle S_p \rangle}{J_N}  \widetilde a_p +  \frac{\langle S_n \rangle}{J_N}  \widetilde a_n \pR^2, \label{cmxSDa}
\eeq
where the kinematic suppression goes like $v_\perp^2$.

\subsection{Real Scalar dark matter, spin one mediator}
\label{rsone}

The real scalar equivalent of \Eq{LscmxV} vanishes identically, so that real scalar dark matter cannot couple to a vector at tree-level.

%\noindent\makebox[\linewidth]{\rule{\paperwidth}{0.5pt}} \textcolor{ForestGreen}{\checkmark}
\subsection{Complex Vector dark matter, spin zero mediator} 
\label{cvzero}

Consider the following interactions for a complex vector dark matter particle, $X_\mu$, and a spin-0 mediator, $A$:
\beq
\mathcal{L} \supset \left[ \mu_X X^\mu X_\mu^\dagger  + \bar{f}( \lambda_{f s} + \lambda_{f p} i \gamma^5 ) f \right] A \,.
\eeq
This has an annihilation cross section given by
\beq
\sigma = \frac{n_c \mu _X^2  }{72 \pi  \bL \left(s-m_A^2\right)^2 +m_A^2 \Gamma_A^2 \bR } \sqrt{\frac{1-4 m_f^2/s}{1-4 m_{\chi }^2/s}} \left[\frac{s}{m_X^2} \left(\frac{s}{4 m_X^2}-1\right)+3\right] \left[\lambda _{fs}^2 \left(1-\frac{4m_f^2}s\right)+ \lambda _{fp}^2\right] \label{sigscvS},
\eeq
which can be expanded in powers of $v^2$ as
\alg{
\sigma v \approx& \frac{n_c \mu _X^2 \sqrt{1-m_f^2/m_X^2} \left[\lambda _{fp}^2+\lambda _{fs}^2\pL1-m_f^2/m_X^2 \pR \right]}{12 \pi  \left(m_A^2-4 m_X^2\right)^2}
\\
&+\frac{n_c \mu _X^2 v^2}{288 \pi m_X^4 \left(4 m_X^2-m_A^2\right)^3 \sqrt{1-m_f^2/m_X^2}} \Bigg[\lambda _{fp}^2 m_X^2\bigg\{4 m_X^2 \left(7 m_f^2-2 m_A^2\right) 
+5 m_A^2 m_f^2-16 m_X^4\bigg\}+
\\
&\lambda _{fs}^2 \left(m_f^2-m_X^2\right) \bigg\{m_A^2 \left(m_f^2+8m_X^2\right)-52 m_f^2 m_X^2+16 m_X^4\bigg\} \Bigg]  \label{sigVscvS}.
}

In the limit that $n_c \to3$, $m_f \to 0$, and taking all couplings equal to $\lambda$ (with $\mu_X \to m_X \lambda/2$), we have
\beq
\langle \sigma v \rangle \simeq \frac{\lambda^4 }{2\pi} \frac{m_X^2}{\pL m_A^2-4m_X^2 \pR^2} \bL1+ \frac{3v^2}4 \frac{1+2m_X^2/m_A^2}{1-4m_X^2/m_A^2} \bR. 
\eeq

Now consider very low momentum exchange for DM elastic scattering with nuclei, so that we we can integrate out the mediator. If we take $\mu_X=\lambda_X m_X$, we can read off the scattering cross section from \Eqs{cmxSIs}{cmxSIp}. In the case with  $\lambda _{ fp}=0$, we have
\beq
\sigma^{\rm SI}_{(s)} \simeq \frac{\mu_{X N}^2 \lambda_{X }^2}{4 \pi m_A^4} \bL  Z \widetilde f_p +  (A-Z) \widetilde  f_n \bR^2, \label{cvSIs}
\eeq
whereas for $\lambda _{ fs}=0$, we find a $q^2$ momentum-suppressed spin-dependent scattering cross section:
\beq
\sigma^{\rm SD}_{(p)} \simeq   \frac{2 \mu_{X N}^2 v^2}{4 m_N^2}  \frac{ \mu_{X N}^2 \lambda_{X}^2}{\pi m_A^4} J_N(J_N+1) \pL \frac{\langle S_p \rangle}{J_N}  \widetilde t_p +  \frac{\langle S_n \rangle}{J_N}  \widetilde t_n \pR^2, \label{cvSIp}
\eeq
where $\widetilde t_{n}$ are dimensionless quantities defined  in \Eq{pseudoscalar-couplings}. They are related to the traditional dimensionful couplings by  $ t_{n} \equiv \lambda_X \widetilde t_{n} /m_A^2 $.

%\noindent\makebox[\linewidth]{\rule{\paperwidth}{0.5pt}} \textcolor{ForestGreen}{\checkmark}
\subsection{Real Vector dark matter, spin zero mediator} 
\label{rvzero}

Similar to the case above, we may write down the Lagrangian for a real vector DM particle, $X_\mu$, that interacts with the SM via a spin-0 mediator, $A$:
\beq
\mathcal{L} \supset \left[ \frac{1}{2} \mu_X X^\mu X_\mu  + \bar{f}( \lambda_{f s} + \lambda_{f p} i \gamma^5 ) f \right] A \, .
\eeq
With the factor of $1/2$ written here, this gives the identical annihilation and scattering cross cross sectionsections as in the complex case. These are given in \Eqm{sigscvS}{cvSIp}.

%\noindent\makebox[\linewidth]{\rule{\paperwidth}{0.5pt}} \textcolor{Red}{X}
\subsection{Complex Vector dark matter, spin one mediator} 
\label{cvone}

Consider the Lagrangian for complex vector dark matter, $X_\mu$, interacting with the SM via a spin-1 mediator, $V_\mu$:
\beq
\mathcal{L} \supset   \left[ g_X\left(X^{\dagger \nu} \partial_\nu X^\mu + \text{h.c.} \right)+\bar{f} \gamma^\mu \left( g_{f v} + g_{f a} \gamma^5 \right) f  \right] V_\mu. 
\eeq
This has an annihilation cross section
\alg{
\sigma &= \frac{g_X^2 \left(s- 4 m_X^2\right)  }{72 \pi  m_X^4 m_v^4 \bL \Gamma_v^2 m_v^2+\left(m_v^2-s\right)^2 \bR} \sqrt{\frac{1-4m_f^2/s}{1- 4m_X^2/s}} 
\bigg\{ 2 g_v^2 m_X^2 m_v^4 \left(2 m_f^2+s\right)
\\&\qquad + g_a^2 \bL 2 m_X^2 \left(s m_v^4 - 2 m_f^2 \left(5 m_v^4-6 m_v^2 s+3 s^2\right)\right) + 3 m_f^2 s \left(m_v^2-s\right)^2 \bR \bigg\}\,,
\label{sigscvV}
}
which can be expanded in powers of velocity as
\beq
\sigma v \approx \frac{n_c v^2 g_X^2 \sqrt{1-m_f^2/m_X^2} }{27 \pi   \left(m_v^2-4 m_X^2\right)^2} \bL m_f^2 \left(g_{fv}^2-2 g_{fa}^2\right)+2 m_X^2 \left(g_{{fa}}^2+g_{{fv}}^2\right)\bR. \label{sigVscvV}
\eeq
Although this is strictly an $s$-wave annihilation, there is no velocity-independent term. This is because the annihilation cross section carries an overall proportionality of the time-like polarizations of the incoming particles \cite{priv-com-Jason}. In the limit that $n_c \to3$, $m_f \to 0$, and taking all couplings equal to $\lambda$, we have
\beq
\langle \sigma v \rangle \simeq \frac{\lambda ^4 v^2 }{\pi}\frac{m_X^2}{ \left( m_v^2-4 m_X^2\right)^2}
\eeq

Now consider very low momentum exchange for DM elastic scattering with nuclei, so that we we can integrate out the mediator. If we take the low-momentum limit, we have $g_X\left(X^{\dagger \nu} \partial_\nu X^\mu + \text{h.c.} \right) \simeq 2 g_X m_X$. Thus, we can read off scattering cross section by rescaling \Eqs{cmxSIV}{cmxSDa}. If we take $g _{ f a}=0$,  we have
\beq
\sigma^{\rm SI}_{(v)} \simeq \frac{\mu_{X N}^2 g_{X }^2}{4 \pi  m_v^4} \bL Z \pL 2 \widetilde b_u+\widetilde b_d \pR + (A-Z) \pL \widetilde b_u+ 2 \widetilde b_d \pR \bR^2. \label{cvSIV}
\eeq
If we instead take $g _{f v}=0$, we can find the rescaled spin-dependent scattering cross section, in analogy with the above. This gives
\beq
\sigma^{\rm SD}_{(a)} \simeq  \frac{ \mu_{X N}^4 g_X^2 v^2}{2\pi \mu^2_{X n} m_v^4} J_N(J_N+1) \pL \frac{\langle S_p \rangle}{J_N}  \widetilde a_p +  \frac{\langle S_n \rangle}{J_N}  \widetilde a_n \pR^2. \label{cvSDa}
\eeq

\subsection{Real Vector dark matter, spin one mediator} 
\label{rvone}

Consider the Lagrangian for real vector dark matter, $X_\mu$, interacting with the SM via a spin-1 mediator, $V_\mu$:
\beq
\mathcal{L} \supset   \left[ \half g_X\left(X^{ \nu} \partial_\nu X^\mu + \text{h.c.} \right)+\bar{f} \gamma^\mu \left( g_{f v} + g_{f a} \gamma^5 \right) f  \right] V_\mu.
\eeq
With the factor of $1/2$ written here, this gives the identical annihilation and direct-detection cross sections as in the complex case. These are given in \Eqm{sigscvV}{cvSDa}.

%\noindent\makebox[\linewidth]{\rule{\paperwidth}{0.5pt}}

\section{$t/u$-Channel Interactions}\label{app:t}
%\noindent\makebox[\linewidth]{\rule{\paperwidth}{0.5pt}}

We list all annihilation and direct detection cross sections for dark matter that interacts through $t/u$-channel Lagrangians. Phenomenologically, we will only be interested in the chiral limit where the interaction vertices are proportional to projection operators, but we present calculations here for fully general couplings. Simplified models with annihilations that are unsuppressed in the chiral limit and taking $v \to 0$ are presented in Secs.~\ref{tdirS},~\ref{tdirV},~\ref{tcvF}, and \ref{trvF}. For completeness, we also include simplified models that annihilate with $v^2$ suppression in the chiral limit in Secs.~\ref{tmajS},~\ref{tmajV}, and \ref{tcmxF}. A process that is $d$-wave ($v^4$-suppressed) in the chiral limit is examined in Sec.~\ref{trealF}.

\subsection{Dirac dark matter, spin zero mediator} 
\label{tdirS}

Consider the following interactions for a Dirac dark matter particle, $\chi$, and a spin-$0$ mediator, $A$:
\beq \label{LtdirS}
\mathcal{L} \supset  \bar{\chi} ( \lambda_s + \lambda_p \gamma^5 ) f A +  \bar{f} ( \lambda_s - \lambda_p  \gamma^5 ) \chi A^\dagger
\eeq
Since there are no resonances in $t$-channel annihilation, the annihilation cross section is always well approximated by the first two terms of its Taylor series expansion. This cross section expanded in powers of $v^2$ is given by
\alg{
\sigma v \approx& \frac{n_c \sqrt{1-m_f^2/m_\chi^2} \Big[\lambda _p^2 \left(m_{\chi }-m_f\right)+\lambda _s^2 \left(m_f+m_{\chi }\right)\Big]^2}{8 \pi \left(m_A^2-m_f^2+m_{\chi }^2\right)^2} 
\\
&-\frac{n_c v^2}{192\pi  m_{\chi }^2 \sqrt{1-m_f^2/m_{\chi }^2} \left(m_A^2-m_f^2+m_{\chi}^2\right)^4} \Bigg[m_A^4 \bigg\{6 m_f^3 m_{\chi } \left(\lambda_p^4-\lambda _s^4\right)+m_f^2 m_{\chi }^2 \left(13 \lambda _p^4+2 \lambda _p^2 \lambda _s^2+13 \lambda _s^4\right)
\\
&+m_f^4 \left(-11 \lambda _p^4+14\lambda _p^2 \lambda _s^2-11 \lambda _s^4\right)-8 m_{\chi }^4 \left(\lambda _p^2+\lambda _s^2\right)^2\bigg\} \\
&+2 m_A^2 \left(m_f^2-m_{\chi}^2\right) \bigg\{\lambda _p^4 \left(m_f-m_{\chi }\right)^2 \left(8 m_f m_{\chi }+11 m_f^2-12 m_{\chi }^2\right)
\\
&-2 \lambda _p^2 \lambda _s^2\left(-19 m_f^2 m_{\chi }^2+7 m_f^4+12 m_{\chi }^4\right)+\lambda _s^4 \left(m_f+m_{\chi }\right)^2 \left(-8 m_f m_{\chi }+11 m_f^2-12 m_{\chi}^2\right)\bigg\}
\\
&-\lambda _p^4 \left(m_f-m_{\chi }\right)^4 \left(m_f+m_{\chi }\right)^2 \left(11 m_f^2-8 m_{\chi }^2\right)+2 \lambda _p^2\lambda _s^2 \left(7 m_f^2-8 m_{\chi }^2\right) \left(m_f^2-m_{\chi }^2\right)^3
\\
&-\lambda _s^4 \left(m_f-m_{\chi }\right)^2 \left(m_f+m_{\chi}\right)^4 \left(11 m_f^2-8 m_{\chi }^2\right)\Bigg] 
.}

In the limit that $n_c \to3$, $m_f \to 0$, and taking all couplings equal to $\lambda$, we have
\beq
\langle \sigma v \rangle \simeq \frac{3\lambda^4 }{2\pi} \frac{m_\chi^2}{\pL m_A^2 + m_\chi^2 \pR^2} \bL1+ \frac{3v^2}4 \frac{1-3m_\chi^2/m_A^2-m_\chi^4/m_A^4}{1+m_\chi^2/m_A^2} \bR. \nonumber
\eeq
If instead we take the low-velocity limit, we can integrate out the mediator as above, and we can find scattering rates. To evaluate the cross section we must first put the matrix element in canonical $s$-channel form using Fierz transformations. As described in Ref.~\cite{Agrawal:2010fh}, for generic values of the couplings, we generate all possible effective $s$-channel scattering processes. Taking $\lambda _s = \pm \lambda _p$ allows us to cancel the spin-0 mediated pieces, leaving
\beq
 \bar{\chi} ( 1 + \gamma^5 ) f   A +  \bar{f} ( 1 -  \gamma^5 ) \chi A^\dagger   \to - \frac1{2m_A^2} \bar \chi \gamma^\mu \pL 1 + \gamma^5  \pR \chi \bar f \gamma_\mu \pL 1- \gamma^5 \pR f  ,
\eeq
where the cross sections for these interactions may be read off from \Eqs{cvSIV}{cvSDa}. This is dominated by the spin-independent piece.

%\noindent\makebox[\linewidth]{\rule{\paperwidth}{0.5pt}}
\subsection{Majorana dark matter, spin zero mediator} 
\label{tmajS}

Consider the following interactions for a Majorana dark matter particle, $\chi$, and a spin-$0$ mediator, $A$:
\beq
\mathcal{L} \supset  \bar{\chi} ( \lambda_s + \lambda_p \gamma^5 ) f A +  \bar{f} ( \lambda_s - \lambda_p  \gamma^5 ) \chi A^\dagger \, .
\eeq
The annihilation cross section at low relative velocity is given by
\alg{
\sigma v \approx& \frac{n_c \sqrt{1-m_f^2/m_{\chi }^2} \Big[\lambda _p^2 \left(m_f-m_{\chi }\right)+\lambda _s^2 \left(m_f+m_{\chi }\right)\Big]^2}{8 \pi  \left(m_A^2-m_f^2+m_{\chi }^2\right)^2}
\\
&+\frac{n_c v^2}{192 \pi  m_{\chi }^2 \sqrt{1-m_f^2/m_\chi^2}\left(m_A^2-m_f^2+m_{\chi }^2\right)^4} \Bigg[m_A^4 \bigg\{\lambda _p^4 \left(m_f-m_{\chi }\right)^2\left(40 m_f m_{\chi }+23 m_f^2+20 m_{\chi }^2\right)
\\
&+6 \lambda _p^2 \lambda _s^2 \left(-5 m_f^2 m_{\chi }^2+m_f^4+4 m_{\chi }^4\right)+\lambda_s^4 \left(m_f+m_{\chi }\right)^2 \left(-40 m_f m_{\chi }+23 m_f^2+20 m_{\chi }^2\right)\bigg\}
\\
&-2 m_A^2 \left(m_f^2-m_{\chi }^2\right)\bigg\{\lambda _p^4 \left(m_f-m_{\chi }\right)^2 \left(24 m_f m_{\chi }+23 m_f^2-8 m_{\chi }^2\right)+2 \lambda _p^2 \lambda _s^2 \left(-11 m_f^2m_{\chi }^2+3 m_f^4+8 m_{\chi }^4\right)
\\
&+\lambda _s^4 \left(m_f+m_{\chi }\right)^2 \left(-24 m_f m_{\chi }+23 m_f^2-8 m_{\chi}^2\right)\bigg\} \\
&+\lambda _p^4 \left(m_f-m_{\chi }\right)^4 \left(m_f+m_{\chi }\right)^2 \left(8 m_f m_{\chi }+23 m_f^2+4 m_{\chi }^2\right)
\\
&+2\lambda _p^2 \lambda _s^2 \left(3 m_f^2-28 m_{\chi }^2\right) \left(m_f^2-m_{\chi }^2\right)^3+\lambda _s^4 \left(m_f-m_{\chi }\right)^2\left(m_f+m_{\chi }\right)^4 \left(-8 m_f m_{\chi }+23 m_f^2+4 m_{\chi }^2\right)\Bigg] .
}

In the limit that $n_c \to3$, $m_f \to 0$, and taking all couplings equal to $\lambda$, we have
\beq
\langle \sigma v \rangle \simeq \frac{9\lambda^4 v^2}{4\pi} \frac{m_\chi^2 \pL 1+m_\chi^4 /m_A^4 \pR}{m_A^4 \pL 1+  m_\chi^2/m_A^2 \pR^4} \, .
\eeq
Taking instead the low-velocity limit, we can integrate out the mediator and rearrange via the Fierz identities described above to get the interaction in canonical form. Since the vector current for a Majorana fermion vanishes, we find
\beq
 \bar{\chi} ( 1 + \gamma^5 ) f   A +  \bar{f} ( 1 -  \gamma^5 ) \chi A^\dagger   \to - \frac1{2m_A^2} \bar \chi \gamma^\mu \gamma^5 \chi \bar f \gamma_\mu \pL 1- \gamma^5 \pR f  ,
\eeq
and evaluating the direct-detection scattering cross section shows that this gives spin-dependent scattering.

%\noindent\makebox[\linewidth]{\rule{\paperwidth}{0.5pt}} \textcolor{ForestGreen}{\checkmark}
\subsection{Dirac dark matter, spin one mediator}
\label{tdirV} 

Consider the following interactions for a Dirac dark matter particle, $\chi$, and a spin-$1$ mediator, $V_\mu$:
\beq
\mathcal{L} \supset  \bar{\chi} \gamma^\mu ( g_{\chi v} + g_{\chi a}\gamma^5 ) f V_\mu + \bar{f} \gamma^\mu ( g_{\chi v} + g_{\chi a}\gamma^5 ) \chi V_\mu^\dagger \, . 
\eeq
The annihilation cross section at low relative velocity is given by
\alg{
\sigma v \approx& \frac{ n_c \sqrt{1-m_f^2/m_{\chi }^2}}{8 \pi  m_v^4 \left(-m_f^2+m_v^2+m_{\chi }^2\right)^2}
\bigg\{
m_f^6 \left(g_{\chi a}^2-g_{\chi v}^2\right)^2+2 m_f^5 m_{\chi } \left(g_{\chi a}^4-g_{\chi v}^4\right)
\\&+m_f^4 \left[2 g_{\chi a}^2 g_{\chi v}^2 \left(4 m_v^2+3 m_{\chi }^2\right)+g_{\chi a}^4 \left(-m_{\chi }^2\right)-g_{\chi v}^4 m_{\chi }^2\right]-4 m_f^3 \left[m_{\chi } \left(g_{\chi a}^4-g_{\chi v}^4\right) \left(m_v^2+m_{\chi }^2\right)\right]
\\&+m_f^2 \left[-2 g_{\chi a}^2 g_{\chi v}^2 \left(8 m_v^2 m_{\chi }^2+2 m_v^4+3 m_{\chi }^4\right)+g_{\chi a}^4 \left(-4 m_v^2 m_{\chi }^2+2 m_v^4-m_{\chi }^4\right)+g_{\chi v}^4 \left(-4 m_v^2 m_{\chi }^2+2 m_v^4-m_{\chi }^4\right)\right]
\\&+2 m_f m_{\chi } \left(g_{\chi a}^4-g_{\chi v}^4\right) \left(2 m_v^2 m_{\chi }^2+2 m_v^4+m_{\chi }^4\right)
\\&+m_{\chi }^2 \left[4 m_v^2 m_{\chi }^2 \left(g_{\chi a}^2+g_{\chi v}^2\right)^2+4 g_{\chi a}^2 g_{\chi v}^2 m_v^4+6 g_{\chi a}^4 m_v^4+6 g_{\chi v}^4 m_v^4+m_{\chi }^4 \left(g_{\chi a}^2+g_{\chi v}^2\right)^2\right]
\bigg\}
\\& +\frac{n_c v^2}{192 \pi  m_v^4 m_{\chi }^2 \sqrt{1-m_f^2/m_{\chi }^2} \left(-m_f^2+m_v^2+m_{\chi }^2\right)^4}
\bigg\{
 m_f^{12} \left[11 \pL g_{\chi a}^4 + g_{\chi v}^4 \pR -14 g_{\chi v}^2 g_{\chi a}^2\right]
\\&+22m_f^{11}m_{\chi }  \left(g_{\chi a}^4-g_{\chi v}^4\right) - m_f^{10} \left[ \left(46 m_v^2+41 m_{\chi }^2\right) \pL g_{\chi a}^4 + g_{\chi v}^4  \pR -2 g_{\chi v}^2 \left(26 m_v^2+43 m_{\chi }^2\right) g_{\chi a}^2 \right]
\\&-8m_f^9 m_{\chi } \left(g_{\chi a}^4-g_{\chi v}^4\right)  \left(11 m_v^2+13 m_{\chi }^2\right)
\\&+m_f^8 \Big[ \left(105 m_v^4+136 m_{\chi }^2 m_v^2+46 m_{\chi }^4\right) \pL  g_{\chi a}^4 +g_{\chi v}^4 \pR -2 g_{\chi v}^2   g_{\chi a}^2\left(61 m_v^4+128 m_{\chi }^2 m_v^2+110 m_{\chi }^4\right)\Big] 
\\&  +2 m_f^7 m_{\chi }  \left(g_{\chi a}^4-g_{\chi v}^4\right) \left(85 m_v^4+164 m_{\chi }^2 m_v^2+98 m_{\chi }^4\right) 
\\&- m_f^6 \Big[\left(116 m_v^6+269 m_{\chi }^2 m_v^4+108 m_{\chi }^4 m_v^2-6 m_{\chi }^6\right) \pL g_{\chi a}^4 + g_{\chi v}^4 \pR 
\\& - 2 g_{\chi v}^2 g_{\chi a}^2\left(72 m_v^6+175 m_{\chi }^2 m_v^4+252 m_{\chi }^4 m_v^2+150 m_{\chi }^6\right)  \Big] 
\\&-4  m_f^5 m_{\chi }  \left[\left(g_{\chi a}^4-g_{\chi v}^4\right) \left(29 m_v^6+117 m_{\chi }^2 m_v^4+114 m_{\chi }^4 m_v^2+46 m_{\chi }^6\right)\right]
\\&+ m_f^4 \Big[\left(46 m_v^8+332 m_{\chi }^2 m_v^6+231 m_{\chi }^4 m_v^4-32 m_{\chi }^6 m_v^2-49 m_{\chi }^8\right) \pL g_{\chi a}^4 + g_{\chi v}^4  \pR
\\&-2 g_{\chi v}^2  g_{\chi a}^2 \left(30 m_v^8+96 m_{\chi }^2 m_v^6+159 m_{\chi }^4 m_v^4+248 m_{\chi }^6 m_v^2+115 m_{\chi }^8\right) \Big]
\\&+2 m_f^3 m_{\chi } \left(g_{\chi a}^4-g_{\chi v}^4\right)  \left(6 m_v^8+122 m_{\chi }^2 m_v^6+213 m_{\chi }^4 m_v^4+140 m_{\chi }^6 m_v^2+43 m_{\chi }^8\right) 
\\&+m_f^2 m_{\chi }^2 \Big[\left(-86 m_v^8-232 m_{\chi }^2 m_v^6-75 m_{\chi }^4 m_v^4+74 m_{\chi }^6 m_v^2+35 m_{\chi }^8\right) \pL g_{\chi a}^4+g_{\chi v}^4\pR
\\& +2 g_{\chi v}^2 g_{\chi a}^2 \left(30 m_v^8-24 m_{\chi }^2 m_v^6+37 m_{\chi }^4 m_v^4+122 m_{\chi }^6 m_v^2+47 m_{\chi }^8\right)  \Big]
\\& -16  m_f m_{\chi }^5\left(g_{\chi a}^4-g_{\chi v}^4\right)  \left(8 m_v^6+8 m_{\chi }^2 m_v^4+4 m_{\chi }^4 m_v^2+m_{\chi }^6\right)
\\&+8 m_{\chi }^4 \Big[\left(8 m_v^8+2 m_{\chi }^2 m_v^6+m_{\chi }^4 m_v^4-3 m_{\chi }^6 m_v^2-m_{\chi }^8\right) \pL g_{\chi a}^4 +g_{\chi v}^4 \pR
\\& -2 g_{\chi v}^2 g_{\chi a}^2 m_{\chi }^2 \left(-6 m_v^6-m_{\chi }^2 m_v^4+3 m_{\chi }^4 m_v^2+m_{\chi }^6\right) \Big]
\bigg\}
.
}

In the limit that $n_c \to3$, $m_f \to 0$, and taking all couplings equal to $\lambda$, we have
\beq
\langle \sigma v \rangle \simeq \frac{3\lambda^4}{2\pi} \frac{ m_\chi^2 }{\left(m_v^2+m_{\chi }^2\right)^2} \bL \pL 2+ \frac{m_\chi^2}{m_v^2} \pR^2 + \frac{3 v^2}4 \frac{4 + 4 m_\chi^2/m_v^2 +m_\chi^4/m_v^4 -3 m_\chi^6/m_v^6 - m_\chi^8/m_v^8}{\left(1+m_\chi^2/m_v^2 \right)^2} \bR \, .
\eeq
Taking instead the low-velocity limit, we can integrate out the mediator and rearrange via the Fierz identities described above to get the interaction in canonical form, whereupon we find
\beq
 \bar{\chi} \gamma^\mu ( 1 + \gamma^5 ) f  V_\mu +  \bar{f} \gamma^\mu  ( 1 +  \gamma^5 ) \chi V_\mu^\dagger  \to \frac1{m_v^2} \bar \chi \gamma^\mu \pL 1+\gamma^5 \pR \chi \bar f \gamma_\mu \pL 1+\gamma^5 \pR f   ,
\eeq
which is dominated by spin-independent scattering.

%\noindent\makebox[\linewidth]{\rule{\paperwidth}{0.5pt}} \textcolor{ForestGreen}{\checkmark}
\subsection{Majorana dark matter, spin one mediator} 
\label{tmajV}

Consider the following interactions for a Dirac dark matter particle, $\chi$, and a spin-$1$ mediator, $V_\mu$:
\beq
\mathcal{L} \supset \bar{\chi} \gamma^\mu ( g_{\chi v} + g_{\chi a}\gamma^5 ) f V_\mu + \bar{f} \gamma^\mu ( g_{\chi v} + g_{\chi a}\gamma^5 ) \chi V_\mu^\dagger \, .
\eeq
The annihilation cross section at low relative velocity is given by
\alg{
\sigma v \approx& \frac{n_c \sqrt{1-m_f^2/m_\chi^2}}{8 \pi  m_v^4  \left(-m_f^2+m_v^2+m_{\chi }^2\right)^2}  
\bigg\{
2 m_f^6 \left(g_{\chi a}^2+g_{\chi v}^2\right)^2+m_f^5 m_{\chi } \left(g_{\chi a}^4-g_{\chi v}^4\right)
\\& +m_f^4 \left[-2 g_{\chi a}^2 g_{\chi v}^2 \left(4 m_v^2+3 m_{\chi }^2\right) - \pL g_{\chi a}^4 + g_{\chi v}^4 \pR \left(4 m_v^2+m_{\chi }^2\right) \right]
\\&-4 m_f^3 \left[m_{\chi } \left(g_{\chi a}^4-g_{\chi v}^4\right) \left(3 m_v^2+m_{\chi }^2\right)\right]
\\&+m_f^2 \left[\left(g_{\chi a}^2+g_{\chi v}^2\right)^2 \left(2 m_v^2+m_{\chi }^2\right)^2-2 m_{\chi }^2 \left(g_{\chi a}^2-g_{\chi v}^2\right)^2 \left(4 m_v^2+m_{\chi }^2\right)\right]
\\& +2 m_f m_{\chi } \left(g_{\chi a}^4-g_{\chi v}^4\right) \left(6 m_v^2 m_{\chi }^2+8 m_v^4+m_{\chi }^4\right)+m_{\chi }^2 \left(g_{\chi a}^2-g_{\chi v}^2\right)^2 \left(4 m_v^2+m_{\chi }^2\right)^2
\bigg\}
 \\
&+\frac{n_c v^2}{192 \pi  m_v^4m_\chi^2 \sqrt{1-m_f^2/m_\chi^2} \left(-m_f^2+m_v^2+m_{\chi }^2\right)^4}
 \bigg\{
m_f^{12} \left(23 g_{\chi a}^4+6 g_{\chi v}^2 g_{\chi a}^2+23 g_{\chi v}^4\right) + 54 m_{\chi } m_f^{11} \left(g_{\chi a}^4-g_{\chi v}^4\right)
\\& + m_f^{10} \left[-\left(162 m_v^2+49 m_{\chi }^2\right) g_{\chi a}^4-2 g_{\chi v}^2 \left(58 m_v^2+43 m_{\chi }^2\right) g_{\chi a}^2-g_{\chi v}^4 \left(162 m_v^2+49 m_{\chi }^2\right)\right]
\\& -40 m_f^9 \left[\left(g_{\chi a}^4-g_{\chi v}^4\right) m_{\chi } \left(8 m_v^2+5 m_{\chi }^2\right)\right] 
\\& +m_f^8 \Big[\left(419 m_v^4+456 m_{\chi }^2 m_v^2-30 m_{\chi }^4\right) \pL g_{\chi a}^4+g_{\chi v}^4 \pR +2 g_{\chi v}^2 g_{\chi a}^2 \left(47 m_v^4+404 m_{\chi }^2 m_v^2+170 m_{\chi }^4\right) \Big]
\\& +2 m_f^7 \left(g_{\chi a}^4-g_{\chi v}^4\right) m_{\chi } \left(311 m_v^4+568 m_{\chi }^2 m_v^2+130 m_{\chi }^4\right)
\\& - m_f^6 \Big[\left(444 m_v^6+1139 m_{\chi }^2 m_v^4+348 m_{\chi }^4 m_v^2-150 m_{\chi }^6\right) \pL g_{\chi a}^4+g_{\chi v}^4 \pR 
\\& - 2 g_{\chi v}^2 g_{\chi a}^2 \left(68 m_v^6-437 m_{\chi }^2 m_v^4-1040 m_{\chi }^4 m_v^2-310 m_{\chi }^6\right) \Big] 
\\&-4m_f^5 \left[\left(g_{\chi a}^4-g_{\chi v}^4\right) m_{\chi } \left(101 m_v^6+459 m_{\chi }^2 m_v^4+372 m_{\chi }^4 m_v^2+30 m_{\chi }^6\right)\right] 
\\& + m_f^4 \Big[\left(164 m_v^8+1164 m_{\chi }^2 m_v^6+1009 m_{\chi }^4 m_v^4-72 m_{\chi }^6 m_v^2-125 m_{\chi }^8\right) \pL g_{\chi a}^4 + +g_{\chi v}^4 \pR 
\\& +2 g_{\chi v}^2  g_{\chi a}^2 \left(-60 m_v^8-4 m_{\chi }^2 m_v^6+1161 m_{\chi }^4 m_v^4+1276 m_{\chi }^6 m_v^2+295 m_{\chi }^8\right)\Big]
\\& +2 m_{\chi } m_f^3 \left(g_{\chi a}^4-g_{\chi v}^4\right)\left(24 m_v^8+458 m_{\chi }^2 m_v^6+903 m_{\chi }^4 m_v^4+424 m_{\chi }^6 m_v^2-5 m_{\chi }^8\right) 
\\& +m_{\chi }^2m_f^2 \Big[\left(-328 m_v^8-816 m_{\chi }^2 m_v^6-277 m_{\chi }^4 m_v^4+174 m_{\chi }^6 m_v^2+27 m_{\chi }^8\right)\pL  g_{\chi a}^4+g_{\chi v}^4  \pR 
\\& +2 g_{\chi v}^2 \left(120 m_v^8-352 m_{\chi }^2 m_v^6-1199 m_{\chi }^4 m_v^4-758 m_{\chi }^6 m_v^2-143 m_{\chi }^8\right) g_{\chi a}^2\Big] 
\\& +16 m_{\chi }^5  m_f \left(g_{\chi a}^4-g_{\chi v}^4\right) \left(-32 m_v^6-37 m_{\chi }^2 m_v^4-11 m_{\chi }^4 m_v^2+m_{\chi }^6\right) 
\\& +4 m_{\chi }^4 \Big[\left(56 m_v^8+24 m_{\chi }^2 m_v^6-3 m_{\chi }^4 m_v^4-12 m_{\chi }^6 m_v^2+m_{\chi }^8\right) \pL g_{\chi a}^4 +g_{\chi v}^4 \pR
\\& +2 g_{\chi v}^2  g_{\chi a}^2\left(-24 m_v^8+72 m_{\chi }^2 m_v^6+107 m_{\chi }^4 m_v^4+44 m_{\chi }^6 m_v^2+7 m_{\chi }^8\right) \Big]
 \bigg\}.
}

In the limit that $n_c \to3$, $m_f \to 0$, and taking all couplings equal to $\lambda$, we have
\beq
\langle \sigma v \rangle \simeq \frac{9\lambda^4 v^2}{ \pi }\frac{ m_{\chi }^2 m_v^4}{  \left(m_v^2+m_{\chi }^2\right)^4} \left(1 + \frac{3m_\chi^2}{m_v^2} +\frac{13  m_{\chi }^4}{4m_v^4}+ \frac{ m_{\chi }^6}{m_v^6}+\frac{m_{\chi }^8}{4m_v^8} \right) \, .
\eeq
Taking instead the low-velocity limit, we can integrate out the mediator and rearrange via the Fierz identities described above to get the interaction in canonical form. We find that this rearranges to
\beq
 \bar{\chi} \gamma^\mu ( 1 + \gamma^5 ) f  V_\mu +  \bar{f} \gamma^\mu  ( 1 +  \gamma^5 ) \chi V_\mu^\dagger  \to \frac1{m_v^2} \bar \chi \gamma^\mu \gamma^5  \chi \bar f \gamma_\mu \pL 1+\gamma^5 \pR f   ,
\eeq
which is dominated by spin-independent scattering.

%\noindent\makebox[\linewidth]{\rule{\paperwidth}{0.5pt}}
\subsection{Complex Scalar dark matter, spin half mediator} 
\label{tcmxF}

Consider a complex scalar dark matter particle, $\phi$, that interacts with the SM via $t$-channel exchange of the fermion, $\psi$:
\beq
\mathcal{L} \supset  \bar{\psi} \left( \lambda_s + \lambda_p \gamma^5 \right) f \phi^\dagger +  \bar{f} \left( \lambda_s - \lambda_p \gamma^5 \right) \psi \phi.
\eeq
The annihilation cross section at low relative velocity is given by
\alg{
\sigma v \approx& \frac{n_c \left(1-m_f^2/m_{\phi }^2\right)^{3/2}}{4 \pi \left(-m_f^2+m_{\psi }^2+m_{\phi }^2\right)^2}  \Big[\lambda _p^2 \left(m_f-m_{\psi }\right)+\lambda _s^2 \left(m_f+m_{\psi}\right)\Big]^2
\\
&+\frac{n_c \sqrt{1-m_f^2/m_{\phi }^2}v^2}{96 \pi  m_{\phi }^2 \left(-m_f^2+m_{\psi }^2+m_{\phi }^2\right)^4} \Bigg[18 m_f^7 m_{\psi } \left(\lambda _s^4-\lambda _p^4\right)+m_f^6 \bigg\{-9 m_{\psi }^2 \left(\lambda _p^4+6 \lambda _p^2 \lambda _s^2+\lambda_s^4\right)
 \\
&-16 m_{\phi }^2 \left(2 \lambda _p^2+\lambda _s^2\right) \left(\lambda _p^2+2 \lambda _s^2\right)\bigg\}+4 m_f^5 m_{\psi } \left(9m_{\psi }^2+13 m_{\phi }^2\right) \left(\lambda _p^4-\lambda _s^4\right)
\\
&+m_f^4 \bigg\{-9 m_{\psi }^4 \left(\lambda _p^4-6 \lambda _p^2 \lambda_s^2+\lambda _s^4\right)+4 m_{\psi }^2 m_{\phi }^2 \left(9 \lambda _p^4+44 \lambda _p^2 \lambda _s^2+9 \lambda _s^4\right)
\\
&+m_{\phi }^4 \left(41\lambda _p^4+130 \lambda _p^2 \lambda _s^2+41 \lambda _s^4\right)\bigg\} -2 m_f^3 m_{\psi } \left(m_{\psi }^2+5 m_{\phi }^2\right) \left(9 m_{\psi}^2+5 m_{\phi }^2\right) \left(\lambda _p^4-\lambda _s^4\right)
\\
&+m_f^2 \bigg\{9 m_{\psi }^6 \left(\lambda _p^2-\lambda _s^2\right)^2-5 m_{\psi }^2m_{\phi }^4 \left(7 \lambda _p^4+34 \lambda _p^2 \lambda _s^2+7 \lambda _s^4\right)+4 m_{\psi }^4 m_{\phi }^2 \left(11 \lambda _p^4-24 \lambda _p^2\lambda _s^2+11 \lambda _s^4\right)
\\
&-2 m_{\phi }^6 \left(11 \lambda _p^4+46 \lambda _p^2 \lambda _s^2+11 \lambda _s^4\right)\bigg\}+16 m_f m_{\psi }m_{\phi }^4 \left(4 m_{\psi }^2+m_{\phi }^2\right) \left(\lambda _p^4-\lambda _s^4\right)+9 m_f^8 \left(\lambda _p^2+\lambda _s^2\right)^2
\\
&+4m_{\phi }^4 \bigg\{m_{\psi }^4 \left(-5 \lambda _p^4+18 \lambda _p^2 \lambda _s^2-5 \lambda _s^4\right)+2 m_{\psi }^2 m_{\phi }^2 \left(\lambda_p^4+6 \lambda _p^2 \lambda _s^2+\lambda _s^4\right)+m_{\phi }^4 \left(\lambda _p^4+6 \lambda _p^2 \lambda _s^2+\lambda_s^4\right)\bigg\}\Bigg] \, .
}
In the limit that $n_c \to3$, $m_f \to 0$, and taking all couplings equal to $\lambda$, we have
\beq 
\langle \sigma v \rangle \simeq \frac{9 \lambda^4v^2}{4\pi} \frac{m_\phi^2}{\pL m_\phi^2 +m_\psi^2 \pR^2} \, .
\eeq
Taking instead the low-velocity limit, we can integrate out the mediator. We find
\beq\label{tcmxFdd}
 \bar \psi ( 1 \pm \gamma^5 ) f \phi^\dagger  +   \bar f  ( 1 \mp  \gamma^5 ) \psi  \phi  \to -\frac2{m_\psi} \bar f  f \phi^\dagger \phi,
\eeq
which mediates spin-independent scattering.

%\noindent\makebox[\linewidth]{\rule{\paperwidth}{0.5pt}}
\subsection{Real Scalar dark matter, spin half mediator} 
\label{trealF}

Consider the following interactions for a real scalar dark matter particle, $\phi$, and a spin-$1/2$ mediator, $\psi$:
\beq
\mathcal{L} \supset  \left[ \bar{\psi} \left( \lambda_s + \lambda_p \gamma^5 \right) f  +  \bar{f} \left( \lambda_s - \lambda_p \gamma^5 \right) \psi \right] \phi \, .
\eeq
The annihilation cross section at low relative velocity is given by
\alg{
\sigma v \approx& \frac{n_c \left(1-m_f^2/m_{\phi }^2\right)^{3/2} }{\pi  \left(-m_f^2+m_{\psi }^2+m_{\phi }^2\right)^2} \Big[\lambda _p^2 \left(m_f-m_{\psi }\right)+\lambda _s^2 \left(m_f+m_{\psi}\right)\Big]^2
\\
&+\frac{n_c  \sqrt{1-m_f^2/m_{\phi }^2} v^2}{24 \pi  m_{\phi }^2 \left(-m_f^2+m_{\psi }^2+m_{\phi }^2\right)^4}  \Big[\lambda _p^2 \left(m_f-m_{\psi }\right)+\lambda _s^2 \left(m_f+m_{\psi }\right)\Big] \times
\\
&\Bigg[ 8 m_{\phi }^2 \left(m_{\phi }^2-m_f^2\right)\bigg\{m_f^2 m_{\psi } \left(\lambda _s^2-\lambda _p^2\right)-2 m_f \left(2 m_{\psi }^2+m_{\phi }^2\right) \left(\lambda _p^2+\lambda _s^2\right)+2m_f^3 \left(\lambda _p^2+\lambda _s^2\right)
\\
&+m_{\psi } \left(3 m_{\psi }^2+m_{\phi }^2\right) \left(\lambda _p^2-\lambda _s^2\right)\bigg\} +9 m_f^2\left(-m_f^2+m_{\psi }^2+m_{\phi }^2\right)^2 \bigg\{\lambda _p^2 \left(m_f-m_{\psi }\right)+\lambda _s^2 \left(m_f+m_{\psi}\right)\bigg\}\Bigg].
}
In the limit that $m_f \to 0$ and with all couplings are equal, we find that this velocity-averaged annihilation cross section is completely suppressed to second order in $v$. This is a unique $d$-wave annihilation cross section.

Taking instead the low-velocity limit, we can integrate out the mediator as in \Eq{tcmxFdd} to find the direct detection cross section. We find the same result as above:
\beq
 \bar \psi ( 1 \pm \gamma^5 ) f \phi^\dagger  +   \bar f  ( 1 \mp  \gamma^5 ) \psi  \phi  \to -\frac2{m_\psi} \bar f  f \phi^\dagger \phi \, .
\eeq

%\noindent\makebox[\linewidth]{\rule{\paperwidth}{0.5pt}} \textcolor{ForestGreen}{\checkmark}
\subsection{Complex Vector dark matter, spin half mediator} 
\label{tcvF}

Consider the following interactions for a complex vector dark matter particle, $X_\mu$, and a spin-$1/2$ mediator, $\psi$:
\beq
\mathcal{L} \supset  \bar{\psi} \gamma^\mu \left( g_v+g_a \gamma^5 \right) f X_\mu^\dagger + \bar{f}  \gamma^\mu \left( g_v+g_a \gamma^5 \right) \psi X_\mu \, .
\eeq
The annihilation cross section at low relative velocity is given by
\alg{
\sigma v \approx& \frac{n_c \left(1-m_f^2/m_X^2\right)^{3/2} }{36 \pi  \left(-m_f^2+m_X^2+m_{\psi }^2\right)^2} \Bigg[-2 g_a^2 g_v^2 \left(3 m_f^2-12 m_X^2+5 m_{\psi }^2\right)
\\
&+g_a^4 \left(6 m_f m_{\psi }+5m_f^2+4 m_X^2+5 m_{\psi }^2\right)+g_v^4 \left(-6 m_f m_{\psi }+5 m_f^2+4 m_X^2+5 m_{\psi }^2\right)\Bigg]
\\
&+\frac{n_c v^2 \sqrt{1-m_f^2/m_X^2} }{864 \pi  m_X^2 \left(-m_f^2+m_X^2+m_{\psi}^2\right)^4} \Bigg[3 m_{\psi }^6 \left(g_a^2-g_v^2\right)^2\left(15 m_f^2+16 m_X^2\right)
\\
&+6 m_f m_{\psi }^5 \left(g_a^4-g_v^4\right) \left(13 m_f^2-16 m_X^2\right)+m_{\psi }^4 \bigg\{2 g_a^2 g_v^2 \left(-452m_f^2 m_X^2+87 m_f^4+500 m_X^4\right)
\\
&+g_a^4 \left(80 m_f^2 m_X^2-37 m_f^4+92 m_X^4\right)+g_v^4 \left(80 m_f^2 m_X^2-37 m_f^4+92m_X^4\right)\bigg\}
\\
&-12 m_f m_{\psi }^3 \left(g_a^4-g_v^4\right) \left(-37 m_f^2 m_X^2+13 m_f^4+24 m_X^4\right)
\\
&-m_{\psi }^2 \left(m_f^2-m_X^2\right)\bigg\{-m_f^2 m_X^2 \left(658 g_a^2 g_v^2+239 g_a^4+239 g_v^4\right)
\\
&+m_f^4 \left(78 g_a^2 g_v^2+61 g_a^4+61 g_v^4\right)+8 m_X^4 \left(50 g_a^2g_v^2+11 g_a^4+11 g_v^4\right)\bigg\}
\\
&+6 m_f m_{\psi } \left(g_a^4-g_v^4\right) \left(13 m_f^2-8 m_X^2\right)\left(m_f^2-m_X^2\right)^2
\\
&+\left(m_f^2-m_X^2\right)^2 \bigg\{-2 m_f^2 m_X^2 \left(66 g_a^2 g_v^2+41 g_a^4+41 g_v^4\right)
\\
&+m_f^4 \left(-6 g_a^2g_v^2+53 g_a^4+53 g_v^4\right)+20 m_X^4 \left(6 g_a^2 g_v^2+g_a^4+g_v^4\right)\bigg\}\Bigg] \, .
}
In the limit that $n_c \to3$, $m_f \to 0$, and taking all couplings equal to $\lambda$, we have
\beq
\langle \sigma v \rangle \simeq \frac{8\lambda^4}{3\pi} \frac{m_X^2}{\pL m_X^2 +m_\psi^2 \pR^2} \bL 1+ \frac{3v^2}{32}\frac{37+ 18m_X^2/m_\psi^2 +5 m_X^4/m_\psi^4}{\pL 1+m_X^2/m_\psi^2 \pR^2} \bR \, .
\eeq
Taking instead the low-velocity limit, we can integrate out the mediator. We find
\beq \label{tcvFdd}
 \bar \psi \gamma^\mu ( 1 \pm \gamma^5 ) f X_\mu^\dagger  +\bar f   \gamma^\mu ( 1 \pm  \gamma^5 ) \psi  X_\mu \to \frac2{m_\psi} \bar f  f X^\dagger_\mu X^\mu,
\eeq
which mediates spin-independent scattering.

%\noindent\makebox[\linewidth]{\rule{\paperwidth}{0.5pt}} \textcolor{ForestGreen}{\checkmark
\subsection{Real Vector dark matter, spin half mediator} 
\label{trvF}

Consider the following interactions for a real vector dark matter particle, $X_\mu$, and a spin-$1/2$ mediator, $\psi$:
\beq
\mathcal{L} \supset  \left[ \bar{\psi} \gamma^\mu \left( g_v+g_a \gamma^5 \right) f + \bar{f}  \gamma^\mu \left( g_v+g_a \gamma^5 \right) \psi \right] X_\mu \, .
\eeq
The annihilation cross section at low relative velocity is given by
\alg{
\sigma v \approx& \frac{n_c \left(1-m_f^2/m_X^2\right)^{3/2}}{9 \pi  \left(-m_f^2+m_X^2+m_{\psi }^2\right)^2} \Bigg[-2 g_a^2 g_v^2 \bigg\{5 m_f^2+3 \left(m_{\psi }^2-4 m_X^2\right)\bigg\}+g_a^4 \left(2 m_fm_{\psi }+3 m_f^2+4 m_X^2+3 m_{\psi }^2\right)
\\
&+g_v^4 \left(-2 m_f m_{\psi }+3 m_f^2+4 m_X^2+3 m_{\psi }^2\right)\Bigg]
\\
&+\frac{n_c v^2 \sqrt{1-m_f^2/m_X^2}}{216\pi  m_X^2 \left(-m_f^2+m_X^2+m_{\psi }^2\right)^4} \Bigg[3 m_{\psi }^6 \left(g_a^2-g_v^2\right)^2\left(m_f^2+8 m_X^2\right)-6 m_f m_{\psi }^5 \left(g_a^4-g_v^4\right) \left(m_f^2-4 m_X^2\right)
\\
&+m_{\psi }^4 \bigg\{-2 g_a^2 g_v^2 \left(46 m_f^2m_X^2+39 m_f^4-160 m_X^4\right)+5 g_a^4 \left(14 m_f^2 m_X^2+m_f^4\right)+5 m_f^2 g_v^4 \left(m_f^2+14 m_X^2\right)\bigg\}
\\
&+12 m_f m_{\psi }^3\left(g_a^4-g_v^4\right) \left(-3 m_f^2 m_X^2+m_f^4+2 m_X^4\right)
-m_{\psi }^2 \left(m_f^2-m_X^2\right) \bigg\{m_f^2 m_X^2 \left(474 g_a^2 g_v^2-29g_a^4-29 g_v^4\right)
\\
&+m_f^4 \left(-174 g_a^2 g_v^2+19 g_a^4+19 g_v^4\right)-80 m_X^4 \left(6 g_a^2 g_v^2+g_a^4+g_v^4\right)\bigg\}
\\
&-2 m_f m_{\psi }\left(g_a^4-g_v^4\right) \left(3 m_f^2-32 m_X^2\right) \left(m_f^2-m_X^2\right)^2+\left(m_f^2-m_X^2\right)^2 \bigg\{8 m_f^2 m_X^2 \left(44 g_a^2g_v^2+g_a^4+g_v^4\right)
\\
&+m_f^4 \left(-90 g_a^2 g_v^2+11 g_a^4+11 g_v^4\right) -56 m_X^4 \left(6 g_a^2 g_v^2+g_a^4+g_v^4\right)\bigg\}\Bigg] \, .
}
In the limit that $n_c \to3$, $m_f \to 0$, and taking all couplings equal to $\lambda$, we have
\beq
\langle \sigma v \rangle \simeq \frac{32\lambda^4}{3\pi} \frac{m_X^2}{\pL m_X^2 +m_\psi^2 \pR^2} \bL 1+ \frac{3 v^2}{16}\frac{5 - 10m_X^2/m_\psi^2 -7m_X^4/m_\psi^4}{\pL 1+ m_X^2/m_\psi^2 \pR^2} \bR
\eeq
Taking instead the low-velocity limit, we can integrate out the mediator as in \Eq{tcvFdd}. We find
\beq
 \bar \psi \gamma^\mu ( 1 \pm \gamma^5 ) f X_\mu^\dagger  +   \bar f   \gamma^\mu ( 1 \mp  \gamma^5 ) \psi  X_\mu  \to \frac2{m_\psi} \bar f  f X^\mu X_\mu,
\eeq
which, as above, mediates spin-independent scattering.

\end{widetext}

\section{Relic Abundance}
\label{relic}

The abundance of dark matter particles which survive the big bang as a thermal relic is found by solving the Boltzmann equation:
\begin{equation}
\frac{dn}{dt}=-3Hn-\langle \sigma v \rangle [n^2-n^2_{\rm eq}],
\label{boltzmann}
\end{equation}
where $H=\dot{a}/a$ is the Hubble parameter, $n$ is the number density of dark matter particles, and $n_{\rm eq}$ is the equlibrium number density. In the non-relativistic limit, $n_{\rm eq}=g (mT/2\pi)^{3/2}\, \exp(-m/T)$, where $g$ is the number of internal degrees of freedom of the DM particle, $m$ is the mass of the DM particle, and $T$ is the temperature. 

The solution to Eq.~\ref{boltzmann} yields a present-day dark matter abundance given by:
\begin{equation}
\Omega h^2 \cong \frac{1.07 \times 10^9 \,{\rm GeV}^{-1}}{J \, g^{1/2}_{\star} \, m_{\rm Pl}},
\end{equation}
where $h$ is present-day Hubble parameter (in units of 100 km/s/Mpc), $m_{\rm Pl}=1.22\times 10^{19}$ GeV, and $g_{\star}$ is the number of effective relativistic degrees-of-freedom at the time of freeze-out (for $g_{\star}$, we adopt the values given in Refs.~\cite{Steigman:2012nb,2006PhRvD..73h5009L}). The quantity $J$ is given by:
\begin{equation}
J=\int^\infty_{x_f} \frac{\langle \sigma v \rangle}{x^2} dx,
\end{equation}
where $x\equiv m/T$ and $x_f$ is the value at the freeze-out temperature, which is found by iterating the following:
\begin{equation}
x_f= \ln \frac{0.038 g m_{\rm Pl} m \langle \sigma v \rangle}{g^{1/2}_{\star}x^{1/2}_f}.
\end{equation}

When not near a resonance, we can expand the annihilation cross section in a Taylor series, $\sigma v  \simeq a + b v^2+\cO(v^4)$, yielding the following relic abundance:
\begin{equation}
\Omega h^2 \cong \frac{x_f \, 1.07 \times 10^9 \,{\rm GeV}^{-1}}{g^{1/2}_{\star} \, m_{\rm Pl} \, (a+3b/x_f)}.
\end{equation}
Near a resonance, however, the Taylor series expansion breaks down, and we instead determine $J(x_f)$ by solving the integrals in \Eq{GGEqs} numerically~\cite{Griest:1990kh}.

\end{document}